\documentclass[11pt]{amsart}
\usepackage{amsmath}
\usepackage{amsxtra}
\usepackage{amscd}
\usepackage{amsthm}
\usepackage{amsfonts}
\usepackage{amssymb}
\usepackage{eucal}
\usepackage{epsfig}

\usepackage{latexsym,amsthm}

\DeclareFontEncoding{OT2}{}{}

\input epsf

\def\figin{\epsfcheck\figin}\def\figins{\epsfcheck\figins}
\def\epsfcheck{\ifx\epsfbox\UnDeFiNeD
\message{(NO epsf.tex, FIGURES WILL BE IGNORED)}
\gdef\figin##1{\vskip2in}\gdef\figins##1{\hskip.5in}
\else\message{(FIGURES WILL BE INCLUDED)}%
\gdef\figin##1{##1}\gdef\figins##1{##1}\fi}
\def\DefWarn#1{}
\def\figinsert{\goodbreak\topinsert}
\def\ifig#1#2#3#4{\DefWarn#1\xdef#1{fig.~\the\figno}
\writedef{#1\leftbracket fig.\noexpand~\the\figno}%
\figinsert\figin{\centerline{\epsfxsize=#3mm \epsfbox{#2}}}
\bigskip\medskip\centerline{\vbox{\baselineskip12pt
\advance\hsize by -1truein\noindent\footnotefont{\sl Fig.~\the\figno:}\sl\ #4}}
\bigskip\endinsert\noindent\global\advance\figno by1}

\theoremstyle{remark}

\theoremstyle{definition}

\numberwithin{equation}{section}

\newcommand{\Ref}[1]{{$($\ref{#1}$)$}}
\newcommand{\bean}{\begin{eqnarray}}
\newcommand{\eean}{\end{eqnarray}}
\newcommand{\be}{\begin{displaymath}}
\newcommand{\ee}{\end{displaymath}}
\newcommand{\bea}{\begin{eqnarray*}}
\newcommand{\eea}{\end{eqnarray*}}

\newcommand{\secref}[1]{Section~\ref{#1}}

\newcommand{\nc}{\newcommand}
\nc{\on}{\operatorname}
\nc{\p}{{\partial}}
\nc{\pa}{{\p}}
\nc{\ch}{\mbox{ch}}
\nc{\Z}{{\mathbb Z}}
\nc{\C}{{\mathbb C}}
\nc{\pone}{{\mathbb C\mathbb P}^1}

\nc{\CA}{{\mathcal A}}
\nc{\CB}{{\mathcal B}}
\nc{\CC}{{\mathcal C}}

\nc{\CE}{{\mathcal E}}
\nc{\CF}{{\mathcal F}}
\nc{\CG}{{\mathcal G}}
\nc{\CH}{{\mathcal H}}
\nc{\CI}{{\mathcal I}}
\nc{\CJ}{{\mathcal J}}
\nc{\CK}{{\mathcal K}}
\nc{\CL}{{\mathcal L}}
\nc{\CM}{{\mathcal M}}
\nc{\CN}{{\mathcal N}}
\nc{\CO}{{\mathcal O}}
\nc{\CP}{{\mathcal P}}
\nc{\CQ}{{\mathcal Q}}
\nc{\CR}{{\mathcal R}}
\nc{\CS}{{\mathcal S}}
\nc{\CT}{{\mathcal T}}
\nc{\CU}{{\mathcal U}}
\nc{\CV}{{\mathcal V}}
\nc{\CW}{{\mathcal W}}
\nc{\CX}{{\mathcal X}}
\nc{\CY}{{\mathcal Y}}
\nc{\CZ}{{\mathcal Z}}

\nc{\zb}{\ol{z}}
\nc{\xb}{{\bar x}}
\nc{\yb}{{\bar y}}
\nc{\ub}{{\bar u}}
\nc{\pb}{\bar\partial}
\nc{\wb}{\ol{w}}
\nc{\qb}{\ol{q}}
\nc{\nb}{\ol{n}}
\nc{\phb}{\ol\ph}

\nc{\ph}{p}

\nc{\bi}{{\bf i}}
\nc{\bj}{{\bf j}}
\nc{\bk}{{\bf k}}
\nc{\bq}{{\bf q}}
\nc{\bv}{{\bf v}}

\nc{\dirac}{D\hspace*{-2.5mm}\slash}

\nc{\al}{\alpha}
\nc{\bt}{{\beta}}
\nc{\dl}{{\delta}}
\nc{\la}{\lambda}
\nc{\m}{\mu}
\nc{\ep}{\epsilon}
\nc{\si}{\sigma}
\nc{\om}{\omega}

\nc{\De}{\Delta}
\nc{\Ga}{\Gamma}
\nc{\La}{\Lambda}

\nc{\el}{\ell}

\nc{\arr}{\rightarrow}
\nc{\larr}{\longrightarrow}
\nc{\ri}{\rangle}
\nc{\lef}{\langle}
\nc{\su}{\widehat{{\mathfrak s}{\mathfrak l}}_2}
\nc{\sw}{{\mathfrak s}{\mathfrak l}}
\nc{\g}{{\mathfrak g}}
\nc{\h}{{\mathfrak h}}
\nc{\n}{{\mathfrak n}}
\nc{\N}{\widehat{\n}}
\nc{\G}{\widehat{\g}}
\nc{\gt}{\widetilde{\g}}
\nc{\one}{{\mathbf 1}}
\nc{\z}{{\mathfrak Z}}
\nc{\wt}{\widetilde}
\nc{\wh}{\widehat}
\nc{\cri}{_{\kappa_c}}
\nc{\kk}{h^\vee}
\nc{\sun}{\widehat{\sw}_N}
\nc{\ol}{\overline}
\nc{\ds}{\displaystyle}
\nc{\dzz}{\frac{dz}{z}}
\nc{\Res}{\on{Res}}
\nc{\mc}{\mathcal}
\nc{\Cal}{\mathcal}
\nc{\bb}{{\mathfrak b}}
\nc{\ot}{\otimes}
\nc{\R}{{\mathbb R}}
\nc{\yy}{{\mc Y}}
\nc{\ga}{\gamma}

\nc{\us}{\underset}
\nc{\opl}{\oplus}
\nc{\beq}{\begin{equation}}
\nc{\Rep}{\on{Rep}}
\nc{\sssec}{\subsubsection}
\nc{\ssec}{\subsection}
\nc{\lan}{\langle}
\nc{\ran}{\rangle}

\nc{\Vect}{\on{Vect}}
\nc{\ghat}{\G}
\nc{\T}{\mc T}
\nc{\Tloc}{\T^\g_{\on{loc}}}
\nc{\vac}{|0\ran}
\nc{\Wick}{{\mb :}}
\nc{\mb}{\mathbf}
\nc{\delz}{\partial_z}
\nc{\cali}{\mathcal}
\nc{\li}{\mathfrak l}
\nc{\lt}{\widetilde{\li}}
\nc{\astar}{a^*}
\nc{\cA}{{\mc A}}
\nc{\ka}{\kappa}

\nc{\OO}{{\mc O}}
\nc{\AutO}{\on{Aut} O}
\nc{\DerO}{\on{Der} O}
\nc{\DerpO}{\on{Der}_+ O}
\nc{\Au}{{\mc A}ut}
\nc{\mf}{\mathfrak}
\nc{\V}{{\mc V}}
\nc{\hh}{\wh{\h}}

\nc{\pp}{{\mathfrak p}}
\nc{\mm}{{\mathfrak m}}
\nc{\rr}{{\mathfrak r}}
\nc{\ket}{\rangle}
\nc{\zz}{{\mathfrak z}}
\nc{\gr}{\on{gr}}
\nc{\Spe}{\on{Spec}}
\nc{\rv}{\crho}
\nc{\can}{\on{can}}
\nc{\MOp}{\on{MOp}_G(D)}
\nc{\Db}{{\mathbb D}}
\nc{\ww}{w}

\nc{\Con}{\on{Conn}(\Omega^{\crho})_D}
\nc{\ConD}{\on{Conn}(\Omega^{\crho})_{\Db}}
\nc{\ConDL}{\on{Conn}(\Omega^{\rho})_{\Db}}
\nc{\ConDtL}{\on{Conn}(\Omega^{\rho})_{\Db^\times}}
\nc{\OpD}{\on{Op}_G(\Db)}
\nc{\crho}{\check{\rho}}
\nc{\chal}{\check{\al}}
\nc{\cchi}{\check{\chi}}
\nc{\cLa}{\check\Lambda}
\nc{\cla}{\check\la}
\nc{\cmu}{\check\mu}

\nc{\PP}{{\mathbb P}}
\nc{\TT}{{\mathbb T}}

\nc{\bone}{{\mb 1}}

\nc{\bs}{\backslash}

\def\tr{{\rm tr}}

\nc{\zzb}{z \zb}

\nc{\pf}{\int\hspace*{-3.5mm}\bs}

\nc{\inn}{\on{in}}
\nc{\out}{\on{out}}
\nc{\covac}{\langle 0|}
\nc{\ptwo}{{\mathbb C}{\mathbb P}^2}
\nc{\BS}{{\mathbb S}}

\begin{document}

\title{Instantons beyond topological theory I}\thanks{Supported by the
DARPA grant HR0011-04-1-0031}

\author{E. Frenkel}

\address{Department of Mathematics, University of California,
       Berkeley, CA 94720, USA}

\author{A. Losev}

\address{Institute of Theoretical and Experimental Physics,
       B. Cheremushkinskaya 25, Moscow 117259, Russia}

\author{N. Nekrasov}

\address{Institut des Hautes \'Etudes Scientifiques, 35, Route de
Chartres, Bures-sur-Yvette, F-91440, France}

\centerline{\hfill ITEP-TH-40/06 \ IHES-P/06/42}
\centerline{\hphantom{void}}
\centerline{\hphantom{void}}

\date{October 2006}

\begin{abstract}
Many quantum field theories in one, two and four dimensions possess
remarkable limits in which the instantons are present, the
anti-instantons are absent, and the perturbative corrections are
reduced to one-loop. We analyze the corresponding models as full
quantum field theories, beyond their topological sector. We show that
the correlation functions of all, not only topological (or BPS),
observables may be studied explicitly in these models, and the
spectrum may be computed exactly. An interesting feature is that the
Hamiltonian is not always diagonalizable, but may have Jordan blocks,
which leads to the appearance of logarithms in the correlation
functions. We also find that in the models defined on K\"ahler
manifolds the space of states exhibits holomorphic factorization. We
conclude that in dimensions two and four our theories are logarithmic
conformal field theories.

In Part I we describe the class of models under study and present our
results in the case of one-dimensional (quantum mechanical) models,
which is quite representative and at the same time simple enough to
analyze explicitly. Part II will be devoted to supersymmetric
two-dimensional sigma models and four-dimensional Yang-Mills
theory. In Part III we will discuss non-supersymmetric models.
\end{abstract}

\maketitle

\tableofcontents

\newpage

\vspace*{20mm}

\section{Introduction}

\bigskip

For a large class of models of quantum field theory there is a
particular limit in which the theory may be analyzed exactly in the
presence of instanton effects. The simplest are the (twisted)
supersymmetric models, in which the path integral measure is defined
in a straightforward way. Classically, these models are described by
first order Lagrangians. The corresponding path integral localizes on
certain finite-dimensional moduli spaces of classical (instanton)
configurations. Therefore in the path integral description the
correlation functions of the corresponding quantum system are given by
integrals over these moduli spaces. Such correlation functions have
been studied in the literature, but attention has been focused almost
exclusively on the correlation functions of the {\em BPS observables},
which represent cohomology of a supersymmetry charge of the
theory. These correlation functions comprise the BPS (or topological)
sector of the model and give rise to important invariants, such as the
Gromov-Witten and Donaldson invariants (in two and four dimensions,
respectively). However, the knowledge of the topological sector is not
sufficient for understanding the full quantum field theory.

In this paper we go beyond the topological field theory of these
models and investigate the correlation functions of more general --
non-BPS, or ``off-shell'' -- observables in the presence of
instantons, i.e., non-perturbatively. We show that in the special
limit of the coupling constant that we are considering (namely,
$\ol\tau \to \infty$, see below) the quantum model may be analyzed and
solved explicitly, both in the Lagrangian (or path integral) formalism
and the Hamiltonian formalism. We describe the space of states of the
quantum theory and show that a large class of observables (satisfying
certain analytic properties) may be realized as operators acting on
this space. Their correlation functions are then represented by the
matrix elements of these operators. These matrix elements agree with
the path integral representation of the correlation functions (given
by integrals over the moduli spaces of instantons), and they also
satisfy the usual identities, such as factorization over the
intermediate states.

We find some interesting and unexpected features in our models. One of
them is the fact that the Hamiltonian is {\em non-diagonalizable} on
the space of states, but has Jordan blocks. This leads to the
appearance of logarithmic terms in the correlation functions. Another
feature is {\em holomorphic factorization} of the space of states in
models defined on K\"ahler manifolds. In particular, we find that
two-dimensional supersymmetric sigma models and four-dimensional
super-Yang-Mills models are {\em logarithmic conformal field theories}
in our limit.

\subsection{Description of the models}

We begin by describing in more detail the class of models that will be
discussed in this paper. These models appear in one, two and four
space-time dimensions and are described by the actions which are
written below (note that all of our actions are written in Euclidean
signature).

\medskip

\noindent $\bullet$ {\bf 1D}: We start with the supersymmetric quantum
mechanics on a compact K\"ahler manifold $X$, with the K\"ahler metric
$\lambda g$, equipped with a holomorphic vector field $\xi$. We will
assume that $\xi$ comes from a holomorphic $\C^\times$-action on $X$
with a non-empty set of fixed points which are all isolated. We modify
the standard action \cite{W:morse} by adding the topological term $- i
\vartheta \int A$, where $A$ is the one-form obtained by contracting
$g$ with the vector field $\xi+\ol\xi$ (our assumptions imply that
$A=df$, where $f$ is a Morse function on $X$). We allow $\vartheta$ to
be complex.

We then set $\tau = \vartheta + i\la, \ol\tau = \vartheta - i\la$
(note that they are not necessarily complex conjugate to each other
because $\vartheta$ is complex). Consider the limit in which $\ol\tau
\to -i\infty$, but $\tau$ is kept finite (which means that $\la \to
+\infty$ and $\vartheta$ is adjusted accordingly, so that it has a
large imaginary part). In this limit the model is described by the
following first order action on a worldline $I$:
\begin{multline}    \label{1D action}
S = -i \int_{I} \left( p_a \left( \frac{dX^a}{dt} - v^a \right) +
\ol{p_a} \left(\frac{d\ol{X^a}}{dt} - \ol{v^a} \right) + \right. \\
\left. - \pi_a \left( \frac{d \psi^a}{dt} - \frac{\pa v^a}{\pa X^b}
\psi^b \right) -
\ol{\pi_a} \left( \frac{d\ol{\psi^a}}{dt} - \frac{\pa \ol{v^a}}{\pa
\ol{X^b}} \ol{\psi^b} \right) \right) dt - i \tau \int A,
\end{multline}
where the $X^a$'s are complex coordinates on $X$ and $\xi = v^a
\frac{\pa}{\pa X^a}$.

The quantum model is described by the path integral $\int e^{-S}$ over
all maps $I \to X$. This path integral represents the ``delta-form''
supported on the moduli space of {\em gradient trajectories},
satisfying
\begin{equation}
\label{gradient}
\frac{dX^a}{dt} = v^a.
\end{equation}
This is the instanton moduli space in this case.

\medskip

\noindent $\bullet$ {\bf 2D}: We start with the twisted (type A)
${\mathcal N}=(2,2)$ supersymmetric sigma model with the target
compact K\"ahler manifold $X$ with the K\"ahler metric $\lambda g$ and
the $B$-field $B = B_{a\ol{b}} dX^a \wedge dX^{\ol{b}}$, which is a
closed (complex) two-form on $X$.

We then set
$$
\tau_{a\ol{b}} = B_{a\ol{b}} + \frac{i}{2} \la g_{a\ol{b}}, \qquad
\ol\tau_{a\ol{b}} = B_{a\ol{b}} - \frac{i}{2} \la g_{a\ol{b}}.
$$
In the limit when $\ol\tau_{a\ol{b}} \to -i\infty$, but the
$\tau_{a\ol{b}}$ are kept finite (which means that $\la \to \infty$
and the $B_{a\ol{b}}$ have a large imaginary part), this model is
described by the following first order action on a worldsheet
$\Sigma$:
\begin{equation}    \label{2D action}
- i \int_{\Sigma} \left( p_a \pa_{\ol{z}} X^{a} +
\ol{p_{a}} \pa_z \ol{X^a} - \pi_a \pa_{\ol{z}} \psi^{a} -
\ol{\pi_a} \pa_{z} \ol{\psi^a} \right) d^2 z + \int_\Sigma
\tau_{a\ol{b}} dX^a \wedge d\ol{X^b}.
\end{equation}
Thus, this model is a particular modification of the ``infinite radius
limit'', achieved by adding to the conventional second order action of
the sigma model the topological term ($B$-field) with a large
imaginary part.

The path integral $\int e^{-S}$ over all maps $\Sigma \to X$ localizes
on the moduli space of {\em holomorphic maps}, satisfying
$$
\pa_{\ol{z}} X^a = 0.
$$

\medskip

\noindent $\bullet$ {\bf 4D}: We start with twisted ${\mc N} = 2$
supersymmetric gauge theory on a four-manifold ${\mb M}^4$ with
compact gauge group $G$, coupling constant $g_{\on{YM}}$ and
theta-angle $\vartheta$, which we allow to be complex.

We then set
$$
\tau = \dfrac{\vartheta}{2\pi} + \dfrac{4\pi
i}{g_{\on{YM}}^2}, \qquad \ol\tau = \dfrac{\vartheta}{2\pi} -
\dfrac{4\pi i}{g_{\on{YM}}^2}.
$$
In the limit when $\ol\tau \to -i\infty$, but
$\tau$ is kept finite (i.e., $g_{\on{YM}} \to 0$ and $\vartheta$ has a
large imaginary part), this model is described by the
first order action
$$
S = - i \int_{{\mb M}^4} \left( {\tr} P^{+} \wedge F_A + {\tau}\
{\tr} F_A \wedge F_A + \on{fermions} \right).
$$
This model is a particular modification of the ``weak
coupling limit'' achieved by adding to the conventional second order
the topological term $- \frac{i \vartheta}{2\pi} \int_{{\bf
M}^{4}}{\tr} F \wedge F$, where $\vartheta$ has a large imaginary
part.

The path integral $\int e^{-S}$ over all connections on ${\mb M}^4$
localizes on the moduli space of {\em anti-self dual connections}
satisfying the equations
$$
F_A^+ = 0.
$$

\bigskip

Thus, to obtain these models we start with the standard (second order)
action and add to it a {\em topological term}: $- i \vartheta \int_I
df$ in 1D, the B-field $\int_\Sigma B_{a\ol{b}} dX^a \wedge
dX^{\ol{b}}$ in 2D, and the $\vartheta$-angle term $- \frac{i
\vartheta}{2\pi} \int_{{\bf M}^{4}}{\tr} F \wedge F$ in 4D. We then
allow $\vartheta$ and $B_{a\ol{b}}$ to be complex and take the limit
$\ol{\tau} \to \infty$ as described above.\footnote{in 2D sigma models
such limits, with large imaginary $B$-field, have been studied in the
literature since the early days of the theory of instantons, see,
e.g., the papers \cite{David} and references therein} In this limit
the instanton contributions are present while the anti-instanton
contributions vanish, and because of that the correlation functions
simplify dramatically.

The resulting models are described by the first order Lagrangians
written above.\footnote{there are also similar models in three
dimensions, but they fall into the class of non-supersymmetric field
theories, which generically become massive upon inclusion of the
instantons; in this paper we consider such models only briefly in
\secref{nonSUSY}} The corresponding path integral represents the
``delta-form'' supported on the instanton moduli space. This moduli
space has components labeled by the appropriate ``instanton numbers'',
which are finite-dimensional (after dividing by the appropriate gauge
symmetry group).  Therefore the correlation functions are expressed in
terms of integrals over these finite-dimensional components of the
moduli space.

When we move away from the special point $\ol\tau = \infty$ (with
fixed $\tau$), both instantons and anti-instantons start contributing
to the correlation functions. The path integral becomes a
Mathai-Quillen representative of the Euler class of an appropriate
vector bundle over the instanton moduli space, which is ``smeared''
around the moduli space of instantons (like Gaussian distribution),
see, e.g., \cite{Moore}. Therefore general correlation functions are
no longer represented by integrals over the finite-dimensional
instanton moduli spaces and become much more complicated.

There is however an important class of observables, called the {\em
BPS observables} whose correlation functions are independent of
$\ol\tau$.  They commute with the supersymmetry charge $Q$ of the
theory and comprise the {\em topological sector} of the theory. The
perturbation away from the point $\ol\tau = \infty$ (that is back to a
finite radius in 2D or to a non-zero coupling constant in 4D) is given
by a $Q$-exact operator, and therefore the correlation functions of
the BPS observables (which are $Q$-closed) remain unchanged when we
move away from the special point. This is the secret of success of the
computation of the correlation functions of the BPS observables
achieved in recent years in the framework of topological field theory:
the computation is actually done in the theory at $\ol\tau = \infty$,
but because of the special properties of the BPS observables the
answer remains the same for other values of the coupling constant
\cite{W:mirror}. But for general observables the correlation functions
do change in a rather complicated way when we move away from the
special point.

Our goal in this paper is to go {\em beyond the topological sector}
and consider more general correlation functions of non-BPS
observables. We are motivated, first of all, by the desire to
understand non-supersymmetric quantum field theories with
instantons. It is generally believed that realistic quantum field
theories should be viewed as non-supersymmetric phases of
supersymmetric ones. This means that the observables of the original
theory may be realized as observables of a supersymmetric theory, but
they are certainly not going to be BPS observables. Therefore we need
to develop methods for computing correlation functions of such
observables.

In particular (and this was another motivation), developing this
theory in two dimensions may help in elucidating the pure spinor
approach to superstring theory \cite{Berk}, where ``curved
$\beta\gamma$-systems'' play an important role \cite{N:curved}.

The third motivation comes from the realization that the
correspondence between the full quantum field theory and its
topological sector is analogous to the correspondence between a
differential graded algebra (DGA) and its cohomology. The cohomology
certainly contains a lot of useful information about the DGA, but far
from all. For example, there are higher (Massey) operations on the
cohomologies, which can only be detected if we use the full DGA
structure. In particular, the cohomology of a manifold does not
determine its geometric type, but the differential graded algebra of
differential forms does (at least, its rational homotopy
type).\footnote{In the words of \cite{rat-hom}, ``to understand
cohomology and maps on cohomology one need deal only with closed
forms, but to detect the finer homotopy theoretic information one also
needs to use non-closed forms. Differential geometric aspects of this
philosophy have been given by Chern and Simons: ``The manner in which
a closed form which is zero in cohomology actually becomes exact
contains geometric information''.''}  Likewise, we expect that the
passage from the topological sector to the full quantum field theory
will reveal a lot of additional information. In particular, while the
correlation functions in the topological sector give rise to
invariants of the underlying manifold, such as the Gromov-Witten and
Donaldson invariants, the correlation functions of the full quantum
field theory may allow us to detect some finer information about its
geometry.

Since our goal now is to understand the full quantum field theory, and
not just its topological sector, it is reasonable to try to describe
the theory first at the special value of the coupling constant
$\ol\tau = \infty$ (and finite $\tau$), where the correlation
functions simplify so dramatically due to the vanishing of the
anti-instanton contributions. One can then try to extend these results
to a neighborhood of this special value by perturbation theory.

Our approach should be contrasted with the standard perturbation
theory approach to quantum field theory, which consists of expanding
around a Gaussian fixed point. This approach has many virtues, but it
cannot be universally applied. In particular, there are issues with
the zero radius of convergence, but more importantly, such
perturbation theory is unlikely to shed light on hard dynamical
questions such as confinement. One can speculate that the reason for
this is that expansion around a Gaussian point does not adequately
represent the non-linearity of the spaces of fields and symmetries.

The expansion around the point $\ol\tau=\infty$ that we propose in
this paper may be viewed as an alternative to the Gaussian
perturbation theory. Here the topology (and perhaps, even the
geometry) of the space of fields is captured by the appropriate moduli
space of instantons. Therefore this approach may be beneficial for
understanding some of the questions that have proved to be notoriously
difficult in the conventional formalism.

We note that the consideration of the theory at $\ol\tau = \infty$ has
already proved to be very useful in the recent interpretation
\cite{AiB} of mirror symmetry for toric varieties via an intermediate
$I$-model and the recent proof \cite{N:sw} of the Seiberg-Witten
solution of the ${\mathcal N} = 2$ supersymmetric gauge theories.

\subsection{Some puzzles}

Before summarizing our results we wish to motivate them by pointing
out some ``puzzles'' which naturally arise when one considers the
models described above. It is natural to start with the
one-dimensional case of supersymmetric quantum mechanics. It already
contains most of the salient features of the models that we are
interested in, and yet is simple enough to allow us to analyze it
explicitly.

Let us first look at the classical theory described by the action
\eqref{1D action}. We can include it into a one-parameter family of
theories depending on a coupling constant $\la$ by adding the term
$\frac{1}{2} \la^{-1} g_{a\ol{b}} p_a \ol{p_b}$. For finite values of
$\la^{-1}$ we may substitute the corresponding equations of motion and
obtain the second order action
\begin{equation}    \label{second order action}
\int_I \left( \frac{1}{2} \la g_{a\ol{b}} \frac{dX^a}{dt}
\frac{d\ol{X^b}}{dt} + \frac{1}{2} \la g^{a\ol{b}} \frac{\pa f}{\pa
X^a} \frac{\pa f}{\pa \ol{X^b}} - i \vartheta \frac{df}{dt} +
\on{fermions} \right) dt,
\end{equation}
where $\vartheta = \tau - i\la$. Here $f$ is a Morse function on $X$,
whose gradient is equal to $v = \xi + \ol{\xi}$, so that we have
$$
df = g_{a\ol{b}} v^a \ol{d X^b} + g_{\ol{a}b} \ol{v^a} d X^b.
$$
This function is also the hamiltonian of the $U(1)$-vector field
$i(\xi-\ol{\xi})$ with respect to the K\"ahler structure. It is shown
in \cite{Frankel} that $f$ always exists under our
assumption that $\xi$ comes from a holomorphic $\C^\times$-action on
$X$ with a non-empty set of fixed points.

The term $- i \vartheta df$ (which plays the role of the $B$-field of
the two-dimensional sigma model) is very important, as we will see
below. Its role is to distinguish the instanton contributions to the
path integral from the anti-instanton contributions. This allows us to
keep the instantons and at the same time get rid of the
anti-instantons in the limit $\ol\tau = \vartheta - i\la \to -i\infty$
with $\tau = \vartheta + i\la$ being fixed.

The limit $\ol\tau \to -i\infty$ with finite $\tau$ is achieved by
taking $\la \to +\infty$ and $\vartheta \to -i\infty$ in such a way
that $\la - |\vartheta|$ is kept finite and fixed. For simplicity we
will consider now the case when $\vartheta = - i \la$ (so that
$\tau=0$). We will therefore view the limit $\ol\tau \to -i\infty$ as
the limit $\la \to \infty$.

At finite values of $\la$ we have the theory with the action
\eqref{second order action} such that the classical Hamiltonian is
bounded from below. The corresponding quantum Hamiltonian is conjugate
to a second order differential operator equal to $- \frac{1}{2}
(\la^{-1} \De + \la |df|^2 + K_f)$ (the Witten Laplacian
\cite{W:morse}) acting on the Hilbert space which is the completion of
the de Rham complex $\Omega^\bullet(X)$ with respect to the $L_2$
norm. This Hamiltonian has non-negative spectrum (see
\secref{recollections morse} for more details).

Now consider the theory in the limit $\la \to \infty$. Here the
classical Hamiltonian corresponding to the action \eqref{1D action} is
not bounded from below, which seems to indicate trouble: unbounded
spectrum of the quantum Hamiltonian. In fact, the naive quantization
of the Hamiltonian is the first order operator ${\mc L}_v = {\mc
L}_\xi + {\mc L}_{\ol\xi}$ (where ${\mc L}$ denotes the Lie
derivative). Under our assumptions on the vector field $\xi$, the only
smooth eigenstates are the constant functions. It is not clear at all
how ${\mc L}_v$ could possibly be realized as the Hamiltonian of a
quantum mechanical model.

This constitutes the first puzzle that we encounter when analyzing the
model \eqref{1D action} (and its higher dimensional analogues).

The second puzzle has to do with holomorphic factorization. The action
\eqref{1D action} manifestly splits into the sum of holomorphic and
anti-holomorphic terms (unlike the action at $\la^{-1} \neq 0$,
because the term $\la^{-1} g_{a\ol{b}} p_a \ol{p_b}$ is mixed). So
naively one expects the same kind of holomorphic factorization for the
space of states and for the correlation functions. However, the only
globally defined holomorphic functions (for compact $X$) are
constants. There may also be some holomorphic differential forms, but
only a finite-dimensional space of those. So it seems that a
holomorphic factorization is impossible due to the absence of globally
defined holomorphic differential forms.

This makes us wonder that perhaps the structure of the space of states
of the theory with the action \eqref{second order action} should
change when $\la \to \infty$ in such a way that the new space of
states is the tensor product of chiral and anti-chiral sectors, the
Hamiltonian is diagonalizable and has bounded spectrum.

In fact, we will show that this is ``almost'' true: the new space of
states has the following structure
$$
{\mc H} = \bigoplus_{\al \in A} {\mc F}_\al \otimes \ol{\mc F}_{\al}
$$
(where $A$ is the finite set of zeros of $\xi$, or equivalently,
critical points of the Morse function $f$), where ${\mc H}_\al$ and
$\ol{\mc H}_\al$ may be viewed as the chiral and anti-chiral blocks of
the model. They have transparent geometric interpretation as spaces of
``delta-forms'' supported on the ascending manifolds of the Morse
function. We also find that the spectrum of the Hamiltonian is
non-negative, but the Hamiltonian is not diagonalizable: it splits
into a sum of Jordan blocks (of finite sizes bounded by $\dim_{\C} X +
1$). These are the phenomena typically associated with two-dimensional
logarithmic conformal field theory. It is quite curious that we
observe these phenomena already at the level of quantum mechanics!

\subsection{Summary of the results}

We now explain our results in more detail, starting with the
one-dimensional case. Let $X$ be a compact K\"ahler manifold, equipped
with a holomorphic vector field $\xi$ which comes from a holomorphic
$\C^\times$-action $\phi$ on $X$ with isolated fixed points. Let us
denote those points by $x_\al, \al \in A$. We will assume that the set
$A$ is non-empty (it is necessarily finite).

Our first result concerns the structure of the spaces of states of the
quantum theory with the action \eqref{1D action}. Under our
assumptions, there is a {\em Bialynicki-Birula decomposition}
\cite{BB,CS}
$$
X = \bigsqcup_{\al \in A} X_\al
$$
of $X$ into complex submanifolds $X_\al$, defined as follows:
$$
X_\al = \{ x \in X | \lim_{t \to 0} \phi(t) \cdot x = x_\al \}.
$$
Under the above assumptions it is proved in \cite{Frankel}
that there exists a Morse function $f$ on $X$, whose gradient is the
vector field $v = \xi + \ol\xi$. The points $x_\al, \al \in A$, are
the critical points of $f$, and the submanifolds $X_\al$ may be
described as the ``ascending manifolds'' of $f$. Furthermore, each
submanifold $X_\al$ is isomorphic to $\C^{n_\al}$, where the index of
the critical point $x_\al$ is $2(\dim_{\C} X - n_\al)$ (see
\cite{BB,CS}). In what follows we will assume that the above
decomposition of $X$ is a stratification, that is the closure of each
$X_\al$ is a union of $X_\beta$'s.

Now let ${\CH}_\al$ be the space of {\em delta-forms supported on}
$X_\al$. An example of such delta-forms is the distribution (or
current) on the space of differential forms on $X$ which is defined by
the following formula:
\begin{equation}    \label{Delal}
\langle \Delta_\al,\eta \rangle = \int_{X_\al} \eta|_{X_\al}, \qquad
\eta \in \Omega^\bullet(X).
\end{equation}
All other delta-forms supported on $X_\al$ may be obtained by
applying to $\Delta_\al$ differential operators defined on a small
neighborhood of $X_\al$. The space ${\CH}_\al$ is graded by
the degree of the differential form.

We then have a holomorphic factorization
$$
{\CH}_\al = {\mc F}_\al \otimes \ol{\mc F}_\al,
$$
where ${\mc F}_\al$ (resp. $\ol{\mc F}_\al$) is the space of
holomorphic (resp., anti-holomorphic) delta-forms supported on
$X_\al$.  For example, if $n_\al = \dim X$, so that $X_\al \simeq
\C^{n_\al}$ is an open subset of $X$, then ${\mc H}_\al$ is the space
of differential forms on $\C^{n_\al}$, and so it factorizes into the
tensor product of holomorphic and anti-holomorphic differential
forms. On the other hand, if $n_\al = 0$, so that $X_\al = x_\al$ is a
point, then ${\mc H}_\al$ is the space of distributions supported at
$x_\al$. It factorizes into the tensor product of the derivatives with
respect to holomorphic and anti-holomorphic vector fields. In the
intermediate cases the space ${\mc F}_\al$ is generated from the
delta-form $\Delta_\al$ supported on $X_\al$ under the action of
holomorphic differential forms along $X_\al$ and holomorphic vector
fields in the transversal directions. Thus, ${\mc F}_\al$ is the space
of global sections of a ${\mc D}_X$-module, where ${\mc D}_X$ is the
sheaf of holomorphic differential operators on $X$.

Now we set
$$
{\mc H} = \bigoplus_{\al \in A} {\mc H}_\al = \bigoplus_{\al \in A}
{\mc F}_\al \otimes \ol{\mc F}_\al.
$$
We claim that this space ${\mc H}$ is isomorphic to the {\em space of
states} of the quantum mechanical model described by the action
\eqref{1D action}.

The reader may wonder how the space of states of the theory for finite
$\la$ described by the action \eqref{second order action}, which is
essentially the space of differential forms on $X$, could possibly
turn into something like this. We explain this in detail below. Here
we will only point out that the procedure of taking the limit $\la \to
\infty$ is quite non-trivial. Before passing to the limit we need to
multiply the wave functions of the quantum Hamiltonian of the theory
\eqref{second order action} by $e^{\la f}$ (this corresponds to adding
the term $-\la df$ to the action \eqref{second order
action}). Standard semi-classical analysis shows that after this
multiplication the wave functions with eigenvalues that remain finite
in the limit $\la \to \infty$ tend to the delta-forms which give us a
monomial basis in the spaces $\CH_\al$ for different $\al$'s. In
particular, the exact supersymmetric vacua (in other words, the BPS
states), which are known to be in bijection with the critical points
of $f$ \cite{W:morse}, become in the limit $\la \to \infty$ the
delta-forms $\Delta_\al$ on $X_\al$ defined by formula \eqref{Delal}.

Next, we consider the Hamiltonian. Naively we expect that it is equal
to
$$
H_{\on{naive}} = {\mc L}_{\xi} + {\mc L}_{\ol{\xi}},
$$
acting on the above space ${\mc H}$. However, we claim that it is
actually equal to
$$
H = H_{\on{naive}} + 4\pi \sum_{\al,\beta} a_{\al\beta} \;
\delta_{\al\beta} \otimes \ol\delta_{\al\beta},
$$
where the summation is over all $\al,\beta$ such that $X_\beta$ is a
codimension $1$ stratum in the closure of $X_\al$. Here
$\delta_{\al\beta}$ is the {\em Grothendieck-Cousin operator} (GC) and
$\ol\delta_{\al\beta}$ is its complex conjugate (the $a_{\al\beta}$
are some non-zero real numbers). The GC operator
acting from ${\mc F}_\al$ to ${\mc F}_\beta$ corresponds to taking the
singular part of a holomorphic differential form on $X_\al$ along this
divisor (see, e.g., \cite{Kempf}).

In particular, we find that the Hamiltonian is not diagonalizable;
rather, it has Jordan blocks!

In order to test these predictions, we investigate the factorization
of correlation functions over intermediate states. Suppose for
simplicity that $X=\pone$ and $f$ is the standard ``height'' function
(see \secref{case of pone}). In this case there is one non-trivial
component of the moduli space of gradient trajectories, which consists
of the trajectories going from the north pole to the south pole. It is
isomorphic to $\C^\times \subset \pone$, hence its natural
compactification is $\pone$. Typical observables of our theory are
smooth differential forms. We know from the path integral description
of the model that the correlation function of observables
$\wh\omega_1,\ldots,\wh\omega_n$ corresponding to differential forms
$\omega_1,\ldots,\omega_n$ is equal to
$$
\langle \wh\omega_1(t_1) \wh\omega_2(t_2) \ldots \wh\omega_n(t_n)
\rangle = \int_{\pone} \omega_1 \wedge
\phi(e^{-(t_1-t_2)})^*(\omega_2) \ldots \wedge
\phi(e^{-(t_{n-1}-t_n)})^*(\omega_n),
$$
where $\phi$ is the standard $\C^\times$-action on $\pone$ and
$\phi(q)^*$ denotes the pull-back of a differential form under the
action of $q \in \C^\times$. Consider the simplest case when
$\omega_1$ is a smooth two-form $\omega$ and $\omega_2$ is a smooth
function $F$ on $\pone$, which we will assume to be non-constant. Then
we have
$$
\langle \wh\omega(t_1) \wh{F}(t_2) \rangle = \int_{\pone} \omega
\; \phi(e^{-t})^*(F),
$$
where $t=t_1-t_2$.

On the other hand, we expect this two-point function to factorize into
the sum of one-point functions over all possible intermediate states:
\begin{equation}    \label{fac}
\langle \wh\omega(t_1) \wh{F}(t_2) \rangle = \langle \wh\omega e^{-tH}
\wh{F} \rangle = \sum_\nu \langle 0|\wh\omega e^{-tH} |\Psi_\nu
\rangle \langle \Psi_\nu^*| \wh{F} |0 \rangle.
\end{equation}
If the hamiltonian were diagonalizable, the right hand side would be
the sum of monomials $q^\al$, where $q = e^{-t}$ and $\al$ runs over the
spectrum of the Hamiltonian (in our case it consists of non-negative
integers). However, consider the following simple example: let
$$
F = \frac{1}{1+|z|^2}, \qquad \omega = \frac{1}{(1+|z|^2)^2} \frac{d^2
z}{\pi}.
$$
We find that
$$
\int_{\pone} \omega \; \phi(q)^*(F) = \frac{1}{1-q^2} +
\frac{2q^2}{(1-q^2)^2} \log q.
$$
The appearance of the logarithmic function indicates that the operator
$H$ is not diagonalizable, but has Jordan blocks of length two, in
agreement with our prediction.

Thus, the logarithmic nature of our model is revealed by elementary
calculation of an integral over the simplest possible moduli space of
instantons. But it is important to stress that in order to see the
logarithm function in a correlation function it is necessary that at
least one of the observables involved not be $Q$-closed. The action of
$Q$ on the above observables $\wh\omega_i$ corresponds to the action
of the de Rham differential on the differential forms
$\omega_i$. Thus, in the above calculation the two-form $\wh\omega$ is
necessarily $Q$-closed, and so it is a BPS observable. But $F$ is not
$Q$-closed due to our assumption that $F$ is not constant. Therefore
$\wh{F}$ is not a BPS observable. If both $F$ and $\omega$ were
$Q$-closed, then the one-point functions appearing on the right hand
side of formula \eqref{fac} would be non-zero only when the
intermediate states are vacuum states. On such states the Hamiltonian
is diagonalizable, so we would not be able to observe the logarithmic
terms. The same argument applies to $n$-point correlation
functions. Thus, we can discover the structure of the space of states
of the theory, and in particular, the existence of the Jordan blocks
of the Hamiltonian, {\em only} if we consider correlation functions of
non-BPS observables. It is impossible to see these structures within
the topological sector of our model. This is yet another reason why it
is important to go beyond the topological sector.

Part I of our article contains a detailed and motivated exposition of
our results describing the structure of our quantum mechanical models
at the special point $\la=\infty$ (or, equivalently, $\ol\tau =
\infty$). We hope that the models corresponding to finite values of
$\la$ may be studied by $\la^{-1}$-perturbation theory around the
point $\la=\infty$. We will present some sample calculations below
which provide some evidence that this is indeed possible.

\medskip

In Part II we will apply our approach to two-dimensional
${\mc N}=(2,2)$ supersymmetric sigma models and four-dimensional ${\mc
N}=2$ supersymmetric Yang-Mills theory. Part III will be devoted to
generalization to non-supersymmetric models.

We now discuss briefly what happens in dimensions two and four, thus
giving a preview of the Part II of this article.

\subsection{Two-dimensional sigma models}

Let us consider first the supersymmetric (type A twisted)
two-dimensional sigma model \cite{W:tsm,W:mirror} in the $\ol\tau \to
\infty$ limit (these limits have been previously discussed in
\cite{BLN,LMZ,AiB}). The first step is to recast these models in the
framework of the quantum mechanical models that we have studied
above. For a fixed Riemann surface $\Sigma$ the space of bosonic
fields in the supersymmetric sigma model with the target manifold $X$
is $\on{Maps}(\Sigma,X)$, the space of maps $\Sigma \to X$. If we
choose $\Sigma$ to be the cylinder $I \times \BS^1$, then we may
interpret $\on{Maps}(\Sigma,X)$ as the space of maps from the interval
$I$ to the loop space $LX = \on{Maps}(\BS^1,X)$. Thus, we may think of
the two-dimensional sigma model on the cylinder with the target $X$ as
the quantum mechanical model on the loop space $LX$. Hence it is
natural to try to write the Lagrangian of the sigma model in such a
way that it looks exactly like the Lagrangian of the quantum
mechanical model on $LX$ with a Morse function $f$.

It turns out that if $X$ is a K\"ahler manifold, this is ``almost''
possible. However, there are two caveats. First of all, the
corresponding function $f$ has non-isolated critical points
corresponding to the constant loops in $LX$, so it is in fact a
Bott-Morse function. We can deal with this problem by deforming this
function so that it only has isolated critical points, corresponding
to the constant loops whose values are critical points of a Morse
function on $X$. The second, and more serious, issue is that our
function $f$ is not single-valued on $LX$, but becomes single-valued
only after pull-back to the universal cover $\wt{LX}$ of $LX$. In
other words, it is an example of a Morse-Novikov function, or, more
properly, Bott-Morse-Novikov function. Because of that, the instantons
are identified with gradient trajectories of the pull-back of $f$ to
$\wt{LX}$.

The universal cover $\wt{LX}$ may be described as the space of
equivalence classes of maps $\wt\ga: D \to X$, where $D$ is a
two-dimensional unit disc, modulo the following equivalence relation:
we say that $\wt\ga \sim \wt\ga'$ if $\wt\ga|_{\pa D} = \wt\ga'|_{\pa
D}$ and $\wt\ga$ is homotopically equivalent to $\wt\ga'$ in the space
of all maps $D \to X$ which coincide with $\wt\ga$ and $\wt\ga'$ on
the boundary circle $\pa D$. We have the obvious map $\wt{LX} \to LX$,
which realizes $\wt{LX}$ as a covering of $LX$. The group of deck
transformations is naturally identified with $H_2(X,\Z)$. The
corresponding Morse function, which goes back to the work of A. Floer
\cite{Floer} is given by the formula
\begin{equation}    \label{f on LX}
f(\wt\gamma) = \int_D \wt\ga^*(\omega_K),
\end{equation}
where $\omega_K$ is the K\"ahler form.

Suppose now that $I = \R$ with a coordinate $t$. In the limits $t \to
\pm \infty$ a gradient trajectory tends to the critical points of $f$
on $\wt{LX}$, which are the preimages of constant maps in
$LX$. Therefore a gradient trajectory may be interpreted as a map of
the cylinder, compactified by two points at $\pm \infty$, to $X$, or
equivalently, a map $\pone \to X$. Moreover, the condition that it
corresponds to a gradient trajectory of $f$ simply means that this map
is {\em holomorphic}. Thus, we obtain that the instantons of the
two-dimensional sigma model are holomorphic maps $\pone \to X$, and
more generally, $\Sigma \to X$, where $\Sigma$ is an arbitrary compact
Riemann surface.

In our infinite radius limit (which corresponds to $\ol\tau \to -i
\infty$, as explained above) we obtain the theory governed by first
order action \eqref{2D action}. Therefore the corresponding path
integral localizes on the moduli space of holomorphic maps $\Sigma \to
X$. Because we are dealing with a Morse-Novikov function, this moduli
space now has infinitely many connected components labeled by $\beta
\in H_2(X,\Z)$, all of which are finite-dimensional (the component
is non-empty only if the integral of the K\"ahler class $\omega$ of
$X$ over $\beta$ is non-negative). The simplest observables of this
model are the evaluation observables corresponding to differential
forms on $X$. Their correlation functions are given by integrals of
their pull-backs to the moduli spaces of holomorphic maps (more
precisely, the Kontsevich moduli spaces of stable maps) under the
evaluation maps.

Such correlation functions have been studied extensively in the
literature in the case of the {\em BPS observables}, corresponding to
closed differential forms on $X$. They are expressed in terms of the
{\em Gromov-Witten invariants} of $X$. Our goal is to go beyond the
BPS (or topological) sector of the model and study correlation
function of more general, non-BPS, observables. These observables
include evaluation observables corresponding to differential forms on
$X$ that are not closed, as well as differential operators on $X$. The
correlation functions of the non-BPS observables reveal a lot more
about the structure of the theory. In particular, as we have
seen previously in quantum mechanical models, they essentially
allow us to reconstruct the spaces of states of the theory in the
limit $\ol\tau \to -i \infty$.

In Part II we will describe in detail the structure of the spaces of
states of the supersymmetric two-dimensional sigma models in our
infinite radius limit. First, we will generalize our results obtained
in Part I to the case of multivalued Morse functions. (Actually,
examples of such functions arise already for finite-dimensional real
manifolds with non-trivial fundamental groups.) We will show how to
modify our results in this case. Essentially, this amounts to
considering the universal cover of our manifold, which is $\wt{LX}$ in
the case of two-dimensional sigma model. One also needs to impose an
equivariance condition on the states of the model corresponding to the
action of the (abelianized) fundamental group $H_1(X,\Z)$ on
$\wt{LX}$. Because of this the corresponding spaces of states acquire
an additional parameter which is familiar from the construction of
``$\vartheta$-vacua''.

Besides those changes, the structure of the space of states in the
limit $\ol\tau \to \infty$ is similar to the one that we have observed
above in our analysis of quantum mechanical models. There are spaces
of ``in'' and ``out'' states, and each of them is isomorphic to the
direct sum of certain spaces of ``delta-forms'' supported on the
strata of the decomposition of $\wt{LX}$ into the ascending and
descending manifolds (however, this direct sum decomposition is not
canonical). Let us modify our function $f$ given by formula \eqref{f
on LX}, as follows: $f \mapsto f_H$, where
$$
f_H(\wt\ga) = \int_D \wt\ga^*(\omega_K) - \int_{\BS^1} \ga^*(H)
d\sigma,
$$
where $H$ is a Morse function on $X$. Then $f_H$ is a Morse
function on $\wt{LX}$, with isolated critical points: constant maps
with values at the critical points $x_\al, \al \in A$, of $H$ on $X$.

The corresponding ascending manifolds in $\wt{LX}$ are isomorphic to
infinite-dimensional vector spaces, which are roughly of half the
dimension of the entire loop space. Let us denote them by
$X_{\al,\frac{\infty}{2}+i}, \al \in A, i \in \Z$. Our space of states
$\wh{\mc H}_\tau$, which now depends on the choice of $\tau \in \C$,
is realized as the subspace of those states $\Psi$ in
$$
\wh{\mc H} = \bigoplus_{\al \in A, i \in \Z} {\mc
F}_{\al,\frac{\infty}{2}+i} \otimes \ol{\mc F}_{\al,\frac{\infty}{2}+i},
$$
where ${\mc F}_{\al,\frac{\infty}{2}+i}$ is the space of holomorphic
``delta-forms'' supported on $X_{\al,\frac{\infty}{2}+i}$, and ${\mc
F}_{\al,\frac{\infty}{2}+i}$ is its complex conjugate, which
satisfy the condition $T(\Psi) = e^{i\tau} \Psi$. Here $T$ is the
shift operator ${\mc H}_{\al,\frac{\infty}{2}+i} \to {\mc
H}_{\al,\frac{\infty}{2}+i+1}$. Thus, $\wh{\mc H}_\tau$ is
(non-canonically) isomorphic to the direct sum as above with fixed
$i$.

The spaces of delta-forms may then be identified with the familiar
Fock representations of the chiral and anti-chiral
$\beta\gamma,bc$-systems. The Hamiltonian and the supersymmetry
charges may be identified with explicit operators acting on the spaces
of states. Thus, the spaces of states are essentially isomorphic to
the direct sums of finitely many tensor products of this form, like in
a conformal field theory. In particular, there is a large chiral
algebra, which is nothing but the chiral de Rham complex of
$X$. However, we find that the Hamiltonian is not diagonalizable and
computing matrix elements of observables acting on the space of
states, we see the appearance of the logarithm function. Thus, we
find that the two-dimensional supersymmetric model in our infinite
radius limit is a {\em logarithmic conformal field theory}.

We stress again that to see this structure it is crucial that we
consider the correlation functions of non-BPS observables. The
hamiltonian is diagonalizable (in fact, is identically equal to zero)
on the BPS states. Therefore correlation functions of the BPS
observables which have been extensively studied in the literature (and
which are closely related to the Gromov-Witten invariants) do not
contain logarithms. The hamiltonian is also diagonalizable on
all purely chiral (and anti-chiral) states; thus, the chiral algebra
of the theory is free of logarithms. Logarithmic CFTs of this type
have been considered, e.g., in \cite{Saleur}.

We remark that in the case when $X$ is the flag manifold of a simple
Lie group, the above semi-infinite stratification of $\wt{LX}$ and the
corresponding spaces of holomorphic delta-forms have been considered
in \cite{FF:si}. These spaces are representations of the affine
Kac-Moody algebra $\ghat$ with level $0$, which are closely related to
the Wakimoto modules. In the non-supersymmetric version the level $0$
algebra $\ghat$ gets replaced by the $\ghat$ at the critical level
$-h^\vee$ (see \cite{FF:si,F:rev}). Therefore we expect that the
corresponding models are closely related to the geometric Langlands
correspondence. This will be discussed in Part III of this article.

\subsection{Four-dimensional Yang-Mills theory}

Finally, we discuss (twisted) ${\mc N}=2$ supersymmetric Yang-Mills
theory with gauge group $G$ on a four-dimensional manifold ${\mathbf
M}^4$ \cite{W:tft}. Suppose that ${\bf M}^{4} = {\R} \times M^3$,
where $M^3$ is a compact three-dimensional manifold. Let $t$ denote
the coordinate along the $\R$ factor. Then the Yang-Mills theory may
be interpreted as quantum mechanics on the space ${\CA}/{\CG}$ of
gauge equivalence classes of $G$-connections on $M^3$, with the
Morse-Novikov function being the Chern-Simons functional
\cite{Atiyah,W:tft}. However, there is again a new element, compared
to the previously discussed theories, and that is the appearance of
gauge symmetry. The quotient ${\CA}/{\CG}$ has complicated
singularities because the gauge group ${\CG}$ has non-trivial
stabilizers in the space ${\CA}$. For this reason we should consider
the {\em gauged Morse theory} on the space ${\CA}$ of connections
itself.

This theory is defined as follows. Let $X$ be a manifold equipped with
an action of a group $G$ and a $G$-invariant Morse function $f$. Then
the gradient vector field $v^{\m} \pa_{x^{\m}}$, where $v^{\m} =
h^{\m\nu} {\p}_{x^\nu}f$ commutes with the action of $G$. Denote by
$\CV_a^{\m} \pa_{x^{\m}}$ the vector fields on $X$ corresponding to
basis elements $J^a$ of the Lie algebra $\g = \on{Lie}(G)$. We define
a gauge theory generalization of the gradient trajectory: it is a pair
$(x(t): \R \to X,A_t(t)dt \in \Omega^1(\R,\g^*))$, which is a solution
of the equation
\begin{equation}    \label{gauged Morse}
\frac{d{x}^{\m}}{dt} = v^{\m} (x(t))+ {\CV}^{\m}_{a}(x(t))
A^{a}_{t}(t).
\end{equation}
The group of maps $g(t): \R \to G$ acts on the space of solutions by
the formula 
$$
g: \left( x(t),A_t(t)dt \right) \mapsto \left( g(t) \cdot x(t), \
g^{-1}(t) {\p}_{t} g (t)+ g^{-1}(t) A_{t}(t) g(t) \right),
$$
and the moduli space of gradient trajectories is, naively, the
quotient of the space of solutions of \eqref{gauged Morse} by this
action. However, because this action has non-trivial stabilizers and
the ensuing singularities of the quotient, it is better to work
equivariantly with the moduli space of solutions of the equations
\eqref{gauged Morse}.

In Part II we will develop a suitable formalism of equivariant
integration on the moduli space of gradient trajectories of the gauged
Morse theory. We then apply this formalism to the case when $X = \CA$,
the space of connections on a three-manifold $M^3$ and $f$ is the
Chern-Simons functional (note that this formalism is also used in
gauged sigma models in two dimensions). In this case the corresponding
equivariant integrals give us the correlation functions of evaluation
observables of the Yang-Mills theory in our weak coupling limit
$\ol\tau \to -i \infty$. In the case of the BPS observables these
correlation functions are the {\em Donaldson invariants}
\cite{W:tft}. They comprise the topological (or BPS) sector of the
theory. We will obtain more general (off-shell) correlation functions
by considering more general, i.e., non-BPS, evaluation observables. We
will present some sample computations of these off-shell correlation
functions which exhibit the same effects as in one- and
two-dimensional models considered above. In particular, we will
observe the appearance of the logarithm (and more generally,
polylogarithm) function in the correlation functions. This indicates
that, just like the two-dimensional sigma models, the four-dimensional
supersymmetric Yang-Mills theory in the $\ol\tau \to \infty$ limit is
a logarithmic conformal field theory.

\subsection{Plan of the paper}

The paper is organized as follows. In \secref{susy qm} we give a
pedagogical description of the Lagrangian formalism of our quantum
mechanical models.  We then discuss the path integral in the limit
when the metric of the manifold and the Morse function are both
multiplied by the same constant $\la$ which tends to infinity (this
corresponds to the limit $\ol\tau \to \infty$ with $\tau = 0$). We
show that in this limit the path integral localizes on the moduli
spaces of gradient trajectories of the Morse function (the instantons
of the quantum mechanical models). We introduce the observables of the
theory and discuss the analogy between their correlation functions and
the Gromov-Witten theory.

In \secref{Hamiltonian formalism} we start developing the Hamiltonian
formalism for our models. We are interested, in particular, with the
structure of the space of states of the model in the limit $\la \to
\infty$. This limit is highly singular and the description of the
spaces of states requires special care. We discuss in detail the
examples of the flat space $\C$ and of the simplest ``curved''
manifold $\pone$.  We show that the space of states decomposes into
the spaces of ``in'' and ``out'' states, each having a simple and
geometrically meaningful description in terms of the stratifications
of our manifold by the ascending and descending manifolds of our Morse
function. Furthermore, we show that the spaces of states exhibit
holomorphic factorization that is absent for finite values of
$\la$. This leads to a great simplification of the correlation
functions in the limit $\la \to \infty$.

Next, in \secref{structure} we give a more precise description of the
spaces of states. We show that states are naturally interpreted as
{\em distributions} (or {\em currents}) on our manifold $X$. Because
some of these distributions require regularization (reminiscent of the
Epstein-Glaser regularization \cite{EG} in quantum field theory), the
action of the Hamiltonian on them becomes non-diagonalizable. We
explain this in detail in the case of the $\pone$ model. For a general
K\"ahler manifold we compute the action of the Hamiltonian on the
spaces of ``in'' and ``out'' states, as well as the action of the
supercharges, in terms of the so-called {\em Grothendieck-Cousin
operators} associated to the stratification of our manifold by the
ascending and descending manifolds. We also compute the cohomology of
the supercharges using the Grothendieck-Cousin (GC) complex.

In \secref{action} we will realize the observables of the model as
linear operators acting on the spaces of states. We will then be able
to obtain the correlation functions as matrix elements of these
operators and to test our predictions by comparing these matrix
elements with the integrals over the moduli spaces of gradient
trajectories which were obtained in the path integral approach of
\secref{susy qm}. We will see that analytic properties of the
observables play an important role in the limit $\la \to \infty$. We
will also see that factorization of the correlation functions over
intermediate states leads to some non-trivial identities on analytic
differential forms. In particular, the appearance of logarithm in the
correlation functions will be seen as the manifestation of the
non-diagonal nature of the Hamiltonian and as the ultimate validation
of our description of the space of states.

Finally, in \secref{generalizations} we discuss possible
generalizations of our results. We consider the question of how to
relate the spaces of states of our models for finite and infinite
values of $\la$, first in the case when $X=\C$ and then for
$X=\pone$. We then discuss the computation of correlation functions in
$\la^{-1}$ perturbation theory. Next, we consider non-supersymmetric
analogues of our models. We discuss, in particular, the computation of
the cohomology of the anti-chiral supercharge $\ol\pa$ in the
``half-supersymmetric'' models, which are one-dimensional analogues of
the $(0,2)$ supersymmetric two-dimensional sigma models. We make
contact with the GC complexes of arbitrary (holomorphic) vector
bundles on K\"ahler manifolds and the results of \cite{Witten:hol,Wu}
on holomorphic Morse theory. We also discuss briefly the
generalization in which a Morse functions is replaced by a Morse-Bott
function having non-isolated critical points.

\subsection{Acknowledgments}

EF thanks M. Zworski for useful discussions and references.

This project was supported by DARPA through its Program ``Fundamental
Advances in Theoretical Mathematics''. We are grateful to DARPA for
generous support which enabled us to carry out this project.

In addition, research of EF was partially supported by the NSF grant
DMS-0303529; of AL by the grants RFFI 04-02-17227, INTAS 03-51-6346
and NSh-8065.2006.2; and of NN by European RTN under the contract
005104 "ForcesUniverse", by ANR under the grants
ANR-06-BLAN-3$\_$137168 and ANR-05-BLAN-0029-01, and by the grants
RFFI 06-02-17382 and NSh-8065.2006.2.

EF and NN thank KITP at UCSB for hospitality during the 2005 program
``Mathematical structures in string theory'' where part of the work
had been done; EF and AL thank IHES for hospitality during various
visits in 2006; and NN thanks MSRI and UC Berkeley, NHETC at Rutgers
University, the Institute for Advanced Study at Princeton, and
University of Pennsylvania for hospitality.

\section{Supersymmetric quantum mechanics}    \label{susy qm}

In this section we begin our investigation of the quantum mechanical
models in the limit $\ol\tau \to \infty$. The natural context for
these models is the physical realization of Morse theory due to
E.~Witten \cite{W:morse}, which we recall briefly at the beginning of
the section. We describe the Lagrangians of these models and the
corresponding path integral. We then discuss the path integral in the
limit when the metric of the manifold and the Morse function are both
multiplied by the same constant $\la$ which tends to infinity (which
corresponds to the limit $\ol\tau \to \infty$ with $\tau=0$, discussed
in the Introduction). We show that in this limit the path integral
localizes on the moduli spaces of gradient trajectories of the Morse
function. We introduce the observables of the theory and discuss the
analogy between their correlation functions and the Gromov-Witten
theory.

\ssec{Recollections on Morse theory}    \label{recollections morse}

Morse theory associates to a compact smooth Riemannian manifold $X$
and a Morse function $f$ (i.e., a function with isolated
non-degenerate critical points) a complex $C^{\bullet}$, whose
cohomology coincides with the de Rham cohomology $H^{\bullet}(X)$.
The $i$th term $C^i$ of the complex is generated by the critical
points of $f$ of index $i$ (the index of a critical point is the
number of negative squares in the Hessian quadratic form at the
critical point). The differential $d: C^{i} \to C^{i+1}$ is obtained
by summing over the gradient trajectories connecting critical points.


E.~Witten \cite{W:morse} has given the following interpretation of
Morse theory. Consider the supersymmetric quantum mechanics on a
Riemannian manifold $X$ (in other words, quantum mechanics on $\Pi
TX$). The space of states is the Hilbert space ${\Omega}^{\bullet}
(X)$, the space of complex-valued $L_2$ differential forms on $X$ with
the hermitean inner product
\begin{equation}
\label{norm}
\langle {\al} \vert {\bt} \rangle = \int_{X} (\star {\overline{\al}})
\wedge {\bt}
\end{equation}
where $\overline{\left( \ldots \right)}$ denotes the complex
conjugation, and $\star$ is the Hodge star operator.

The supersymmetry algebra is generated by the operators:
\begin{align} \label{susy1}
{\mathcal Q} &= d_\la = e^{-\la f} d e^{\la f} = d + {\la}\ df \wedge
\\ \label{susy2} {\mathcal Q}^{*} &= (d_\la)^* = \frac{1}{\la} e^{\la
f} d^* e^{-\la f} = \frac{1}{\la} d^{*} + \iota_{{\nabla} f}.
\end{align}
Here the operator $d^*$ is defined as the adjoint of $d$ with respect
to a fixed metric $g$ on $X$. But ${\mathcal Q}^{*}$ is the adjoint of
$\CQ = d_\la$ with respect to the metric $\la g$, which explains the
overall factor $\la^{-1}$.

Their anti-commutator $H = \frac{1}{2} \{ {\mathcal Q} , {\mathcal
Q}^{*} \}$ is the Hamiltonian\footnote{In our conventions the
Laplacian ${\Delta} = - \{ d , d^{*} \}$ is negative
definite}:
\begin{equation}
\label{ham}
H = H_\la = \frac{1}{2} \left( - {\la}^{-1} {\Delta} + \la \Vert df
\Vert^2 + K_{f} \right)
\end{equation}
where $K_{f} = ({\mathcal L}_{{\nabla} f} +
{\mathcal L}_{{\nabla} f}^{*})$. Recall that for a vector field $\xi$
we denote by ${\mc L}_\xi$ the Lie derivative acting on differential
forms.

The supersymmetry generators $\mathcal Q$, $\mathcal Q^{*}$ are
nilpotent, ${\mathcal Q}^{2} = 0, \mathcal (\CQ^{*})^2 = 0$. For
compact $X$, the cohomology of the operator $\mathcal Q$ coincides
with the cohomology of the operator $d$, since they are related by the
similarity transformation. Standard Hodge theory argument shows that
the span of the ground states of $H$ (i.e., those annihilated by $H$)
is isomorphic to the cohomology of ${\mc Q}$, and hence to
$H^{\bullet}(X)$.

Indeed, among the ${\mathcal Q}$-closed differential forms $\al$
representing a given cohomology class we choose a representative
${\al}_{\on{har}}$ which minimizes the norm $\Vert {\al} \Vert^2$ with
respect to the inner product \eqref{norm}. This representative
${\al}_{\on{har}}$ is annihilated by ${\mathcal Q}^{*}$, in addition
to $\mathcal Q$. As a consequence, $\al_{\on{har}}$ is annihilated by
$H$. Conversely, if $\al$ is annihilated by $H$, then
\begin{equation}
\label{hamqq}
0 = \langle {\al} \vert H \vert {\al} \rangle = \Vert \mathcal Q \al
\Vert^2 + \Vert {\mathcal Q}^{*} {\al} \Vert^2,
\end{equation}
hence ${\mathcal Q} \al = 0$, ${\mathcal Q}^{*} {\al} = 0$.

The first step in Witten's approach to Morse theory \cite{W:morse} is
constructing the approximate ground states of $H$ in the limit ${\la}
\to \infty$. According to the semi-classical analysis, they are given
by the differential forms localized near the critical points $x$, such
that $df_{x} = 0$. Near such a point the Hamiltonian \eqref{ham} may
be approximated by that of supersymmetric harmonic oscillator. We will
discuss this in more detail below. Now we just mention that for each
critical point of index $i$ one finds a ground state of the
Hamiltonian, which is a differential form $\omega_i$ of degree $i$
approximately equal to a Gaussian distribution around this critical
point. The simplest of these are the $0$-form $C_\la e^{-\la f}$
localized at the absolute minimum of $f$ and the top form $C'_\la
e^{\la f} d\mu$, localized at the absolute maximum of $f$ (here $d\mu$
is the volume form induced by the metric and $C_\la, C'_\la$ are the
constants making the norms of these forms equal to $1$).

The eigenvalues of $H$ on these approximate ground states $\omega_i$
tend to zero very fast as $\la \to \infty$. Therefore for large $\la$
their span ``splits off'' as a subcomplex of the de Rham complex,
equipped with the twisted differential ${\mc Q} = d_\la$. Since
cohomology classes may be represented by ground states, as we have
seen above, we obtain that the cohomology of this subcomplex is equal
to the cohomology of the entire de Rham complex. This homology is in
turn isomorphic the cohomology of $X$. By construction, the dimension
of the $i$th group of this subcomplex is equal to the number of
critical points of $f$ of index $i$, and using this fact we obtain
estimates on the ranks of the cohomology groups of $X$, i.e., the
Betti numbers of $X$. This way Witten proved in \cite{W:morse} the
Morse inequalities relating the Betti numbers of $X$ to the numbers of
critical points of $f$ of various indices (see also
\cite{Helffer1,Helffer2}).

Perturbatively, each $\omega_i$ is annihilated by the supersymmetry
charge $\mathcal Q$. The next step in Witten's construction is the
computation of the instanton corrections to ${\mc Q}$ on $\omega_i$'s
due to the tunneling transitions. Witten argued that this way one
obtains the Morse complex of $f$. (This is indeed the case if a
certain ``Morse-Smale transversality condition'' is satisfied.) Thus,
one obtains an interpretation of Morse theory in terms of
supersymmetric quantum mechanics.

\ssec{Important special case}

A special case of this construction occurs when $X$ is a compact
K\"ahler manifold, and the Morse function $f$ is the hamiltonian
corresponding to a $U(1)$-action on $X$. Its complexification gives us
a $\C^\times$-action on $X$. In this case, the gradient vector field
$v = {\nabla}f$ may be split as the sum of a holomorphic vector field
$\xi$ and its complex conjugate $\ol\xi$. On the other hand, the vector
field $i ( {\xi} - {\ol\xi} )$ generates the $U(1)$ action.

In the main body of this paper we will focus exclusively on this case,
because it is the one-dimensional analogue of the two-dimensional
supersymmetric sigma models and the super-Yang-Mills theory that we
are interested in. Some important simplifications occur in this
case. For example, all ground states in the limit $\la=\infty$ may be
deformed to ground states for finite $\la$. In other words, there are
no instanton corrections to the action of the supercharge $\CQ$ (see
\secref{top sector} for more details). These are also the models
exhibiting holomorphic factorization in the limit $\la=\infty$, as we
will see below.

\ssec{Path integral and gradient trajectories}    \label{pi and grad}

Let us discuss the Lagrangian version of the theory. The space of
states of our theory is the space of functions on the supermanifold
${\Pi}TX$. Introduce the corresponding coordinates $x^{\m},
{\psi}^{\m}$ and the momenta $p_\mu, \pi_\mu$. The configuration space
is the space of maps $I \to X$, where $I = I_{t_i,t_f}$ is the
``worldline'', which could be a finite interval $[t_i,t_f]$, or
half-line $(-\infty,t_f), [t_i,+\infty)$ or the entire line
$(-\infty,+\infty)$. The standard action is given by the formula
\cite{W:morse}
\begin{multline}   \label{first action}
S = \int_I \left( \frac{1}{2} \la g_{\mu\nu} \frac{dx^\mu}{dt}
\frac{dx^\nu}{dt} + \frac{1}{2} \la g^{\mu\nu} \frac{\pa f}{\pa x^\mu}
\frac{\pa f}{\pa x^\nu} \right. + \\ \left. i \pi_\mu D_t \psi^\mu - i
g^{\mu\nu} \frac{D^2 f}{D x^\nu D x^\al} \pi_\mu \psi^\al +
\frac{1}{2} \la^{-1} R^{\mu\nu}_{\al\beta} \pi_\mu \pi_\nu \psi^\al
\psi^\beta \right) dt.
\end{multline}
Here $D/Dx^\mu$ is the covariant derivative on $X$ corresponding to
the Levi-Civita connection, and $D_t$ is its pull-back to $I$ under
the map $x: I \to X$.  It is easy to the see that the hamiltonian
corresponding to this action is the quasi-classical limit of the
hamiltonian \eqref{ham}.

The correlation functions of the theory are given by path integrals:
\begin{multline}    \label{path int}
\langle x_f| e^{(t_n-t_f)H} \OO_n e^{(t_{n-1}-t_n)H} \ldots
         e^{(t_1-t_2)H} \OO_1 e^{(t_i-t_1)H} |x_i \rangle = \\
         \underset{I \to X; x(t_i) = x_i, x(t_f) = x_f}\int \OO_1(t_1)
         \ldots \OO_n(t_n) e^{-S}.
\end{multline}
Here $\OO_1,\ldots,\OO_n$ are observables which we will discuss in
detail below. More precisely, the right hand side gives the integral
kernel for this correlation function with respect to $x_i$ and $x_f
\in X$. In other words, if $\Psi, \Psi^*$ are states in the Hilbert
space of the theory (which is the space of $L_2$ differential forms on
$X$), then
\begin{multline}    \label{path int1}
\langle \Psi^* | e^{(t_n-t_f)H} \OO_n e^{(t_{n-1}-t_n)H} \ldots
         e^{(t_1-t_2)H} \OO_1 e^{(t_i-t_1)H} | \Psi \rangle = \\
         \int_{X^2} \Psi^*(x_f) \Psi(x_i) \; \underset{I \to X; x(t_i) =
         x_i, x(t_f) = x_f}\int \; \OO_1(t_1) \ldots \OO_n(t_n) e^{-S}.
\end{multline}

We will now discuss in detail how to pass to the limit $\la \to
\infty$ in such a way that we keep the instanton contributions, but
get rid of the anti-instantons. The procedure will be similar in two
and four dimensions.

We start with the trivial, but crucial observation (sometimes called
the ``Bogomolny trick'') that the bosonic part of the action may be
rewritten as follows:
\begin{equation}    \label{trick}
\int_{I} \left( \frac{1}{2} \la \left| \dot x \mp \nabla f \right|^2
\pm \la \frac{df}{dt} \right) dt,
\end{equation}
where
$$
(\nabla f)^\mu = g^{\mu\nu} \frac{\pa f}{\pa x^\nu}.
$$
It is clear from this formula that the absolute minima of the action,
with fixed boundary conditions $x(t_i) = x_i, x(t_f) = x_f$, will be
achieved on the gradient trajectories of $f$ (appearing below with the
$+$ sign) or the gradient trajectories of $-f$ (with the $-$ sign):
$\dot x = \pm \nabla f$, or equivalently,
$$
\frac{dx^\mu}{dt} = \pm g^{\mu\nu} \frac{\pa f}{\pa x^\nu}
$$
(provided that gradient trajectories connecting $x_i$ and $x_f$
exist). These are the {\em instantons} and {\em anti-instantons} of
our model, respectively. The former realize maps for which $f(x(t_f)) >
f(x(t_i))$ and the latter realize maps for which $f(x(t_f)) <
f(x(t_i))$. Both contribute to the path integral with the same weight
factor $e^{-\la |f((x(t_f)) - f(x(t_i))|}$. As $\la \to \infty$ this
factor goes to $0$ exponentially fast, and this is the reason why
instanton and anti-instanton contributions are negligible compared to
the contributions of small fluctuations around the constant maps.

Now we wish to modify our Lagrangian in such a way that we retain the
instantons and make anti-instantons disappear altogether in the $\la
\to \infty$ limit. This is achieved by adding the term
\begin{equation}    \label{the term}
- \int_{I} \la df = - \int_{I} \la \frac{df}{dt} dt = \la(f(x_i) -
    f(x_f))
\end{equation}
to the action \eqref{first action}. The resulting action reads
\begin{equation}    \label{second action}
\int_{I} \left( \la \left| \dot x - \nabla f \right|^2 +
\on{fermions} \right) dt.
\end{equation}
The effect is that now the instantons, i.e., the gradient trajectories
$\dot x = \nabla f$, become the absolute minima of the action. The
action on them is equal to $0$, so all of them make contributions to
the path integral of the finite order (independent of $\la$). In
contrast, the action on anti-instantons is now $2\la |f(x_f) -
f(x_i)|$. They do not correspond to the absolute minima of the action
any more. Accordingly, their contribution to the path integral is even
more suppressed than before: now they occur in the path integral with
the weight factor $e^{-2 \la |f(x_f) - f(x_i)|}$. Therefore in the
limit $\la \to \infty$ instantons will make finite contributions to
the path integral (on par with the fluctuations around the constant
maps), but anti-instantons will not contribute to the path integral at
all.\footnote{note that by adding to the Lagrangian the term $\la df$
instead, we would retain the anti-instantons and get rid of the
instantons}

No matter how large $\lambda$ is though, both instantons and
anti-instantons make contributions to general correlation
functions.\footnote{with the exception of some $\la$-independent
correlation functions of the topological sector of the theory
discussed below} Therefore if we want to eliminate completely the
anti-instanton contributions from the correlation functions, we really
have to take the limit $\la \to \infty$.

In order to achieve that, we first rewrite the action in terms of a
first order Lagrangian as follows:
\begin{multline}    \label{second action prime}
S_\la = \int_{I} \left( -i p_\mu \left( \frac{dx^\mu}{dt} + g^{\mu\nu}
\frac{\pa f}{\pa x^\nu} \right) + \frac{1}{2} \la^{-1} g^{\mu\nu}
p_\mu p_\nu \right. \\ \left. + i \pi_\mu \left(D_t \psi^\mu -
g^{\mu\nu} \frac{D^2 f}{D x^\nu D x^\al} \psi^\al \right) +
\frac{1}{2} \la^{-1} R^{\mu\nu}_{\al\beta} \pi_\mu \pi_\nu \psi^\al
\psi^\beta \right) dt.
\end{multline}
For finite values of $\la$, by eliminating the momenta variables using
the equations of motion, we obtain precisely the action \eqref{second
action}. Therefore the two actions are equivalent for finite values of
$\la$. But now we can take the limit $\la \to \infty$ in the new
action. The resulting action is
\begin{equation}    \label{third action}
S_\infty = -i \int_{I} \left( p_\mu
\left( \frac{dx^\mu}{dt} + g^{\mu\nu} \frac{\pa f}{\pa x^\nu} \right)
- \pi_\mu \left( D_t \psi^\mu - g^{\mu\nu} \frac{D^2 f}{D x^\nu D x^\al}
\psi^\al \right) \right) dt.
\end{equation}

Now the equations
\begin{equation}    \label{grad traj}
\frac{dx^\mu}{dt} = g^{\mu\nu} \frac{\pa f}{\pa x^\nu}
\end{equation}
are the equation of motion. Thus, the instantons (gradient
trajectories of $f$), which in the original theory corresponded to
absolute minima of the action, but were {\em not} the equations of
motion, have now become ones. At the same time anti-instantons
(gradient trajectories of $-f$) have disappeared.

We note that in the sum of the terms
$$
- \int_I g^{\mu\nu} \left( p_\mu \frac{\pa f}{\pa x^\nu}
- \pi_\mu g^{\mu\nu} \frac{D^2 f}{D x^\nu D
x^\al} \psi^\al \right) dt
$$
in the above action we can replace $dt$ by an arbitrary connection
$A_t$ on a principal $\R$-bundle on $I$. This observation will be very
useful in the context of two-dimensional sigma models.

We now describe how the coordinate invariance is realized in the above
action. The bosonic variables $x^\mu$ and $p_\mu$ transform as
functions and one-forms on $X$, respectively. The fermionic variables
${\pi}_{\m}$, ${\psi}^{\m}$ transform as sections of the cotangent and
tangent bundles to $X$, respectively. Note that we have
$$
D_t \psi^\la = \frac{d\psi^\la}{dt} + {\Gamma}_{\mu\nu}^{\la}
\frac{dx^\mu}{dt} \psi^\nu,
$$
where ${\Gamma}$ is the Levi-Civita connection on the tangent bundle
$TX$. Therefore if we redefine $p_\mu$ as follows:
\begin{equation}
\label{altp}
p'_{\m} = p_{\m} + {\Gamma}_{\m\nu}^{\la} {\psi}^{\nu}
{\pi}_{\la},
\end{equation}
we absorb the connection operators into $p'_\mu$ and obtain the
following formula for the action:\footnote{more generally, we could
use another connection in formula \eqref{altp}}
\begin{equation}    \label{fourth action}
S_\infty = -i \int_{I} \left( p'_\mu \left(
\frac{dx^\mu}{dt} - g^{\mu\nu} \frac{\pa f}{\pa x^\nu} \right) -
\pi_\mu \left( \frac{d\psi^\mu}{dt} - \frac{\pa}{\pa x^\al}
\left(g^{\mu\nu} \frac{\pa f}{\pa x^\nu} \right) \psi^\al \right)
\right) dt.
\end{equation}
However, the new momenta $p'_\mu$ no longer transform as one-forms.

Indeed, we have under the coordinate transformation $x^\mu \mapsto
{\tilde x}^\nu$:
$$
{\psi}^\mu \mapsto {\tilde\psi}^\mu = {\psi}^\nu \frac{{\partial
        \tilde     x^\nu}}{{\partial x}^\mu} \ , \qquad {\pi}_\mu \mapsto
        {\tilde\pi}_\mu = {\pi}_\nu \frac{{\partial x}^\nu}{{\partial
        \tilde x}^\mu}.
$$
This transformation law forces $p'_{\m}$ to transform inhomogeneously:
\begin{equation}    \label{tr for p}
p'_\mu \mapsto {\tilde p}'_\mu = p'_\nu \frac{\partial
      x^\nu}{\partial\tilde x^\mu} +
\frac{{\partial^{2} x^\al}}{\partial {\tilde x}^\mu \partial
{\tilde x}^\nu} \frac{\partial {\tilde x}^\nu}{\partial x^\beta}
      \pi_\al \psi^\beta.
\end{equation}

The action \eqref{fourth action} is invariant under the supersymmetry
generated by the supercharges $Q$ and $Q^*$ defined by the formulas
\begin{eqnarray*}
{Q} x^{\m} &= {\psi}^{\m} , \qquad {Q} {\psi}^{\m} &= 0\\
{Q} {\pi}_{\m} &= p'_{\m}, \qquad {Q} p'_{\m}
&= 0, \\
Q^* x^\m &= 0, \qquad Q^* \psi^\mu &= g^{\mu\nu} \frac{\pa f}{\pa
       x^\nu}, \\
Q^* \pi_\mu &= 0, \qquad Q^* p'_\mu &= 0.
\label{modifiedq}
\end{eqnarray*}
They correspond to the de Rham differential and the contraction
operator $\imath_{\nabla f}$, respectively.

In particular, we find that the Lagrangian is $Q$-exact:
$$
L = -i Q \cdot \pi_\mu \left( \frac{dx^\mu}{dt} - g^{\mu\nu}
\frac{\pa f}{\pa x^\nu} \right).
$$

Recall that the deformation from $\la = \infty$ back to finite $\la$
is achieved by adding the terms
$$
\frac{1}{2} \la^{-1} g^{\mu\nu} p_\mu p_\nu + \frac{1}{2} \la^{-1}
R^{\mu\nu}_{\al\beta} \pi_\mu \pi_\nu \psi^\al \psi^\beta
$$
to Lagrangian \eqref{third action}. It is important to note that, just
like the Lagrangian  \eqref{third action}, this expression is
$Q$-exact and equal to
\begin{equation}    \label{Q-exact}
Q \cdot \frac{1}{2} \la^{-1} g^{\mu\nu} \pi_\mu p'_\nu.
\end{equation}

\ssec{Correlation functions as integrals over moduli spaces of
       instantons}    \label{cor fns as int}

We now discuss the correlation functions in our model in the
Lagrangian, i.e., path integral formalism. It turns out that these
correlation functions may be represented by integrals over
finite-dimensional moduli spaces of gradient trajectories.

The first question to ask is what are the observables of our theory.
Typically, observables in quantum mechanics on a manifold $Y$ are
obtained by quantizing functions on $T^{*} Y$, which in our case is
$T^*\left( {\Pi} T X \right)$. The simplest are the observables
corresponding to the functions on ${\Pi}TX$, which are the same as
differential forms on $X$. These are quantized in a straightforward
way in the coordinate polarization. In the original quantum mechanical
model, for finite $\la$, the operator, corresponding to a differential
$n$-form ${\om}$ on $X$ (which is the same thing as a function on
${\Pi}TX$, which is a degree $n$ polynomial in the fermionic
variables), is the operator of multiplication by $\om$. The
correlation functions of these observables are easiest to compute in
the $\la \to \infty$ limit by using the path integral.

To see that, consider the following finite-dimensional model
situation: a vector space $\R^M$ and functions $f^a, a=1,\ldots,N$,
defining a codimension $N$ submanifold $C \subset \R^M$. Then the
delta-like differential form supported on this subvariety has the
following integral representation:
$$
\delta_C = \int \prod_a dp_a d \pi_a e^{i p_a f^a + i \pi_a df_a}.
$$
This delta-form may be viewed as the limit, when $\la \to \infty$, of
the regularized integral
$$
\delta_{C,\la} = \int \prod_a dp_a d \pi_a
e^{i p_a f^a + i \pi_a df_a - \la^{-1} p_a p_{\ol{a}}}.
$$

Comparing these formulas to \eqref{third action} and \eqref{second
action prime}, we see that the path integral
$$
\int Dp D\pi e^{-S_\infty}
$$
looks like the delta-like form supported on the gradient trajectories,
solutions of the equations \eqref{grad traj}, while $\int Dp D\pi
e^{-S_\la}$ may be viewed as its regularized version. More precisely,
the integral $\int Dp D\pi e^{-S_\la}$ should be viewed as the
Mathai-Quillen representative of the Euler class of an appropriate
vector bundle over the space of gradient trajectories (see
\cite{BS,Moore,KL}).

In a similar way one shows that in the limit $\la \to \infty$ the
correlation functions in our theory will be equal to integrals of
differential forms over the moduli spaces of gradient
trajectories. Note that the integral over fluctuations around the
instanton solutions contributes only the one-loop determinants, which
cancel each other out for bosonic and fermionic degrees of freedom, up
to a sign. Moreover, this sign disappears in the case that we are most
interested in: when $X$ is a K\"ahler manifold and the Morse function
$f$ satisfies the conditions listed in \secref{ground} below.

In particular, the kernel of the evolution operator in our theory is
just the delta-form supported on the submanifold
of those pairs $(x_i,x_f) \in X \times X$ which are connected by the
gradient trajectories $x(t): I_{t_i,t_f} \to M$ such that $x(t_i) =
x_i$ and $x(t_f) = x_f$.

\smallskip

{}From now on we will focus on the case of the infinite line $I =
\R$.

\smallskip

The gradient trajectories $\R \to X$ necessarily start and end at the
critical points of $f$ (recall that we have assumed that they are
isolated). The corresponding moduli space is therefore a a union of
connected components labeled by pairs of critical points of $f$,
$x_{-}$ and $x_{+}$, which play the role of the boundary conditions in
the path integral. Let ${\CM}_{x_{-}, x_{+}}$ be the moduli space of
the gradient trajectories, that is solutions $x(t)$ to (\ref{grad
traj}), which obey:
\begin{equation}
\label{pmcnd}
x(t) \longrightarrow x_{\pm}, \qquad t \to \pm \infty.
\end{equation}
We have evaluation maps
\begin{equation}
\label{eval}
{\rm ev}: {\CM}_{x_{-}, x_{+}} \times {\R} \longrightarrow X, \qquad
\on{ev}_{t} : {\CM}_{x_{-}, x_{+}} \longrightarrow X
\end{equation}
$$
{\rm ev} ( m , t) = x_{m}(t), \qquad \on{ev}_{t} ( m ) = x_{m} ( t) \\
$$

The simplest observables of our theory correspond to differential
forms on $X$. They are called the {\em evaluation observables}. The
correlation function of the evaluation observables $\wh\omega_i$
corresponding to differential forms $\omega_i, i=1,\ldots,n$, on $X$,
in the sector corresponding to the boundary conditions $x_\pm$ is
given by the integral
\begin{equation}    \label{corrf}
\langle_{\kern -.2in x_{-}} \ \wh{\om}_{1} ( t_{1}) \
\wh{\om}_{2}(t_{2}) \ \ldots \ {\om}_{k}( t_{k})
\rangle_{x_{+}} =
\int_{{\CM}_{x_{-}, x_{+}}} \ \on{ev}_{t_{1}}^{*}{\om}_{1}
\wedge \on{ev}_{t_{2}}^{*}{\om}_{2} \wedge \ldots \wedge
\on{ev}_{t_{k}}^{*} {\om}_{k}
\end{equation}

If the forms $\omega_{i}$ have definite cohomological degrees, then,
according to formula \eqref{corrf}, the above integral is
non-vanishing only if the following selection rule\footnote{fermionic
charge conservation} is obeyed:
\begin{equation}
\sum_{i=1}^{n} {\rm deg} \; {\om}_{i} = {\rm dim}{\CM}_{x_{-},
x_{+}} = {n}_{x_{+}} - {n}_{x_{-}}
\label{indi}
\end{equation}
where ${n}_{x}$ is the index of the critical point $x$.

Let us note that the correlation function \Ref{corrf} is invariant
with respect to the time shift $t \mapsto t + {\rm const}$. This
invariance is verified by the expression (\ref{corrf}) due to the
fact that the time shifts act on ${\CM}_{x_{-}, x_{+}}$. Indeed, if
$x_{m}(t)$ is the gradient trajectory corresponding to a point $m \in
{\CM}_{x_{-}, x_{+}}$, then so is
$$
x_{m^{s}}(t) = x_{m}(t+s).
$$
Thus, we obtain an action of the transformations $g^{s}: m \mapsto
m^{s}, s \in \R$, on ${\CM}_{x_{-}, x_{+}}$.

Since the integral \eqref{corrf} is not changed by the changes of the
integration variables, the simultaneous time shift $t_{i} \mapsto
t_{i} +s$, which can be absorbed into the change of moduli $m \mapsto
m^s$, does not affect the correlation function.

Note that if we wish to distinguish contributions of different types
of instantons, running between different critical points, we may also
add a finite term
\begin{equation}    \label{Delta S theta}
\Delta S_\tau = -i \tau \int_{-\infty}^{+\infty} df
\end{equation}
to the action \eqref{fourth action}. Then the above correlation
function will get multiplied by the factor $e^{i {\tau} (f(x_+) -
f(x_-))}$.

A natural way to obtain the term \eqref{Delta S theta} is as follows:
introduce an additional parameter $\vartheta$, the ``$\vartheta$
angle''. Let us set
$$
\tau = \vartheta + i\la, \qquad \ol\tau = \vartheta - i\la.
$$
Let us add to the second order action
\eqref{first action}, instead of \eqref{the term}, the term
$$
-i\vartheta \int_I df = - \la \int_I df - i \tau \int_I df.
$$
Consider the limit when $\la \to +\infty, \vartheta \to -i\infty$ so
that $\tau = i(\la-|\vartheta|)$ remains finite, but $\ol\tau \to
-i\infty$. In this limit we recover the first order Lagrangian
\eqref{fourth action} that we have previously obtained in the $\la \to
\infty$ limit, but with the term \eqref{Delta S theta} added.

The additional coupling constant $\vartheta$ is the precursor of the
$B$-field in two-dimensional sigma models and the $\vartheta$ angle in
four-dimensional Yang-Mills theory (this will be discussed in detail
in Part II). The interpretation viewing the $\la \to \infty$ limit as
the limit $\ol\tau \to \infty$ has a direct generalization to quantum
field theories in two and four dimensions, which we consider in Part
II. However, in the context of quantum mechanics, this does not add
much extra value. Indeed, since the instanton moduli space has
finitely many components (labeled by pairs of critical points
$x_-,x_+$), separating the contributions of different components with
the weight factor $e^{i {\tau} (f(x_-) - f(x_+))}$ does not make much
of a difference (unlike the two-dimensional and four-dimensional
models, where the instanton moduli spaces have infinitely many
components).

\ssec{Topological sector}    \label{top sector}

Let us suppose now that our forms $\omega_i$ are closed, $d
{\om}_{i} = 0$. Since the supersymmetry charge $Q$ of our
model at $\la = \infty$ corresponds to the de Rham differential, this
means that the corresponding observables $\wh\omega_i$ are
$Q$-closed. The correlation functions (\ref{corrf}) simplify
considerably in this case.

This simplification is particularly drastic in the case when $X$ is a
K\"ahler manifold, and so we will focus on this case from now
on. There are two reasons for that. The first reason is that in the
calculations below we would like to use the fact that the integral of
an exact differential form over ${\CM}_{x_{-}, x_{+}}$ is equal to
$0$. But ${\CM}_{x_{-}, x_{+}}$ is not compact, and so this statement
is not true in general. Note that ${\CM}_{x_{-}, x_{+}}$ is the
intersection of the descending manifold $X^{x_+}$ of $x_+$ and the
ascending manifold $X_{x_-}$ of $x_-$. Recall that $X^{x_+}$ consists
of the possible values at $t=0$ of the gradient trajectories
$[0,+\infty) \to X$ whose value at $t=+\infty$ is the critical point
$x_+$. Likewise, $X_{x_-}$ consists of the possible values at $t=0$ of
the gradient trajectories $(-\infty,0] \to X$ whose value at
$t=-\infty$ is $x_-$.

Now suppose that $X$ is a compact K\"ahler manifold and the Morse
function $f$ is the hamiltonian of a vector field corresponding to a
$U(1)$-action. Let us write this vector field as $i(\xi-\ol\xi)$,
where $\xi$ is a holomorphic vector field on $X$. Then the gradient of
$f$ is the vector field $\xi+\ol\xi$. In this case the descending and
ascending manifolds are isomorphic to $\C^n$. Let us assume in
addition that that the manifolds $X^{x_+}$ and $X_{x_-}$ form
transversal stratifications of $X$. Then $X^{x_+}$ has a natural
compactification $\ol{X}^{x_+}$ which is just the closure of $X^{x_+}$
inside $X$. Since the descending manifolds form a stratification of
$X$, this closure is the union of the descending manifolds $X^{x'_+}$,
where $x'_+$ runs over a subset of the critical points which are
``above'' $x_+$. Likewise, $X_{x_-}$ has a compactification which is
the union of $X_{x'_-}$ with $x'_-$ running over the set of critical
points that are ``below'' $x_-$.  But then the moduli space
${\CM}_{x_{-}, x_{+}}$ has a natural compactification obtained by
"gluing in" the moduli spaces ${\CM}_{x'_{-}, x'_{+}}$, where $x'_-$
and $x'_+$ are critical points that lie ``below'' $x_-$ and ``above''
$x_+$, respectively.\footnote{Note that the evaluation maps extend
naturally to this compactification.} Hence the moduli space
${\CM}_{x_{-}, x_{+}}$ is also a complex manifold and its complement
in the compactification has real codimension $2$. Therefore the
integral of an exact form over ${\CM}_{x_{-}, x_{+}}$ is equal to $0$.

The second reason why an additional simplification occurs for K\"ahler
manifolds is that in this case all ground states of the quantum
hamiltonian at the $\la \to \infty$ limit are $Q$-closed, and hence
they deform to true ground states for finite values of $\la$. Indeed,
the action of the supersymmetry charge $Q$ on the ground states is
given by the Morse differential, as shown in \cite{W:morse}. Since the
indices of the critical points are even if $X$ is a K\"ahler manifold,
its action is equal to $0$ in this case.

Let us now derive some properties of the $Q$-closed observables
corresponding to closed differential forms on a K\"ahler manifold
$X$. First of all, they are independent of the individual times $t_i,
i=1,\ldots,n$. Indeed, using the Cartan formula
\begin{equation}    \label{cartan formula}
{\mc L}_v = \{d,\imath_v\}
\end{equation}
for the vector field $v = \nabla f$, we find
$$
\frac{d}{dt} \on{ev}^{*}_{t} {\om}_{i} = - \on{ev}^{*}_{t} {\CL}_{v}
{\om}_{i} = - d \left( \on{ev}^{*}_{t} \iota_{v}  {\om}_{i}
\right),
$$
hence the $t$-derivative of the integral \eqref{corrf} is equal to
zero.

Another important property is that if all $\omega_i$'s are closed and
at least one of them is exact: $\omega_j = d \eta_j$, then the
corresponding correlation function vanishes. Indeed, we then find that
$$
\int_{{\CM}_{x_{-}, x_{+}}} \ \on{ev}_{t_{1}}^{*}{\om}_{1}
\wedge \ldots \wedge \on{ev}_{t_{k}}^{*}
{\om}_{{\al}_{k}} = \int_{{\CM}_{x_{-}, x_{+}}} \ d \left(
\on{ev}_{t_{1}}^{*}{\om}_{{\al}_{1}} \wedge \ldots
\on{ev}_{t_{j}}^{*}\eta_j \wedge \ldots \wedge \on{ev}_{t_{k}}^{*}
        {\om}_{k} \right) = 0.
$$
This is also clear from the point of view of the original path
integral, because the Lagrangian of our theory is $Q$-exact.

This has an important consequence: consider the theory at finite
values of $\la$. As we explained above, it can be viewed as a
deformation of the theory at $\la = \infty$ obtained by adding to the
Lagrangian the expression \eqref{Q-exact}. Since this expression is
$Q$-exact, the corresponding correlation function of $Q$-closed
observables will be independent of this deformation, and so the answer
that we obtain in the theory at $\la = \infty$ will remain valid at
finite values of $\la$ (at least in some neighborhood of $\la^{-1} =
0$).

Here it is important to note that the supersymmetry charge $Q$ of the
theory with the action \eqref{second action} is independent of $\la$
and corresponds to the de Rham differential (in particular, it is the
same for finite $\la$ as for $\la=\infty$). But the supersymmetry
charge of the ``physical'' theory with the action \eqref{first action}
differs from it by conjugation with $e^{-{\la}f}$. However, this
conjugation does not change the evaluation observables corresponding
to the differential forms, and therefore these observables are
$Q$-closed in both theories.

Thus, we arrive at the following conclusion: there is a sector of the
``physical'' theory which is independent of $\la$. It comprises the
$Q$-closed observables, corresponding to closed differential forms on
$X$. The correlation functions of these observables (on the infinite
line and with the boundary conditions $x_i = x_-, x_f = x_+$) are
given, for all values of $\la$, by integrals over the
finite-dimensional moduli spaces of instantons ${\CM}_{x_{-},
x_{+}}$.\footnote{more precisely, because the Lagrangian \eqref{second
action} differs from the ``physical'' Lagrangian \eqref{first action}
(for finite values of $\la$) by the term $\la df$, the correlation
functions of the ``physical'' theory will be equal to the correlation
functions of the first order theory at $\la = \infty$ times
$e^{-\la(f(x_+)-f(x_-))}$} The correlation functions of the $Q$-closed
observables $\wh\omega_i$ do not depend on the closed forms $\omega_i$
themselves, but only on their cohomology classes. For this reason this
sector of the theory is called the ``topological sector'' and the
corresponding theory is referred to as ``topological field
theory''. Alternatively, the $Q$-closed observables are referred to as
the ``BPS observables'' and the topological sector is called the ``BPS
sector''.

\ssec{Analogy with the Gromov-Witten theory}

It is instructive to note the analogy between the topological sector
of the Morse quantum mechanical model considered above and the
Gromov-Witten theory. This analogy will become more clear in Part II
of this article when we discuss the two-dimensional sigma models. This
material is discussed in more detail in \cite{Cohen}.

Let $X$ be a compact K\"ahler manifold and $\Sigma$ a compact Riemann
surface. The analogues of the moduli spaces $\CM_{x_-,x_+}$ in
Gromov-Witten theory are the moduli spaces ${\mathcal
M}_\Sigma(X,\beta)$ of holomorphic maps $\Phi: \Sigma \to X$ of a
fixed degree $\beta \in H_2(X)$. For a point $p \in \Sigma$ we have
evaluation maps $\on{ev}_p: {\mathcal M}_\Sigma(X,\beta) \to X$. Now,
given an $n$-tuple of points $p_1,\ldots,p_n$ and a collection of
differential forms $\omega_1,\ldots,\omega_n$ on $X$, we can consider
the integral
\begin{equation}    \label{fixed Sigma}
\int_{{\mathcal M}_\Sigma(X,\beta)} \on{ev}_{p_1}^*(\omega_1) \wedge
\ldots \wedge \on{ev}^*_{p_n}(\omega_n).
\end{equation}
We will assume for simplicity that $(\Sigma,(p_i))$ does not admit any
continuous automorphisms. This integral\footnote{the moduli space
${\mathcal M}_\Sigma(X,\beta)$ is not compact, but for compact $X$
this integral is well-defined for smooth differential forms $\omega_i$
on $X$ under the above assumption on $(\Sigma,(p_i))$} is analogous to
the integrals \eqref{corrf}. They are equal to correlation functions
of evaluation observables of the two-dimensional supersymmetric sigma
model with the target $X$ in the infinite radius limit (we will
consider this model in more detail in Part II).

Let ${\mathcal M}_{g,n}(X,\beta)$ be the moduli space of data
$(\Sigma,(p_i),\Phi)$, where $\Sigma$ is a genus $g$ Riemann
surface. Then we have a projection $\pi_{g,n}: {\mathcal
M}_{g,n}(X,\beta) \to {\mathcal M}_{g,n}$, and ${\mathcal
M}_\Sigma(X,\beta)$ is the fiber of $\pi_{g,n}$ at $(\Sigma,(p_i)) \in
{\mathcal M}_{g,n}$. We have natural evaluation maps $\on{ev}_i:
{\mathcal M}_{g,n}(X,\beta) \to X$. The general Gromov-Witten
invariants are the integrals
\begin{equation}    \label{GW}
\int_{{\mathcal M}_{g,n}(X,\beta)} \on{ev}_1^*(\omega_1) \wedge
\ldots \wedge \on{ev}^*_n(\omega_n).
\end{equation}
These are the correlation functions of what is often referred to as
the "sigma model coupled to gravity", and the observables are the
"cohomological descendents" of the evaluation observables.

Instead of integrating over ${\mathcal M}_{g,n}(X,\beta)$, we may take
the push-forward
\begin{equation}    \label{GW inv}
\pi_{g,n*}(\on{ev}_1^*(\omega_1) \wedge \ldots \wedge
\on{ev}^*_n(\omega_n)),
\end{equation}
which is a differential form on ${\mathcal M}_{g,n}$. In particular,
\eqref{fixed Sigma} occurs as a special case when the degree of this
differential form is equal to zero. Then its value at $(\Sigma,(p_i))
\in {\mathcal M}_{g,n}$ is given by \eqref{fixed Sigma}. More general
observables give rise to differential forms of positive degree on
${\mathcal M}_{g,n}$.

More precisely, we need to replace ${\mathcal M}_{g,n}(X,\beta)$ by
the Kontsevich's space of stable maps $\ol{\mathcal M}_{g,n}(X,\beta)$
and ${\mathcal M}_{g,n}$ by its Deligne-Mumford compactification
$\ol{\mathcal M}_{g,n}$.

In the quantum mechanical model the analogues of the moduli spaces of
holomorphic maps are the moduli spaces ${\mc M}_{x_-,x_+}$ of gradient
trajectories, and the analogues of the moduli spaces of stable maps
are compactifications of ${\mc M}_{x_-,x_+}$ discussed above. The
integrals \eqref{corrf} that we have considered so far are the
analogues of the integrals \eqref{fixed Sigma}.

The definition of the analogues of the more general integrals
\eqref{GW} is also straightforward. Let ${\mc M}_{x_-,x_+,n}$ be the
moduli space of data $(p_1,\ldots,p_n,x)$, where $p_1,\ldots,p_n$ are
distinct points of the real line, considered as an affine line, i.e.,
without a fixed coordinate, and $x: \R \to X$ is a gradient
trajectory. If we choose a coordinate $t$ on $\R$, then we can replace
$(p_1,\ldots,p_n)$ by the real numbers $(t_1,\ldots,t_n)$ and $x$ by a
parameterized map $x(t)$. Other coordinates are obtained by a shift $t
\mapsto t+u$. Therefore we may equivalently consider the data
$(t_1,\ldots,t_n,x(t))$ modulo the diagonal action of the group $\R$
of translations:
$$
(t_1,\ldots,t_n,x(t)) \mapsto (t_1+u,\ldots,t_n+u,x(t+u)).
$$

Note that the group of translations plays here the same role that the
group $PGL_2$ of M\"obius transformation of $\Sigma = \pone$ plays in
the Gromov-Witten theory. We have a natural map $\pi_n: {\mc
M}_{x_-,x_+,n} \to \on{Conf}_n$, where
$$
\on{Conf}_n = (\R^n \bs \Delta)/\R_{\on{diag}} \simeq \R_{>0}^{n-1},
$$
is the configuration space of $n$ points on the real line (here
$\Delta$ is the union of all diagonals). It plays the role of ${\mc
M}_{g,n}$. There is also a natural relative compactification $\ol{\mc
M}_{x_-,x_+,n}$ of ${\mc M}_{x_-,x_+,n}$ defined similarly to the
moduli spaces of stable maps (see \cite{Cohen}).

We have the evaluation maps $\on{ev}_i: \ol{\mc M}_{x_-,x_+,n} \to X$
corresponding to evaluating the map $x: \R \to X$ at the point
$p_i$. Now it is clear that the analogues of the general Gromov-Witten
invariants \eqref{GW inv} in Morse theory are obtained as the
push-forwards
\begin{equation}    \label{Morse inv}
\pi_{n*}(\on{ev}_1^*(\omega_1) \wedge \ldots
\wedge \on{ev}^*_n(\omega_n)).
\end{equation}
These "Morse theory invariants" are nothing but differential forms on
the configuration space $\on{Conf}_n$. The simplest examples are the
zero-form components of these differential forms whose values at fixed
points $p_1,\ldots,p_n$ are just the integrals \eqref{corrf}
introduced above. These more general correlation functions may be
interpreted as the correlation functions of the $\wh\omega_i$'s and
their "cohomological descendents", which are constructed following
\cite{W:mirror}.

Note that the observable $\wh\omega_i$ is a $0$-form on the "worldline"
$\R$, i.e., a function. The cohomological descendant of $\wh\omega_i$ is
the $1$-form $\wh\omega_i^{(1)}$ on $\R$ defined by the formula
$$
\wh\omega_i^{(1)} = \wh{\imath_{v} \omega_i} dt,
$$
where $v$ is the gradient vector field $\nabla f$. In particular,
suppose that $\omega_i$ is a closed differential form on $X$. Since
$\wh\omega_i$ is obtained from $\omega_i$ by pulling back with respect to
a gradient trajectory, we have $d\wh\omega_i/dt = \wh{{\mc L}_{v}
\omega_i}$, where ${\mc L}_{v}$ is the Lie derivative with respect to
the gradient vector field $v$. Since the action of $Q$ corresponds
to the action of the de Rham differential $d_X$ along $X$, we obtain,
using the Cartan formula ${\mc L}_{\nabla f} = \{ d_X,\imath_{\nabla
f}\}$, that
$$
Q \cdot \wh\omega_i^{(1)} dt = d_t \wh\omega_i,
$$
where $d_t$ is the de Rham differential along the worldline. This is
analogous to the formula for the cohomological descent in the
Gromov-Witten theory \cite{W:mirror}.

The general "Morse theory invariants" \eqref{Morse inv}
may be interpreted as correlation functions of observables of this
type alongside the $\wh\omega_i$'s considered before.

More precisely, when we take the pull-back of a differential $p$-form
$\omega_i$ on $X$ via $\on{ev}_i: \ol{\mc M}_{x_-,x_+,n} \to X$, we
obtain an $p$-form on $\ol{\mc M}_{x_-,x_+,n}$, which decomposes
locally into the sum of two differential forms. One is a $p$-form
along the fiber of the projection $\pi_n$ and $0$-form along the base:
it corresponds to $\wh\omega_i$, and the other is a $(p-1)$-form along
the fiber and a $1$-form along the base: this is
$\wh\omega_i^{(1)}$. Thus, the correlation function of $m$
"zero-observables" $\wh\omega_i$ and $k$ "one-observables"
$\wh\omega_j^{(1)}$ will pick up precisely the $k$-form component of
the general "Morse theory invariant" on $\on{Conf}_{m+k}$ defined by
formula \eqref{Morse inv}.

{}From the physical perspective, defining these more general
correlation functions corresponds to "coupling our quantum mechanical
model to gravity". In the path integral formalism it is described as
follows. We write our action in the form
$$
S = \int (- i p' \, \dot q + i \pi \dot \psi + H dt),
$$
where $H$ is the classical hamiltonian whose quantization gives ${\mc
L}_v$, and ${Q}^*$ is its superpartner whose quantization gives
$\imath_v$. In order to enforce the invariance under the time
reparameterizations and at the same time preserve the $Q$-symmetry we
add the einbein field $e$ and its superpartner $\chi = Q \cdot e$ and
consider the action
\begin{equation}
S_{\on{top \ grav}} =  \int (- i p' \, \dot q + i \pi \dot \psi + ( e
H - {\chi} {Q}^{*} ) d t). \label{grav1}
\end{equation}
The path integrals corresponding to this action may be expressed in
terms of the "Morse theory invariants" \eqref{Morse inv}.

Note also that more generally we may consider arbitrary graphs instead
of the real line. We then need to assign to each edge of the graph a
Morse function and impose the condition that the sum of the functions
corresponding to the edges coming out of each vertex is zero. The
corresponding integrals are related to the integrals considered by
Fukaya \cite{Fukaya}. They may also be interpreted as the terms of the
perturbative expansion of a particular quantum field theory on
$X$. For instance, if we only allow three-valent graphs, this will be
the perturbative expansion of the (generalized) Chern-Simons theory on
$X$ around a particular background, and can be viewed as a topological
open string on $T^{*}X$, as in \cite{WittenCSs,SchwarzA}. More
precisely, if we allow $N$ different Morse functions $f_1,\ldots,f_N$,
then this will be the Chern-Simons theory with the Lagrangian
$$
L = \on{Tr} (A \wedge dA + \frac{2}{3} A \wedge A \wedge A),
$$
where $A$ is an $N \times N$-valued differential form on $X$, and we
make the perturbation theory expansion around the one-form
$\on{diag}(df_1,\ldots,df_N)$. This is however beyond the scope of the
present paper.

For more on the topological supersymmetric quantum mechanics, see
\cite{Lysov,LP}.

\ssec{More general observables}    \label{other observables}

As we discussed above, the observables in the BPS sector correspond to
closed differential forms on $X$. However, {\em any} differential form
$\omega$ on $X$ gives rise to a legitimate observable in our theory,
and the correlation functions of such observables are still given by
the same integrals over the moduli spaces ${\mc M}_{x_-,x_+}$ (or
${\mc M}_{x_-,x_+,n}$) as in the case of closed forms. The difference
is, of course, that these correlation functions are no longer
independent of $\la$, so this answer is correct only at $\la=\infty$.

Why should we bother considering non-BPS evaluation observables?  We
have already partially answered this question in the Introduction. In
particular, if we only consider the BPS observables, we cannot gain
any insights into the structure of the space of states of our theory
beyond the ground states. Indeed, their correlation functions only
depend on their $Q$-cohomology classes. One can modify any BPS
observable by $Q$-exact terms so as to make it commute with $Q$ and
$Q^*$. Such a representative transforms a ground state into a ground
state. But non-BPS observables transform ground states into excited
states, and, as we will see below, we can understand the structure of
the space of states by considering their correlation functions.

Besides, considering non-BPS observables allows us to bring into play
some important $Q$-exact observables, which are "invisible" in the BPS
sector.

A general local observable of the quantum mechanical model that we are
considering corresponds to the quantization of an arbitrary function
${\CO}( x , p , {\psi} , {\pi})$. Upon quantization they become {\em
differential operators} on ${\Omega}^{\bullet}(X)$:
$$
{\CO} ( x, p , {\psi}, {\pi} ) \mapsto {\hat \CO} = {\CO} ( x , - i
\frac{\p}{{\p}{x}} , {\psi} , - i \frac{\p}{\p \psi} )
$$
Examples are the differential forms themselves, which we have already
considered above, and the Lie derivatives ${\mc L}_v$ with respect to
vector fields on $X$. If we write $v = v^\mu \pa/\pa x^\mu$, then
$$
{\mc L}_v = i \left( v^\mu p'_\mu + \frac{\pa v^\mu}{\pa x^\nu} \psi^\nu
\pi_\mu \right).
$$
These observables have a transparent path integral
interpretation. Namely, inserting this observable at the time $t_0$
corresponds to infinitesimally deforming the gradient trajectory at
the time $t_0$ along the vector field $v$. In particular, our
hamiltonian is included among these observables.

The observables ${\mc L}_v$ are $Q$-closed, but they are also
$Q$-exact, as follows from the Cartan formula ${\mc L}_v = \{
d,\imath_v \}$ that we have already encountered above.

This means that if we insert the operator ${\mc L}_v$ into a
correlation function of BPS observables, then we will obtain zero. But
the observables ${\mc L}_v$, and other differential operators, play a
very important role in the full theory. Indeed, on a K\"ahler manifold
we often have a large Lie algebra of global holomorphic vector fields,
and the corresponding Lie derivatives will be chiral operators of our
theory. The algebra of holomorphic differential operators that they
generate is the precursor of the chiral de Rham complex of
\cite{MSV}. If we wish to understand the role played by the chiral de
Rham complex in the two-dimensional sigma models, it is natural to
consider its quantum mechanical analogue: the algebra of holomorphic
differential operators on $X$ (more precisely, its supersymmetric
analogue: the algebra of holomorphic differential operators on $\Pi
TX$). But in order to obtain non-trivial correlation functions
involving these operators we must consider non-BPS observables.

As we already indicated above, at $\la=\infty$ the correlation
functions of non-BPS evaluation observables are easy to obtain from
the path integral point of view. Our goal now is to give an
interpretation of these correlation functions from the Hamiltonian
point of view. In other words, we wish to describe explicitly the
space of states of the theory at $\la=\infty$, represent the general
observables as operators acting on this space and represent their
correlation functions as matrix elements of these operators. We will
take up this task in the next section.

\section{Hamiltonian formalism}    \label{Hamiltonian formalism}

In the previous section we discussed the Lagrangian (or path integral)
formulation of the supersymmetric quantum mechanical model on a
K\"ahler manifold $X$ governed by the first order action \eqref{third
action} in the limit $\la=\infty$. This formulation is convenient
because it gives a simple answer for the correlation functions of the
observables corresponding to differential forms on $X$: they are given
by integrals over finite-dimensional moduli spaces of gradient
trajectories.

Now we would like to develop the Hamiltonian formalism for this
model. This means that we need to define the space of states of the
model and realize our observables as linear operators acting on this
space of states. The correlation functions are then given by matrix
elements of these operators. Note that the Lagrangian description
provides us with an important testing device: these matrix elements
should reproduce the finite-dimensional integrals described above.

We will see in this section that the Hamiltonian structure of our
model is rather unusual: the quantum Hamiltonian is non-hermitean and
even non-diagonalizable, and the space of states decomposes into the
spaces of ``in'' and ``out'' states. However, these spaces have a
simple and geometrically meaningful description in terms of the
stratifications by the ascending and descending manifolds of our Morse
function. Moreover, the spaces of states exhibit holomorphic
factorization that is absent for finite values of $\la$. This leads to
a great simplification of the correlation functions. In
\secref{action} we will establish the equivalence between the results
obtained in the Hamiltonian and the Lagrangian formalisms, discovering
along the way some interesting identities on integrals of analytic
differential forms.

\ssec{Supersymmetric quantum mechanics at $\la=\infty$}

Our task is to describe the Hamiltonian formalism of the theory with
the first order action \eqref{third action}. Following the most
obvious route, we start with the classical hamiltonian found from this
action:
$$
H_{\on{class}} = i \left( p'_\mu g^{\mu\nu} \frac{\pa f}{\pa x^\nu}
+ \frac{\pa}{\pa x^\al} \left( g^{\mu\nu} \frac{\pa f}{\pa x^\nu}
\right) \pi_\mu \psi^\al \right).
$$
Its naive quantization is the operator
$$
H_{\rm naive} = {\CL}_{v},
$$
which is the Lie derivative with respect to the gradient vector field
$v = \nabla f$.\footnote{the importance of considering such first
order hamiltonians has been emphasized by G. 't Hooft}

The corresponding supersymmetry charges are $Q = d$ and $Q^* =
2\imath_v$. They satisfy the relation
$$
H_{\rm naive} = \frac{1}{2} \{ Q , Q^{*} \}
$$
according to Cartan's formula \eqref{cartan formula}.

The standard Hodge theory, which is at work in the quantum mechanical
model at finite values of $\la$ (see \secref{recollections morse}),
deals with the operators like $d$ and $d^*$, which depend
explicitly on the metric on the manifold $X$.  Here we choose instead
the operators $d$ and $2\iota_{v}$, for some vector field $v$, which
are metric independent. The problem with this definition is that the
Hamiltonian ${\CL}_{v}$ is a first order differential operator, which
naively has an unbounded spectrum.  Also, if we take the time
evolution operator
$$
e^{- t H_{\rm naive}}
$$
for large $t$, it will tend to make the wave functions concentrated
near the critical points of the Morse function (where $v=0$), thus
posing some problems with completeness. Besides, the choice of zero
eigenstates of $H_{\rm naive}$ seems quite ambiguous, for any
differential form (or current) supported on a $v$-invariant
submanifold in $X$ naively leads to such a state.

Finally, in the standard supersymmetric quantum mechanics the
operators $\CQ$ and $\CQ^{*}$ are adjoint to each other, and as the
result, the Hamiltonian is self-adjoint. But this property no longer
holds in our case, and so we cannot expect that ${\CL}_{v}$ is
self-adjoint.

\ssec{Way out: $\la$ regularization}
\label{way out}

In order to make sense of all this, we recall how we got the action
\eqref{third action} in the first place: we started with the second
order action \eqref{first action} for finite values of $\la$, then we
added the topological term
\begin{equation}    \label{addition}
- \int_I \la df = \la(f(x_i) - f(x_f)),
\end{equation}
passed to the first order action \eqref{second action}, and finally
took the limit $\la \to \infty$. This suggests that in order to
develop the correct Hamiltonian formalism of the theory we should
retrace these steps from the Hamiltonian point of view.

We recall from \secref{recollections morse} that the hamiltonian of
the theory with the action \eqref{first action} is given by formula
$$
H = \frac{1}{2} \{ {\mathcal Q} , {\mathcal Q}^{*} \},
$$
where
$$
{\CQ} = d + {\la} df \wedge \  ,  \qquad \ {\CQ}^{*} = \frac{1}{\la}
d^{*} + \iota_{\nabla f}.
$$
This hamiltonian is Witten's Laplacian given by formula
\eqref{ham}. The corresponding space of states is just the space of
$L_2$ differential forms on $X$. This is a Hilbert (super)space with
respect to the hermitean inner product \eqref{norm}.

Here is is important to note that because we rescale by $\la$ both the
Morse function and the {\em metric} on $X$ (which leads to the overall
factor $\la^{-1}$ in the formula for $\CQ^*$), we obtain Witten's
Laplacian multiplied by $\la^{-1}$. This difference in normalization
is important: in the normalization of \cite{W:morse} the smallest
non-zero eigenvalue of the Hamiltonian is of the order of $\la$, and
hence only the ground states survive in the limit $\la \to \infty$
(recall that we are under the assumption that $X$ is K\"ahler, and so
all ground states at $\la=\infty$ deform to true ground states at
finite $\la$). In our normalization not only the ground states, but
also the states with the eigenvalues proportional to $\la$ (in the
normalization of \cite{W:morse}), survive in this limit. These will be
the excited states of our theory at $\la=\infty$, as we will see
below.

The next step is to add the term \eqref{addition} to the action. What
does this correspond to from the Hamiltonian point of view? To see
that, we recall the correspondence between the correlation functions
in the Lagrangian and Hamiltonian formalisms expressed in formula
\eqref{path int1}. When we add the topological term \eqref{addition} to
the action, the right hand side of this formula is multiplied by
$e^{\la(f(x_f)-f(x_i))}$. This means that the correlation function in
the new theory (with the term \eqref{addition}) between the ``in''
state $e^{\la f(x_i)} \Psi(x_i)$ and the ``out'' state $e^{-\la
f(x_f)} \Psi^*(x_f)$ is the same as the correlation function of the
old theory (without the term \eqref{addition}) between the states
$\Psi(x_i)$ and $\Psi^*(x_f)$.

Thus, we obtain that from the Hamiltonian point of view the effect of
the addition of the term \eqref{addition} to the action is the
following:
\begin{itemize}
\item the ``in'' states get multiplied by $e^{\la f(x)}$:
\begin{equation}    \label{inout1}
\Psi \mapsto \wt\Psi = e^{\la f} \Psi;
\end{equation}

\item the ``out'' states get multiplied by $e^{-\la f(x)}$:
\begin{equation}    \label{inout2}
{\Psi}^{*} \mapsto {\wt\Psi}^{*} = e^{-{\la}f} {\Psi}^{*};
\end{equation}

\item the operators get conjugated:
$$
{\mc O} \mapsto \wt{\mc O} = e^{\la f} {\mc O} e^{-\la f}.
$$
\end{itemize}

The last rule applies, in particular, to the operators $\CQ$, $\CQ^*$
and $H$. We find that the new operators are\footnote{Note the
difference between this conjugation and the procedure by which we had
defined the operators $\CQ$ and $\CQ^*$ in formulas \eqref{susy1},
\eqref{susy2}: there we defined $\CQ$ by conjugating $d$ by $e^{-\la
f}$, and then defined $\CQ^*$ as the {\em adjoint} of $\CQ$, so that
$\CQ^*$ was obtained by conjugating $d^*$ by $e^{\la f}$ (and dividing
by $\la$). In particular, their anti-commutator $\{ \CQ,\CQ^* \}$ has
a very different spectrum from $\{ d,d^* \}$. Now we are conjugating
both $\CQ$ and $\CQ^*$ by $e^{\la f}$. The resulting operators are
$Q=d$ and $Q^* = \frac{1}{\la} e^{2\la f} d^* e^{-2\la f}$. They are
not adjoint to each other any more. But the corresponding
anti-commutator $\{ Q,Q^* \}$ has the same spectrum as $\{ \CQ,\CQ^*
\}$ for any finite value of $\la$.}
\begin{align}
Q_\la &= \wt{\CQ} = e^{\la f} \CQ e^{-\la f} = d, \\
Q^*_\la &= \wt{\CQ}^* = e^{\la f} {\CQ}^{*} e^{-\la f} = 2 \iota_{v} +
\frac{1}{\la} d^{*}, \\
\wt{H}_\la &= e^{\la f} H e^{-\la f} = \frac{1}{2} \{ Q_\la,Q^*_\la \} =
{\mc L}_v - \frac{1}{2\la} \Delta.
\label{hlat}
\end{align}

Finally, we may take the limit $\la \to \infty$ and we indeed recover
the operators $Q, Q^*$ and $H_{\on{naive}}$ that we discussed above.

However, now we obtain a more clear picture of what is happening with
the space of states as we perform the above procedure. Namely, to
obtain the ``in'' space of states we need to choose a suitably
normalized basis of eigenfunctions of the Hamiltonian $H$, then
multiply all of them by $e^{\la f}$ and pass to the limit $\la \to
\infty$. Likewise, to obtain the ``out'' space of states we need to
multiply those basis eigenfunctions by $e^{-\la f}$ and then pass to
the limit. The main question is of course what we mean by ``taking the
limit'', in particular, in what ambient space the limiting elements
``live''. As we will argue in \secref{structure}, the proper ambient
space is not the space of functions on $X$ (or, more generally,
differential forms), but the space of {\em generalized functions},
i.e., a suitable space of linear functionals on the space of functions
(or, more generally, {\em currents}, i.e., functionals on the space of
differential forms). We will first explain how this works in the flat
space, namely, for $X=\C$, and then discuss in detail the first
example of a ``curved'' space, namely, $X=\pone$.

Before proceeding with these examples, we wish to comment on the
reason why the spaces of ``in'' and ``out'' states turn out to be
different in the limit $\la \to \infty$. This is easiest to explain
from the Lagrangian point of view. The ``in'' and ``out'' states in a
general quantum mechanical model may be constructed by acting by local
observables on the vacuum and covacuum states, respectively. Namely, let
${\mc O}_i(t_i), i=1,\ldots,n$, be observables with
$t_1<t_2<\ldots<t_n$ and $({\mc O}_n(t_n) \ldots {\mc O}_1(t_1)
\vac)(x)$ the corresponding state, considered as a differential
form on $X$. Then we have the following symbolic representation of
this state using the path integral:
\begin{equation}    \label{in st}
({\mc O}_n(t_n) \ldots {\mc O}_1(t_1) \vac)(x) = \underset{x(t):
(-\infty,0] \to X; \, x(0) = x} \int {\mc O}_1(t_1) \ldots {\mc
O}_n(t_n) e^{-S}.
\end{equation}
The boundary conditions at $t=-\infty$ in the above integral are
determined by the choice of the ground state $\vac$.  Likewise, an
``out'' state of our model is constructed by the formula
\begin{equation}    \label{out st}
(\langle 0| {\mc O}'_1(s_1) \ldots {\mc O}'_m(s_m))(x) =
\underset{x(t): [0,+\infty) \to X; \, x(0) = x}\int {\mc O}'_1(s_1)
\ldots {\mc O}'_m(s_m) e^{-S},
\end{equation}
where the boundary conditions at $t=+\infty$ are determined by the
choice of the covacuum state $\langle 0|$.

This construction allows us to define a natural pairing between the
two spaces: if we denote the state \eqref{in st} by $\Psi$ and the
state \eqref{out st} by $\Psi^*$, then by definition
\begin{equation}    \label{pairing}
\langle \Psi^*,\Psi \rangle = \underset{x(t):
(-\infty,+\infty) \to X}\int {\mc O}_1(t_1) \ldots {\mc O}_n(t_n) {\mc
     O}'_1(s_1) \ldots {\mc O}'_m(s_m) e^{-S},
\end{equation}
with appropriate boundary conditions understood.

In general, the spaces of ``in'' and ``out'' states are
different. However, suppose that the action of our theory is CPT
invariant, that is invariant under the time reversal $t \mapsto -t$
and complex conjugation (this corresponds to the Hamiltonian being a
self-adjoint operator). In this case we have a natural anti-linear map
from the space of ``in'' states to the space of ``out'' states:
namely, by applying the time reversal and complex conjugation we
transform the integral \eqref{in st} into an integral of the form
\eqref{out st}, hence we transform an ``in'' state into an ``out''
state. This allows us to identify the two spaces in the case when the
action is invariant under the CPT symmetry. Combining this
identification and the pairing \eqref{pairing}, we obtain a hermitean
inner product on the resulting (single) space of states.

However, our action is not invariant under the above CPT symmetry. The
original action \eqref{first action} is CPT-invariant, but the
topological term \eqref{addition} that we have added to it breaks this
invariance.\footnote{Note that the invariance would have been
preserved if $\la$ were purely imaginary, but we need $\la$ to be
real!} Therefore there is no natural identification between the spaces
of ``in'' and ``out'' states. For finite values of $\la$, the CPT
invariance is only mildly violated: when we apply the CPT
transformation, we shift the action by $2\la \int df$. This means that
there is still a map from the space of ``in'' states to the space of
``out'' states, but it involves multiplication by $e^{- 2 \la
f}$. More precisely, this map sends
$$
\Psi \mapsto e^{- 2\la f} \star \ol{\Psi}.
$$
However, the operation of multiplication by $e^{- 2 \la f}$ has no
obvious limit $\la \to \infty$, and so at $\la = \infty$ the two
spaces become {\em non-isomorphic}. This is reflected
in the fact that the action \eqref{third action} is not CPT-invariant
and the hamiltonian $H_{\on{naive}}$ is not self-adjoint. Thus, we
arrive at the following conclusion:

\medskip

{\it In the limit ${\la} \to \infty$ our model has two
spaces of states: the space of "in" states ${\CH}^{\on{in}}$, and the
space of "out" states ${\CH}^{\on{out}}$. The transition amplitudes
define a pairing:
$$
{\CH}^{\on{out}} \otimes {\CH}^{\on{in}} \longrightarrow {\C} \ , \
$$
but the two spaces are not canonically isomorphic.}

\ssec{The case of flat space $\C$}    \label{flat space}

Now we analyze the simplest example where we can follow the fate of
the states in the Hilbert space while taking the $\la \to \infty$
limit. This is the case $X = {\C}$,
$$
f =\frac{1}{2} {\om} |z|^2, \qquad g = dzd{\zb},
$$
where $\omega$ is a non-zero real number. Thus, $g_{z\ol{z}} =
\frac{1}{2}, g^{z\ol{z}} = 2$. The corresponding gradient vector field
is
\begin{equation}    \label{grad v f}
v = \nabla f = \omega(z\pa_z + \ol{z} \pa_{\ol{z}}).
\end{equation}
The potential is
$$
\vert df \vert^2 = \omega^2 |z|^2,
$$
and the Hamiltonian has the form
$$
H_\la = - \frac{2}{\la} \pa_z \pa_{\ol{z}} +
\frac{\la}{2} \omega^2 |z|^2 + K_\omega,
$$
where
$$
K_\omega = \omega(F+\ol{F}-1),
$$
where $F$ and $\ol{F}$ are the fermionic left and right charge
operators. Thus, $K_\omega$ is equal to $-\omega$ on $0$-forms, $0$
on $1$-forms, and $\omega$ on $2$-forms. Hence our model is nothing
but the two-dimensional supersymmetric harmonic oscillator.

We have the following orthonormal basis of eigenfunctions of $H_\la$:
the $0$-forms
\begin{equation}    \label{Psi}
\Psi_{n,\ol{n}} = \frac{1}{\sqrt{\pi (\la \omega)^{(n+\ol{n}-1)}
n!  \ol{n}!}} \; e^{\frac{1}{2} \la |\omega| \zzb}
       \; \pa_z^{\ol{n}} \pa_{\ol{z}}^n \; (e^{-\la |\omega| \zzb}),
       \qquad n,\ol{n} \geq 0,
\end{equation}
and the $1$-forms and $2$-forms obtained by multiplying
$\Psi_{n,\ol{n}}$ with $dz$ and $d\ol{z}$. The corresponding
eigenvalues are
\begin{equation}    \label{Enn}
E_{n,\ol{n}} = |\omega|(n+\ol{n}+1) + K_\omega.
\end{equation}
Note that these eigenfunctions have an additional property that they
are also eigenfunctions of the operators of $U(1)$ rotation $z \mapsto
z e^{i \varphi}$.

Now we describe the space of eigenfunctions of the conjugated operator
\begin{equation}    \label{wtHla}
\wt{H}_\la = e^{\la f} H_\la e^{-\la f} = \omega(z\pa_z + \ol{z}
\pa_{\ol{z}}) - \frac{2}{\la} \pa_z \pa_{\ol{z}} + (K_\omega +
\omega)
\end{equation}
and its adjoint. They will be basis elements of the ``in'' and ``out''
spaces of states. The eigenfunctions of $\wt{H}_\la$ are obtained by
multiplying the functions $\Psi_{n,\ol{n}}$ (and the corresponding
differential forms) with $e^{\la f} = e^{\frac{1}{2} \la \omega
z\ol{z}}$, and the eigenfunctions of its adjoint are obtained by
multiplying with $e^{- \la f} = e^{-\frac{1}{2} \la \omega z\ol{z}}$.

At this point the sign of $\omega$ becomes crucial. Let us assume first
that $\omega>0$. This means that the point $0$ is a ``repulsive''
critical point: the gradient trajectories flow away from $0$. In this
case a basis of the ``in'' space is given by the functions
\begin{equation}    \label{Psi in}
\wt\Psi^{\on{in}}_{n,\ol{n}} = \frac{1}{(\la\omega)^{n+\ol{n}}}
     \; e^{\la \omega z \ol{z}} \; \pa_z^{\ol{n}} \pa_{\ol{z}}^n \;
     e^{-\la \omega z \ol{z}}, \qquad n,\ol{n} \geq 0,
\end{equation}
and the differential forms obtained from them by multiplying with
$dz$ and $d\ol{z}$.  We recall that the ``out'' state corresponding to
an ``in'' state $\Psi$ is $e^{- 2\la f} \star \ol{\Psi}$. Therefore a
basis of the ``out'' space is given by the two-forms
\begin{equation}    \label{Psi out}
\wt\Psi^{\on{out}}_{n,\ol{n}} = \frac{\la\omega}{2\pi} \frac{1}{n!
    \ol{n}!} \; \pa_z^{n} \pa_{\ol{z}}^{\ol{n}} \; e^{-\la \omega z
    \ol{z}} dz d\ol{z}, \qquad n,\ol{n} \geq 0,
\end{equation}
and the differential forms obtained from them by contracting with the
vector fields $\pa_z$ and $\pa_{\ol{z}}$. The eigenvalues of
$\wt{H}_\la$ on these functions are given by the same formula
\eqref{Enn} (so they are independent of $\la$).

Here and below we use the notation
\begin{equation}    \label{d2z}
dz d\ol{z} = d^2 z = i dz \wedge d\ol{z} = 2 dx \wedge dy, \qquad z =
x+iy.
\end{equation}

The normalization in formulas \eqref{Psi in} and \eqref{Psi out} is
chosen in such a way that these expressions have well-defined limits
as $\la \to \infty$ (see below) and the pairing between
$\wt\Psi^{\on{in}}_{n,\ol{n}}$ and $\wt\Psi^{\on{out}}_{m,\ol{m}}$ is
equal to $\delta_{n,m} \delta_{\ol{n},\ol{m}}$. To obtain the ``in''
states satisfying this property, we multiply the states
$\Psi_{n,\ol{n}}$ by the function
$$
A_{n,\ol{n}} = \sqrt{\pi n! \ol{n}!} (\la \omega)^{-(n+\ol{n}+1)/2}
e^{\frac{1}{2} \la |\omega| z\ol{z}},
$$
and to obtain the ``out'' states we multiply $\Psi_{n,\ol{n}}$ by
$(A_{n,\ol{n}})^{-1}$. This suggests that the transformation from the
states of the theory at finite $\la$ to the ``in'' states of the new
theory, normalized as above, is achieved not merely by multiplying the
states by $e^{\la f}$, but by applying the operator
$$
\Psi \mapsto \wt\Psi^{\on{in}} = \la^{-\frac{1}{2}} e^{\la f} \left(
     \la^{-H/2|\omega|} \cdot \Psi \right).
$$
Likewise, for ``out'' states we have
$$
\Psi \mapsto \wt\Psi^{\on{out}} = \la^{\frac{1}{2}} e^{-\la f} \left(
     \la^{H/2|\omega|} \cdot \Psi \right).
$$

We now come to the key point of our analysis: finding the limits of
the states \eqref{Psi in} and \eqref{Psi out} as $\la \to
\infty$. First, we find from formula \eqref{Psi in} that in this limit
we have
$$
\wt\Psi^{\on{in}}_{n,\ol{n}} \to z^n \ol{z}^{\ol{n}},
$$
so the wave functions become monomials! On the other have, we find
that
$$
\wt\Psi^{\on{out}}_{n,\ol{n}} \to \frac{1}{n!
    \ol{n}!} \; \pa_z^{n} \pa_{\ol{z}}^{\ol{n}}
\delta^{(2)}(z,\ol{z}) dz d\ol{z}.
$$
Thus, the ``out'' states become the derivatives of the delta-form
supported at $0 \in \C$!

We conclude that the space of ``in'' states of our theory at
$\la=\infty$ is the space of polynomial differential forms on $\C$:
\begin{equation}
{\CH}^{\on{in}}= {\C}[ z , {\zb}] \otimes
{\Lambda}[ dz , d{\zb}] \ , \
\label{ins}
\end{equation}
on which the Hamiltonian $H=\wt{H}_{\infty}$ simply acts by
dilatations:
$$
e^{-tH} {\Psi} ( z, {\zb} , dz , d{\zb}) = {\Psi} ( q z , q{\zb},
q dz, q d{\zb} ),
$$
where $q=e^{-\omega t}$.

There is a unique ground state, $\wt{\Psi}^{\on{in}}_{\on{vac}} = 1$,
and the spectrum of excited states is degenerate, consisting of all
positive integers.

The space of "out" states is the space of "delta-forms" supported at
$z=0$:
\begin{equation}
{\CH}^{\on{out}} = {\Lambda} [ dz, d{\zb}] \otimes {\C} [ {\p}_{z},
{\p}_{\zb} ] \cdot {\dl}^{(2)}(z , {\zb}) \ , \
\label{outs}
\end{equation}
on which the evolution operator acts as:
$$
e^{-tH} {\Upsilon} ( dz , d{\zb}, {\p}_{z}, {\p}_{\zb} ) {\dl}^{(2)}(
z, {\zb}) = q^{2} {\Upsilon} ( q^{-1} dz, q^{-1} d{\zb}, q {\p}_{z}, q
{\p}_{\zb} ) {\dl}^{(2)}(z , {\zb}).
$$
We see that $H$ acts on ${\CH}^{\on{in}}$ as ${\CL}_{v}$, where $v$ is
the gradient vector field given by formula \eqref{grad v f}, and on
${\CH}^{\on{out}}$ as $-{\CL}_{v}$, yet the spectra of these two
seemingly opposite operators are identical. This is how the
"self-adjoint" nature of the Hamiltonian is realized in the $\la \to
\infty$ limit.

There is a natural pairing between the ``in'' and ``out'' spaces
defined by the formula
$$
\langle \wt\Psi^{\on{out}},\wt\Psi^{\on{in}} \rangle = \int
\wt\Psi^{\on{out}} \wedge \wt\Psi^{\on{in}}.
$$
This pairing is well-defined because $\wt\Psi^{\on{out}}$ is a
distribution (more precisely, a current) supported at $0 \in \C$ and
$\wt\Psi^{\on{out}}$ is a differential form that is smooth in the
neighborhood of $0$. This completes the analysis of the spaces of
states in the case when $\omega > 0$.

Now consider the case when $\om < 0$, which corresponds to an
``attractive'' critical point $0 \in \C$. Then the roles of
${\CH}^{\on{in}}$ and ${\CH}^{\on{out}}$ are reversed. Thus,
${\CH}^{\on{in}}$ is the space of delta-forms with support at $0 \in
\C$ and $\CH^{\on{out}}$ is the space of polynomials functions on
$\C$.

\ssec{The kernel of the evolution operator in the limit $\la \to
     \infty$}    \label{evolution}

It is instructive to analyze how the kernel of the evolution operator
at finite $\la$ becomes the delta-form supported on the gradient
trajectories in the limit $\la \to \infty$.

Suppose we study supersymmetric quantum mechanics on an
$n$-dimensional manifold $X$, and the space of states is the space of
$L_2$ differential forms on $X$. Then the kernel $K_t$ of the
evolution operator is an $n$-form on $X \times X$ defined by the
formula
$$
\langle \Psi^*|e^{-tH}|\Psi \rangle = \int_{X \times X} K_t(x,y) \wedge
\Psi(x) \wedge \star \ol{\Psi^*}(y).
$$
It is normalized so that $K_0$ is the delta-form (of degree $n$)
supported on the diagonal in $X \times X$.

Suppose that we have a complete basis $\{ \Psi_\ga \}$ of normalized
eigenfunctions of the Hamiltonian $H$, with the eigenvalues $\{ E_\ga
\}$. Then we have the following formula for $K_t$:
\begin{equation}    \label{int kernel}
K_t = \sum_\ga p_i^*(\Psi_\ga) p_f^*(\star \ol{\Psi^*}(y)) q^{E_\ga},
\end{equation}
where $q=e^{-t}$ and $p_i, p_f$ are the projections $X \times X \to X$
on the first and the second factors, respectively.

Let us compute the kernel of the evolution operator in the theory on
$X = \C$ at finite $\la$, before conjugation by $e^{\la f}$. We have
the complete basis
$$
\Psi_{n,\ol{n},\ph,\ol{\ph}} = \Psi_{n,\nb} (dz)^{\ph} (d\zb)^{\phb},
\qquad n,\nb \geq 0; \ph,\phb=0,1.
$$
where $\Psi_{n,\ol{n}}$ is given by formula \eqref{Psi}. Thus, the
degree of this state considered as a differential form on $\C$ is
$\ph+\ol\ph$.  These states are eigenstates of the hamiltonian
$H_\la$ with the eigenvalues $|\omega|(n+\ol{n}+\ph+\ol\ph)$. In
addition, they are eigenstates of the rotation operator $P$, which
is the Lie derivative with respect to the vector field
$|\omega|(z\pa_z - \ol{z}\pa_{\ol{z}})$. The corresponding eigenvalues
are $|\omega|(n-\ol{n}+\ph-\ol\ph)$.

Instead of considering the kernel $K_t$ of the evolution operator
$e^{-tH}$ we will consider the kernel $K_{t,\ol{t}}$ of the modified
evolution operator $e^{-\frac{1}{2}t(H+P)-\frac{1}{2}\ol{t}(H-P)}$ (it
reduces to $K_t$ if $\ol{t}=t$). Denote $q=e^{-|\omega|t}, \ol{q} =
e^{-|\omega|\ol{t}}$. Then we have an analogue of formula \eqref{int
kernel}, from which we find that
\begin{equation}    \label{KTT}
K_{t,\ol{t}} = \sum_{n,\ol{n}} \Psi_{n,\ol{n},\ph,\ol\ph}(z,\ol{z})
\; \ol{\Psi_{n,\ol{n},\ph,\ol\ph}(w,\ol{w})} \; q^n \ol{q}^{\ol{n}}
\; d(qz-w) \wedge d(\ol{q}\ol{z}-\ol{w}).
\end{equation}
Denote this expression by $U_{t,\ol{t}} d(qz-w) \wedge
d(\ol{q}\ol{z}-\ol{w})$. Using formula \eqref{Psi}, we find that
\begin{align*}
U_{t,\ol{t}} &= \sum_{n,\ol{n}}
\frac{1}{n!\ol{n}! \pi (\la\omega)^{n+\ol{n}-1}} e^{\frac{1}{2} \la
|\omega|(z\ol{z}+w\ol{w})} \left( \pa_z^{\ol{n}} \pa_{\ol{z}}^n
e^{-\la |\omega|z\ol{z}}\right) \pa_{\ol{w}}^{\ol{n}} \pa_w^n e^{-\la
|\omega|w\ol{w}} \\ &= \frac{\la\omega}{\pi} e^{\frac{1}{2}\la
|\omega| (z\ol{z} + w\ol{w})} e^{\frac{q}{\la|\omega|} \pa_{\ol{z}}
\pa_w + \frac{\ol{q}}{\la|\omega|} \pa_z \pa_{\ol{w}}} \cdot
e^{-\la |\omega| (z\ol{z}+w\ol{w})}.
\end{align*}

Substituting the formula
$$
e^{-\la|\omega| z\ol{z}} = \int \frac{dk d\ol{k}}{\la|\omega|\pi}
e^{-\frac{k\ol{k}}{\la|\omega|} + i(k\ol{z}+\ol{k}z)}
$$
in the above expression, we obtain
$$
U_{t,\ol{t}} = \frac{\la|\omega|}{\pi(1-q\ol{q})}
\exp\left( \frac{1}{2}\la|\omega|(z\ol{z}-w\ol{w}) -
\frac{\la|\omega|}{1-q\ol{q}} (z-\ol{q}w)(\ol{z}-q\ol{w}) \right).
$$

Before we pass to the limit $\la \to \infty$, we need to multiply
$U_{t,\ol{t}}$ by $e^{\la(f(z,\ol{z})-f(w,\ol{w}))}$, where
$f(z,\ol{z}) = \frac{1}{2}\omega z\ol{z}$. Suppose that $\omega>0$
(for $\omega<0$ the calculation is similar). Then we find that
$$
U_{t,\ol{t}} \mapsto \wt{U}_{t,\ol{t}}=
\frac{\la|\omega|}{\pi(1-q\ol{q})} \exp\left(-
\frac{\la|\omega|}{1-q\ol{q}} (w-qz)(\ol{w}-\ol{q}\ol{z}) \right).
$$

It is clear that when $\la \to \infty$, this expression tends to the
delta-function supported on the shifted diagonal $w=qz,
\ol{w}=\ol{q}\ol{z}$:
$$
\delta^{(2)}(w-qz,\ol{w}-\ol{q}\ol{z}).
$$
Therefore the kernel $K_{t,\ol{t}}$ of the (modified) evolution
operator tends to
\begin{equation}    \label{tends to}
K_{t,\ol{t}} \to \delta^{(2)}(w-qz,\ol{w}-\ol{q}\ol{z}) \; d(qz-w)
\wedge d(\ol{q}\ol{z}-\ol{w}).
\end{equation}
As expected, this is precisely the delta-form supported on the shifted
diagonal $w=qz$, which corresponds to the flow along the gradient
trajectory $z \mapsto zq$.

This completes our analysis of the spaces of states of the model
defined on $X=\C$.

\ssec{The case of $\pone$: ground states}    \label{case of pone}

Now we consider the first non-trivial ``curved'' manifold, namely, $X
= \pone$. We will choose the Fubini-Study metric
\begin{equation}
\label{metpone}
g = \frac{dzd{\zb}}{( 1 + z{\zb})^2},
\end{equation}
and the Morse function
$$
f = \frac{1}{4} \frac{z{\zb}-1}{z{\zb}+1}.
$$
The corresponding gradient vector field is the Euler vector field
$$
v = z \pa_z + \ol{z} \pa_{\ol{z}},
$$
and so it has the form $v = {\xi} + {\ol\xi}$, where ${\xi} = z
{\p}_{z}$ is the holomorphic vector field on $\pone$. This vector
field generates the standard ${\C}^\times$ action: $z \mapsto zq, q
\in \C^\times$.

The hamiltonian (before conjugation by $e^{\la f}$) is given by
formula \eqref{ham}, which in this case reads
\begin{equation}    \label{Hla for pone}
H_\la = - \frac{2}{\la} (1+z\ol{z})^2 \pa_z \pa_{\ol{z}} +
\frac{\la}{2} \frac{z\ol{z}}{(1+z\ol{z})^2} -
\frac{z\ol{z}-1}{z\ol{z}+1} (F+\ol{F}-1).
\end{equation}
Our Morse function has two critical points, $z = 0$ and $z = \infty$.
Near $z = 0$ we have:
\begin{equation}    \label{exp zero}
f = - \frac{1}{4} + \frac{1}{2} z{\zb} + \ldots,
\end{equation}
while near $z=\infty$ we have
\begin{equation}    \label{exp inf}
f = \frac{1}{4} - \frac{1}{2} w {\wb} + \ldots,
\end{equation}
where $w = z^{-1}$ is a local coordinate near the point $\infty$.

Thus, $z=0$ is a ``repulsive'' critical point, and $z=\infty$ is an
``attractive'' critical point. This indicates that both scenarios
discussed in the case of the flat space $X=\C$ should somehow be
realized in the $\pone$ model.

\begin{figure}
\begin{center}
\epsfig{file=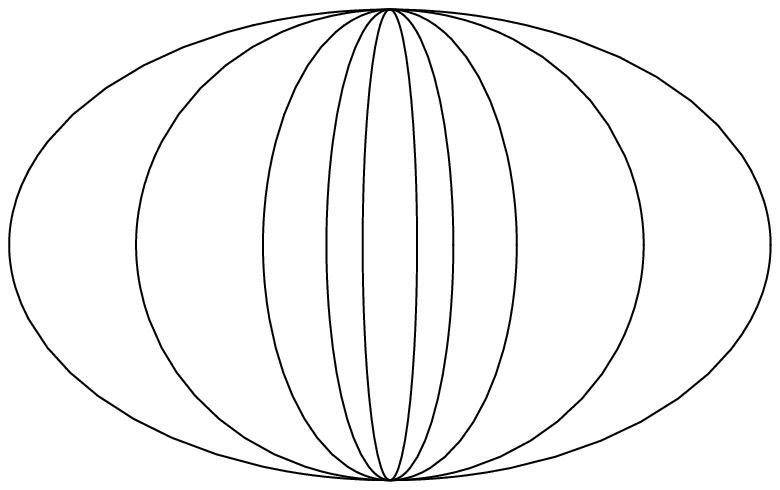,height=80mm,width=85mm}

\begin{picture}(0,0)
\setlength{\unitlength}{1mm}
\thicklines
\put(-3.8,3){$z=\infty$}
\put(-26.8,45){\vector(0,-1){1}}
\put(-13.3,45){\vector(0,-1){1}}
\put(-6.7,45){\vector(0,-1){1}}
\put(-2.7,45){\vector(0,-1){1}}
\put(2.8,45){\vector(0,-1){1}}
\put(6.8,45){\vector(0,-1){1}}
\put(13.4,45){\vector(0,-1){1}}
\put(26.9,45){\vector(0,-1){1}}
\put(-3.8,84){$z=0$}
\end{picture}
\end{center}

\begin{center}
Morse theory on $\pone$
\end{center}

\end{figure}

What is the structure of the spaces of states of our theory? We start
with the ``in'' space ${\mc H}^{\on{in}}$. For finite values of $\la$
the space of states is the space of $L_2$ differential forms on
$\pone$. It is easy to find the ground states of $H_\la$. There are
two of them, and they are localized near the critical points. The
one corresponding to $z=0$ is the function
\begin{equation}    \label{ground zero}
_0\Psi_{\on{vac}} = \sqrt{\frac{\la}{\pi(e^{\la/2}-e^{-\la/2})}} \;
e^{-\la f},
\end{equation}
and the one corresponding to $z=\infty$ is the two-form
\begin{equation}    \label{ground inf}
_\infty\Psi_{\on{vac}} = \sqrt{\frac{\la}{\pi(e^{\la/2}-e^{-\la/2})}}
\; e^{\la f} \omega_{FS},
\end{equation}
where
$$
\omega_{FS} = \frac{dz d{\zb}}{( 1 + z{\zb})^2}
$$
is the Fubini-Study K\"ahler form. To see that these are ground
states, we check that they are annihilated by both supersymmetry
charges. In each case, we obtain that one of the supercharges
obviously annihilates it by counting the degree of the differential
form, and it is a straightforward calculation to show that the other
one does as well. We have normalized these states in such a way that
they have unit norm with respect to our hermitean inner product
\eqref{norm}.

Now we change the theory by adding the term \eqref{addition} to the
action. As explained in \secref{way out}, this amounts to
multiplying the ``in'' states by the function $e^{\la f}$. In the case
of ground states, we obtain the following states of the new theory:
\begin{align*}
_0\wt\Psi^{\on{in}}_{\on{vac}} &= 1, \\
_\infty\wt\Psi^{\on{in}}_{\on{vac}} &=
\frac{\la}{\pi(e^{\la/2}-e^{-\la/2})} e^{2 \la f} \omega_{FS}.
\end{align*}
We have changed the normalization factors so as to ensure that the
above expressions have well-defined limits as $\la \to \infty$ (as
distributions).

The limit of $_0\wt\Psi^{\on{in}}_{\on{vac}}$ is just the constant
function $1$. This is not surprising, because this ground state
corresponds to the point $z=0$, which is a ``repulsive'' critical
point. So by analogy with the case of $X=\C$ we should expect that
this ground state, appropriately rescaled, becomes the constant
function. On the other hand, $_\infty\wt\Psi^{\on{in}}_{\on{vac}}$
corresponds to the ``attractive'' critical point $z=\infty$. Again,
our analysis in the case of $X=\C$ suggests that in this case we
should expect the limit of this ground state to be the delta-form
supported at $z=\infty$ (or, equivalently, $w=0$). This is exactly
what happens: the above formula, interpreted as a distribution on
$\pone$, has a well-defined limit which is equal to
$\delta^{(2)}(w,\ol{w}) dw d\ol{w}$.

Likewise, we obtain the ``out'' ground states by multiplying the
suitably normalized ground states by $e^{-\la f}$. It is clear that
mapping $f \mapsto -f$ we interchange the ``in'' and ``out'' ground
states. However, under this map the critical point $z=0$ becomes
``attractive'' while the critical point $z=\infty$ becomes
``repulsive''. Therefore their roles get interchanged, so that the
``out'' ground state corresponding to $z=0$ is the delta-form
supported at $0$, $\delta^{(2)}(z,\zb) dz d\zb$, while the ``out''
ground state corresponding to $z=\infty$ is the function $1$.

\ssec{Ground states for other K\"ahler manifolds}    \label{ground}

The calculation of the previous section has a natural generalization
to other K\"ahler manifolds. Suppose that we have a compact K\"ahler
manifold $X$ with a holomorphic vector field $\xi$, which comes from a
$\C^\times$-action $\phi$ on $X$ with isolated fixed points. Let us
denote the fixed points of $\phi$ by $x_\al, \al \in A$. We will
assume that the set $A$ is non-empty. According to \cite{Frankel},
there exists a Morse function $f$ whose gradient is the vector field
$v = \xi + \ol\xi$. The critical points of $f$ coincide with the fixed
points of $\phi$ and the zeroes of $\xi$ and $v$.\footnote{note also
that $f$ is the hamiltonian of the vector field $i(\xi-\ol\xi)$,
corresponding to the subgroup $U(1)\subset \C^\times$, with respect to
the K\"ahler structure on $X$}

Under these assumptions, we have the Bialynicki-Birula decompositions
\cite{BB}
\begin{equation}    \label{stratification}
X = \bigsqcup_{\al \in A} X_\al = \bigsqcup_{\al \in A} X^\al
\end{equation}
of $X$ into complex submanifolds $X_\al$ and $X^\al$, defined as
follows:
\begin{align}    \label{Xal1}
X_\al &= \{ x \in X | \lim_{t \to -\infty} \phi(e^{t}) \cdot x = x_\al
\}, \\ \label{Xal2}
X^\al &= \{ x \in X | \lim_{t \to +\infty} \phi(e^{t}) \cdot x = x_\al
\}.
\end{align}
The submanifolds $X_\al$ and $X^\al$ are the {\em ascending} and {\em
descending} manifolds of our Morse function $f$, respectively,
introduced in \secref{top sector}. Each submanifold $X_\al$ is
isomorphic to $\C^{n_\al}$, and $X^\al$ is isomorphic to
$\C^{n-n_\al}$, where the index of the critical point $x_\al$ is
$2(n-n_\al)$. In what follows we will assume, for simplicity, that the
following {\em Morse-Smale condition} holds: the strata $X_\al$ and
$X^\beta$ intersect transversely for all $\al,\beta$. Then, according
to \cite{BB1}, the two decompositions \eqref{stratification} of $X$
are in fact stratifications, that is the closure of each $X_\al$ is a
union of $X_\beta$'s (and similarly for $X^\al$'s).

In the case when $X=\pone$ both stratifications consist of two
``cells'': the ascending manifolds are the one-dimensional cell $\C_0
= \pone \bs \infty$ and the point $\infty \in \pone$, and the
descending manifolds are the one-dimensional cell $\C_\infty = \pone
\bs 0$ and the point $0 \in \pone$ (they satisfy the transversality
condition). The above calculation shows that the ground states of our
theory may be viewed as the delta-forms supported on these cells. For
an open cell it is just the function $1$, and for a one-point cell it
is the delta two-form supported at that point.

In general, for each stratum $X_\al$ or $X^\al$ of our decompositions
we can construct a similar delta-form, which we will denote by
$\Delta_\al$, or $\Delta^\al$, respectively. For the sake of
definiteness, consider the case of ascending manifolds $X_\al$. Then
$\Delta_\al$ is the constant function along $X_\al$, extended as a
delta-form (of degree equal to the codimension of $X_\al$, that is
$2(n-n_\al)$) in the transversal directions. More precisely, this is a
distribution (or current) on the space of differential forms on $X$
which is defined by the following formula:
\begin{equation}    \label{Deltaal}
\langle \Delta_\al,\eta \rangle = \int_{X_\al}
\eta|_{X_\al}, \qquad \eta \in \Omega^\bullet(X)
\end{equation}
(the integral converges because $\eta$ is well-defined on $X$, which
is assumed to be compact).

Under our assumptions, all critical points have even indices, and the
semi-classical analysis of \cite{W:morse} (see also
\cite{Helffer1,Helffer2}) shows that ground states are also in
one-to-one correspondence with the critical points of the Morse
function. As in the case of $X=\pone$, it is easy to write down
explicit formulas for the ground states corresponding to the minimum
and maximum of $f$, which are going to be a function and a top form,
respectively.\footnote{while these are legitimate states in the models
under consideration, we note that in more general models of quantum
field theory (and even in quantum mechanics on a non-compact manifold)
the analogous wave functions do not belong to the physical spectrum,
see \cite{W:kodama}} For the other ground states one can write
approximate semi-classical formulas for large $\la$, which are
essentially given by the Gaussian distributions around the critical
points of the form
\begin{equation}    \label{gr state finite la}
_\al\Psi_{\on{vac}} \sim \exp \left( - \la \sum_{i=1}^{n} |\mu_i|
|z_i|^2 \right) d^2 z_{n_\al+1} \wedge \ldots \wedge d^2 z_{n},
\end{equation}
where the $z_i$'s are normal holomorphic coordinates around $x_\al$
with respect to which the Hessian of $f$ is the diagonal matrix with
the eigenvalues $\mu_i, i=1,\ldots,n$, each occurring with
multiplicity two (recall our convention \eqref{d2z} for the
differentials). We order them in such a way that the eigenvalues
$\mu_i$ are positive for $i=1,\ldots,n_\al$ and negative for
$i=n_\al+1,\ldots,n$.

On a general (real) manifold $X$, because of the instanton
corrections, only some linear combinations of these states give rise
to true ground states for finite values of $\la$. In particular,
because of the instanton corrections the supercharge $Q$, acting on
these states, becomes the differential of the Morse complex
\cite{W:morse}.\footnote{note however that they become ground states
at $\la=\infty$, even though they are not annihilated by $Q$,
reflecting the non-Hodge nature of the algebra of supercharges in this
limit} But in our case, since the critical points have only even
indices, each of the functions $_\al\Psi_{\on{vac}}$ corresponds to a
true ground state of the Hamiltonian $H_\la$ for finite $\la$.

Now we perform the operation of multiplication by the function $e^{\la
f}$. Near $x_\al$ we have
$$
f \sim f(x_\al) + \sum_{i=1}^n \mu_i |z_i|^2 + \ldots
$$
Therefore after multiplication by $e^{\la f}$ the ground states become
(up to an overall factor)
\begin{equation}    \label{ground state alpha}
_\al\Psi^{\on{in}}_{\on{vac}} \mapsto
{}_\al\wt\Psi^{\on{in}}_{\on{vac}} \sim \exp \left( -2 \la
\sum_{i=n_\al+1}^{n} |\mu_i| |z_i|^2 \right) d^2 z_{n_\al+1} \wedge
\ldots \wedge d^2 z_{n}.
\end{equation}
In other words, the terms with positive eigenvalues $\mu_i$ get
canceled, while the terms with negative $\mu_i$ get doubled. The
resulting form (after including an appropriate $\la$-dependent
normalization constant) tends to the delta-form $\Delta_\al$.

For example, if $X=\pone$, then the ``delta-form'' on the open orbit
$\C_0$ is just the function $1$, and the delta-form corresponding to
$\infty$ is $\delta^{(2)}(w,\ol{w}) d^2 w$.

Thus, we claim that the suitably normalized ground states become,
after multiplication by $e^{\la f}$ and taking the limit $\la \to
\infty$, the delta-forms $\Delta_\al$.

It is instructive to derive this result using the path integral
approach. As already noted above, we may construct states of the
theory by using path integral over half-line $(-\infty,0]$, see formula
\eqref{in st}. In particular, the vacuum state
$_\al\wt\Psi^{\on{in}}_{\on{vac}}$, viewed as a differential form on
$X$, may be represented symbolically by the path integral
$$
_\al\wt\Psi^{\on{in}}_{\on{vac}}(x) = \underset{x(t): (-\infty,0] \to
X; \, x(0) = x} \int e^{-S},
$$
(where the action $S$ includes the term \eqref{addition}). The
question is which boundary condition to take at $t=-\infty$. Recall
that in the limit $\la = \infty$ that we are considering the path
integral localizes on the gradient trajectories. But the value of a
gradient trajectory $x(t): (-\infty,0] \to X$ at $t=-\infty$ is
necessarily a critical point. It is clear therefore that the boundary
condition that we need to take in order to obtain the ground state
$_\al\wt\Psi^{\on{in}}_{\on{vac}}(x)$ in the limit $\la=\infty$ is
$x(-\infty) = x_\al$.

Thus, we find that $_\al\wt\Psi^{\on{in}}_{\on{vac}}(x)$ is given by
the integral of $e^{-S}$ over the gradient trajectories $x(t):
(-\infty,0] \to X$ connecting $x_\al$ at $t=-\infty$ and $x$ at
$t=0$. But such trajectories exist only if $x$ belongs to the
ascending manifold $X_\al$, and if it does, then there is exactly one
such trajectory! So from this perspective it is clear that
$_\al\wt\Psi^{\on{in}}_{\on{vac}}(x)$ has to be supported on
$X_\al$. One needs to work a little harder and analyze the
fermionic contribution to the path integral to see that
$_\al\wt\Psi^{\on{in}}_{\on{vac}}(x)$ is in fact the delta-form
$\Delta_\al$, but since we have already obtained this result from the
Hamiltonian perspective, we will not do this here.

The same analysis applies to the ``out'' ground states. Here we need
to consider the path integral on the half-line from $0$ to $+\infty$,
so the ground states are supported on the submanifold of those $x \in
X$ which could be connected to the critical point $x_\al$ by a
gradient trajectory $x(t): [0,+\infty) \to X$ such that $x(0) = x$ and
$x(+\infty) = x_\al$. This is precisely the descending manifold
$X^\al$. From the Hamiltonian point of view this is also clear,
because the ``out'' state corresponding to \eqref{gr state finite la}
is
$$
_\al\wt\Psi^{\on{out}}_{\on{vac}} = e^{-2 \la f} \star {}
\ol{_\al\wt\Psi^{\on{in}}_{\on{vac}}} \sim \exp \left( -2 \la
\sum_{i=1}^{n_\al} |\mu_i| |z_i|^2 \right) d^2 z_{1}
\wedge \ldots \wedge d^2 z_{n_\al},
$$
which tends to the delta form $\Delta^\al$ supported on $X^\al$.

To summarize, we have now determined the ground states of our quantum
mechanical model at $\la = \infty$, for any K\"ahler manifold equipped
with a Morse function satisfying the conditions listed above:

\medskip

{\em The ``in'' ground states are the delta-forms $\Delta_\al$
supported on the closures of the ascending manifolds $X_\al$ of the
Morse function, and the ``out'' ground states are the delta-forms
$\Delta^\al$ supported on the closures of the descending manifolds
$X^\al$.}

\ssec{Back to $\pone$: excited states}    \label{back to pone}

What about the excited states of the theory? For finite values of
$\la$ explicit formulas for those are unknown, but for large $\la$ we
can understand the behavior of ``low-lying'' eigenfunctions of the
hamiltonian $H_\la$ qualitatively using the standard semi-classical
methods. Here ``low-lying'' means that the eigenvalue remains a
finite number as $\la \to \infty$. In general, there will be other
eigenfunctions as well (for example, of order $\la$), but since we
are interested in the limit $\la \to \infty$, we will ignore them.

Let us consider the case of $X=\pone$ first.  Semi-classical
approximation tells us that as far as the low-lying excited states are
concerned, the situation is as follows: the eigenfunctions are
localized in the neighborhoods of the critical points, and the picture
around each critical point is qualitatively the same as in the case of
the flat space $X=\C$ discussed in the previous section. Thus, we have
two sets of eigenfunctions, corresponding to the points $0$ and
$\infty$, which will be indicated by a subscript. In both cases we
also have to take into account the degree of the state considered as a
differential form.

Consider first the states corresponding to the critical point $z=0$
(we will mark them with a left subscript $0$). This is a ``repulsive''
critical point as can be seen from the expansion \eqref{exp zero} of
$f$ near $z=0$. Therefore the structure of the low-lying
eigenfunctions localized near $z=0$ will be the same as that of the
eigenfunctions obtained in the case of $X=\C$ and a ``repulsive''
critical point. Let us start with the $0$-forms. Recalling the
formulas obtained in \secref{flat space}, we find that those are
labeled by two integers $n,\ol{n} \geq 0$, and near zero are
approximately equal to
$$
_0\Psi_{n,\ol{n},0,0} = z^n \ol{z}^{\ol{n}} e^{-\la f} + O(\la^{-1}),
\qquad n,\ol{n} \geq 0.
$$
The first of them, $_0\Psi_{0,0,0,0}$ is in fact the ground state
$_0\Psi_{\on{vac}}$ given by formula \eqref{ground zero}. However, we
have now changed our normalization, so as to make the limits of these
functions, multiplied by $e^{\la f}$, well-defined. We also have
excited states that are $1$-forms and $2$-forms:
$$
_0\Psi_{n,\ol{n},\ph,\ol\ph} = z^n \ol{z}^{\ol{n}}
e^{-\la f} (dz)^\ph (d\ol{z})^{\ol\ph} + \ldots, \qquad
n,\ol{n} \geq 0; \ph,\ol\ph=0,1.
$$

After we multiply these eigenfunctions by $e^{\la f}$, they become
polynomial differential forms on $\C_0$, which is the ascending
manifold corresponding to the critical point $z=0$.

Thus, we conclude that in the $\la \to \infty$ limit the space of
``in'' states contains a piece $${\CH}_{{\C}_{0}}^{\on{in}} =
{\C}[z,{\zb}] \otimes \Lambda[dz, d{\zb}],$$ as in the case of $X=\C$
and a ``repulsive'' critical point.

Next, we look at the excited states corresponding to the critical
point $z=\infty$, or $w = 0$, where $w = z^{-1}$ (we will mark them
with a left subscript $\infty$). This critical point is
``attractive'', according to formula \eqref{exp inf}. Following the
example of $X=\C$ with an ``attractive'' critical point, and recalling
the expansion of $f$ near $\infty$, we find that there will be other
eigenfunctions, which near $w=0$ are equal to
$$
_\infty\Psi_{n,\ol{n},1,1} = \frac{\la}{\pi(e^{\la/2}-e^{-\la/2})}
     e^{-\la f} \frac{1}{n! \ol{n}!} \pa_w^{n} \pa_{\ol{w}}^{\ol{n}}
     e^{2\la f} dw d\ol{w} + O(\la^{-1}), \qquad n,\ol{n} \geq
     0.
$$
The first of them is the ground state \eqref{ground inf}. We have
again changed our normalization so as to obtain a well-defined limit
as $\la \to \infty$.

Multiplying these forms by $e^{\la f}$ and taking the limit $\la \to
\infty$, we obtain the derivatives of the delta-form
$$
_\infty\wt\Psi^{\on{in}}_{n,\ol{n},1,1} = \frac{1}{n! \ol{n}!}
\pa_w^{n} \pa_{\ol{w}}^{\ol{n}} \delta^{(2)}(w,\ol{w}) dw
d\ol{w}, \qquad n,\ol{n} \geq 0.
$$
In addition, there will be one-forms and zero-forms
$_\infty\wt\Psi^{\on{in}}_{n,\ol{n},\ph,\phb}$ obtained by
contracting the above two-forms with the vector fields $\pa_w$ and
$\pa_{\ol{w}}$. Thus, we obtain that this critical point contributes
the piece
$$
{\CH}_{\infty}^{\on{in}} = {\C} [ {\p}_{w} , {\p}_{\wb} ] {\dl}^{(2)}
( w, {\wb} ) \otimes {\Lambda} [ dw , d{\wb} ]
$$
to the space of "in" states of the theory at $\la=\infty$.

We conclude that the space of ``in'' states is isomorphic to the
direct sum of two subspaces attached naturally to the critical points:
$$
{\CH}^{\on{in}} \simeq {\CH}^{\on{in}}_{{\C}_{0}} \oplus
{\CH}^{\on{in}}_{\infty}.
$$
Naively, the hamiltonian is $H_{\on{naive}} = {\mc L}_v$, which
naturally acts on this direct sum. It has as basis of eigenstates the
obvious monomial basis of this space, and the eigenvalues are
non-negative integers.

However, we will see below that ${\CH}^{\on{in}}$ does not canonically
decompose into a direct sum, but is rather an extension
$$
0 \to {\CH}^{\on{in}}_{\infty} \to {\CH}^{\on{in}} \to
{\CH}^{\on{in}}_{{\C}_{0}} \to 0.
$$
Moreover, as we will explain in \secref{structure}, this space is
realized as a canonical subspace of the space of distributions on
$\pone$. The hamiltonian is indeed ${\mc L}_v$, but acting on this
space it is not diagonalizable. It is diagonalizable on the subspace
${\CH}^{\on{in}}_{\infty}$, but has generalized eigenvectors on
${\CH}^{\on{in}}_{{\C}_{0}}$ that are adjoint to eigenvectors in
${\CH}^{\on{in}}_{\infty}$. We will give below explicit formulas for
the action of the hamiltonian.

The same analysis applies to the ``out'' states of our model. Now the
roles of $0$ and $\infty$ get interchanged, so the corresponding space
of states is isomorphic to the direct sum
$$
{\CH}^{\on{out}} \simeq {\CH}^{\on{out}}_{{\C}_{\infty}} \oplus
{\CH}^{\on{out}}_{0},
$$
where
\begin{align*}
{\CH}_{0}^{\on{out}} &= {\C} [ {\p}_{z} , {\p}_{\zb} ] {\dl}^{(2)} (
z, {\zb} ) \otimes {\Lambda} [ dz , d{\zb} ], \\
{\CH}^{\on{out}}_{{\C}_{\infty}} &= {\C} [w , {\wb}] \otimes
\Lambda[dw, d{\wb}].
\end{align*}
The naive hamiltonian is $-{\mc L}_v$ which is diagonalized on the
monomial elements, with the eigenvalues being non-negative integers.

But in fact we will see that ${\CH}_{0}^{\on{out}}$ is an extension
$$
0 \to {\CH}^{\on{out}}_{0} \to {\CH}^{\on{in}} \to
{\CH}^{\on{out}}_{{\C}_{\infty}} \to 0,
$$
which is also realized in the space of distributions on $\pone$. The
hamiltonian $-{\mc L}_v$, acting on this space, has off-diagonal terms
that make it non-diagonalizable.

\ssec{Generalization to other K\"ahler manifolds and holomorphic
     factorization}    \label{hol fact}

The discussion of the previous section is generalized in a
straightforward way to the case of an arbitrary K\"ahler manifold
satisfying the above conditions. Recall that we have the
stratifications of $X$ by descending and ascending manifolds. For each
ascending manifold $X_\al$ we define the space ${\CH}^{\on{in}}_\al$
of all {\em delta-forms supported on} $X_\al$. In particular, it
contains the ground state $\Delta_\al$ constructed in
\secref{ground}. Moreover, the space ${\CH}^{\on{in}}_\al$ is
generated from $\Delta_\al$ under the action of differential operators
defined in the neighborhood of $X_\al$.

To describe the structure of ${\CH}^{\on{in}}_\al$ in more concrete
terms, we recall that the stratum $X_\al$ is isomorphic to
$\C^{n_\al}$, where the index of the corresponding critical point
$x_\al$ is $2(n-n_\al)$. Let us choose holomorphic coordinates $z_i,
i=1,\ldots,n$, in the neighborhood of $X_\al \subset X$ in such a way
that the coordinates $z_1,\ldots,z_{n_\al}$ are holomorphic
coordinates along $X_\al$ and the holomorphic coordinates
$z_{n_\al+1},\ldots,z_n$ are transversal to $X_\al$ (so that $X_\al$
is described by the equations $z_i = 0, \ol{z}_i = 0,
i=n_\al+1,\ldots,n$). Then ${\CH}^{\on{in}}_\al$ is spanned by the
monomials which may schematically be represented in the form
\begin{equation}    \label{states in Hal}
\prod_{1 \leq i,\ol{i},j,\ol{j} \leq n_\al} z_i \ol{z}_{\ol{i}}
dz_j d\ol{z}_{\ol{j}} \prod_{n_\al+1 \leq k,\ol{k},l,\ol{l} \leq n}
\pa_{z_k} \pa_{\ol{z}_{\ol{k}}} \imath_{\pa_{z_l}}
\imath_{\pa_{\ol{z}_{\ol{l}}}} \cdot \Delta_\al
\end{equation}
(here, as before, $\imath_v$ denotes the operator of contraction of a
differential form by a vector field $v$). Thus, we see that
${\CH}^{\on{in}}_\al$ is indeed generated from $\Delta_\al$ under the
action of (super)differential operators.

In addition, the space ${\CH}^{\on{in}}_\al$ exhibits the
following holomorphic factorization:
$$
{\CH}^{\on{in}}_\al \simeq {\mc F}^{\on{in}}_\al \otimes
\ol{\mc F}^{\on{in}}_\al,
$$
where ${\mc F}^{\on{in}}_\al$ (resp. $\ol{\mc
F}^{\on{in}}_\al$) is the space of holomorphic (resp.,
anti-holomorphic) delta-forms supported on $X_\al$.

For example, if $n_\al = \dim X$, so $X_\al \simeq \C^{n_\al}$ is an
open subset of $X$, then ${\mc H}^{\on{in}}_\al$ is the space of
differential forms on $\C^{n_\al}$. Therefore it factorizes into the
tensor product of the spaces ${\mc F}^{\on{in}}_\al$ and $\ol{\mc
F}^{\on{in}}_\al$ of holomorphic and anti-holomorphic differential
forms, respectively.

On the other hand, if $n_\al = 0$, so $X_\al = x_\al$ is a point, then
${\mc H}^{\on{in}}_\al$ is the space of distributions supported at
$x_\al$. It factorizes into the tensor product
$$
{\mc H}^{\on{in}}_\al \simeq
\left( \C[\pa_{z_i}] \otimes \Lambda[\imath_{\pa_{z_i}}]
\right)_{i=1,\ldots,n} \otimes
\left( \C[\pa_{\ol{z}_i}] \otimes \Lambda[\imath_{\pa_{\ol{z}_i}}]
\right)_{i=1,\ldots,n} \cdot \Delta_\al
$$
(note that the operators $\imath_{\pa_{z_i}}$ anti-commute and hence
generate an exterior algebra). Therefore we may write
$$
{\mc H}^{\on{in}}_\al \simeq {\mc F}^{\on{in}}_\al \otimes
\ol{\mc F}^{\on{in}}_\al,
$$
where
$$
{\mc F}^{\on{in}}_\al = \C[\pa_{z_i}] \otimes
\Lambda[\imath_{\pa_{z_i}}]_{i=1,\ldots,n}, \qquad \ol{\mc
F}^{\on{in}}_\al = \C[\pa_{\ol{z}_i}] \otimes
\Lambda[\imath_{\pa_{\ol{z}_i}}]_{i=1,\ldots,n}.
$$
A proper interpretation of the space ${\mc F}^{\on{in}}_\al$ for a
general critical point $x_\al$ is achieved in the framework of
the theory of holomorphic ${\mc D}$-modules (see, e.g., \cite{KS}).

Let us return to the case when $n_\al = \dim X$ and $X_\al \simeq
\C^{n_\al}$ is an open subset of $X$. Then ${\mc F}^{\on{in}}_\al$ is
the space of holomorphic differential forms on $\C^{n_\al}$. In
particular, its subspace of ${\mc F}^{\on{in},0}_\al$ of degree zero
forms consists of holomorphic functions on $X_\al$. The holomorphic
differential operators on $X_\al$ naturally act on ${\mc
F}^{\on{in},0}_\al$. Since $X_\al$ is open and dense in $X$, ${\mc
F}^{\on{in},0}_\al$ is the space of global sections of a holomorphic
${\mc D}_X$-module, where ${\mc D}_X$ is the sheaf of holomorphic
differential operators on $X$. Its generator is the function $1$,
which is annihilated by $\pa_{z_i}, i=1,\ldots,n$.

The entire space ${\mc F}^{\on{in}}_\al$ may be viewed as the space of
global sections of a ${\mc D}_{\Pi TX}$-module, where ${\mc D}_{\Pi
TX}$ is the sheaf of holomorphic differential operators on the
supermanifold $\Pi TX$. The constant function $1$ is again a generator
of this ${\mc D}_X$-module.\footnote{Note that there is no natural
structure of (left) ${\mc D}_X$-module on the subspace ${\mc
F}^{\on{in},i}_\al$ of $i$-forms in ${\mc F}^{\on{in}}_\al$, except
for $i=0$. However, the subspace of top forms has a natural structure
of right ${\mc D}_X$-module.}

Now consider the space ${\mc F}^{\on{in}}_\al$ in the case when $X_\al
= x_\al$. Then the degree zero part ${\mc F}^{\on{in},0}_\al$ of ${\mc
F}^{\on{in}}_\al$ may also be interpreted as the space of global
sections of a ${\mc D}_X$-module, called the ${\mc D}_X$-module of
``holomorphic delta-functions with support at $x_\al$'', or of ``local
cohomology of ${\mc O}_X$ with support at $x_\al$''. It is defined as
follows: its space of sections on any open subset not containing the
point $x_\al$ is zero, and the space of sections on an open subset $U$
containing $x_\al$ is the space ${\mc F}^{\on{in},0}_\al$. To define
the structure of ${\mc D}_X$-module we need to show how to act on
${\mc F}^{\on{in},0}_\al$ by holomorphic differential operators on
$U$. Without loss of generality we may assume that $U$ is a very small
neighborhood of $x_\al$ with coordinates $z_1,\ldots,z_n$. Thus, we
need to show how to act on ${\mc F}^{\on{in},0}_\al$ by functions
$z_i$ and vector fields $\pa_{z_i}$. This is done by realizing ${\mc
F}^{\on{in},0}_\al$ as the module over the algebra of polynomial
differential operators in $z_i,\pa_{z_i}, i=1,\ldots,n$, generated by
a vector annihilated by $z_i, i=1,\ldots,n$. Let us denote this
generating vector by $\delta^{\on{hol}}_\al$. Informally, we may view
$\delta^{\on{hol}}_\al$ as a ``holomorphic delta-function'', because
it satisfies $z_i \cdot \delta^{\on{hol}}_\al = 0$ for all
$i=1,\ldots,n$.

As for the entire space ${\mc F}^{\on{in}}_\al$, it may be viewed as a
${\mc D}_{\Pi TX}$-module.  It is generated by a vector, which we
denote by $\Delta^{\on{hol}}_\al$, which satisfies the relations
$$
z_i \cdot \Delta^{\on{hol}}_\al = 0, \quad dz_i \cdot
\Delta^{\on{hol}}_\al = 0, \qquad i=1,\ldots,n.
$$
It is instructive to think of $\Delta^{\on{hol}}_\al$ as the
``holomorphic delta-form''. Note that $\delta^{\on{hol}}_\al =
\imath_{\pa_{z_1}} \ldots \imath_{\pa z_n} \Delta^{\on{hol}}_\al$
(since $\Delta^{\on{hol}}_\al$ is a top form).

For a more general critical point $x_\al$ with index $2(n-n_\al)$, the
space ${\mc F}^{\on{in}}_\al$ is a ${\mc D}_{\Pi TX}$-module,
generated from a ``holomorphic delta-form'' $\Delta^{\on{hol}}_\al$
supported on $X_\al$. By definition, $\Delta^{\on{hol}}_\al$ is
annihilated by $z_i,dz_i, i=1,\ldots,n_\al$, and by
$\pa_{z_i},\imath_{\pa_{z_i}}, i=n_\al+1,\ldots,n$. The space
${\mc F}^{\on{in}}_\al$ is obtained from $\Delta^{\on{hol}}_\al$
under the action of holomorphic polynomials along $X_\al$ (in the
variables $z_1,\ldots,z_{n_\al}$) and holomorphic vector fields in the
transversal directions (that is
$\pa_{z_{n_\al+1}},\ldots,\pa_{z_{n}}$), as well as the exterior
algebra in $dz_1,\ldots,dz_{n_\al},\imath_{\pa_{z_{n_\al+1}}},\ldots,
\imath_{\pa_{z_n}}$.

In particular, the degree zero part of ${\mc F}^{\on{in}}_\al$ is
the space of global sections of a ${\mc D}_X$-module. This ${\mc
D}_X$-module is in fact the push-forward of the ${\mc
D}_{X_\al}$-module ${\mc O}_{X_\al}$ to $X$ under the embedding $X_\al
\hookrightarrow X$. It may also be realized as the local cohomology
$H^{n-n_\al}_{X_\al}({\mc O}_X)$ of the structure sheaf ${\mc O}_X$ on
$X$ with support on $X_\al$ (for more on this, see \secref{action of
Q}). The entire space ${\mc F}^{\on{in}}_\al$ is identified with the
local cohomology $H^{n-n_\al}_{X_\al}(\Omega_{X,\on{hol}})$, where
$\Omega_{X,\on{hol}}$ is the sheaf of holomorphic differential forms
on $X$. It is naturally a ${\mc D}_{\Pi TX}$-module.

For example, if $X = \pone$, then there are two critical points: $0$
and $\infty$. If $x_\al=0$, then the corresponding stratum $X_\al$ is
$\C_0 = \pone \bs \infty$. The corresponding algebra of differential
operators is generated by $z$ and $\pa_z$. The zero-form part of ${\mc
H}^{\on{in}}_\al$ is in this case the ${\mc D}$-module of functions on
$\C_0$. Its space of global sections is $\C[z]$, and the algebra of
differential operators naturally acts on it.

If $x_\al = \infty$, then we have a coordinate at $\infty$ that we
previously denoted by $w$. The corresponding algebra of differential
operators is generated by $w$ and $\pa_w$. The zero-form part of ${\mc
H}^{\on{in}}_\al$ is the ${\mc D}_{\pone}$-module of holomorphic
delta-functions with support at $\infty$. Its space of sections over
any open subset $U \subset \pone$ containing $\infty$ may be defined
concretely as $\C[w,w^{-1}]/\C[w]$. In other words, it is the quotient
of the space of functions defined on $\C_\infty = \pone \bs 0$ by the
subspace of those functions which are regular at $\infty$ (thus, it is
spanned by the ``polar parts'' of these
functions).\footnote{equivalently, we could have chosen any open
subset $U$ containing $\infty$ and taken the quotient of holomorphic
functions on $U \bs \infty$ by holomorphic functions on $U$; or we
could take the quotient $\C((w^{-1}))/\C[[w]]$ corresponding to
the formal disc around $\infty$} It is generated, under the action of
$\pa_w$, by the vector $1/w$. Inside the quotient
$\C[w,w^{-1}]/\C[w]$ this vector is annihilated by $w$, and
hence it may be thought of as a particular realization of the
``holomorphic delta-function'' $\delta^{\on{hol}}_\infty$ at $\infty$.

We have a similar description of the anti-holomorphic factor $\ol{\mc
F}^{\on{in}}_\al$.

Based on the semi-classical analysis similar to the one performed in
the case of $\pone$, we now describe the space of ``in'' states as
follows:

\medskip

{\em The space of ``in'' states of
the theory at $\la=\infty$ is isomorphic to the direct sum}
\begin{equation}    \label{space in}
{\mc H}^{\on{in}} \simeq \bigoplus_{\al \in A} {\mc
       H}^{\on{in}}_\al = \bigoplus_{\al \in A} {\mc
       F}^{\on{in}}_\al \otimes \ol{\mc F}^{\on{in}}_\al.
\end{equation}

\medskip

Thus, we observe the appearance of ``conformal blocks'' (or
``holomorphic blocks'') in the space of states reminiscent of the
structure of two-dimensional conformal field theory. It is surprising
that we observe this structure at the level of quantum mechanics,
i.e., one-dimensional quantum field theory. This structure is also
reflected in the correlation functions of the observables decomposing
into the product of holomorphic and anti-holomorphic parts: they
decompose into the sum of products of holomorphic and anti-holomorphic
expressions. There are no non-constant smooth observables of this
type, but if we allow singularities of special kind, such observables
can be easily constructed. For example, in the case of $X = \pone$ one
can consider observables of the form
$$
\left( \sum_{i=1}^M \frac{\al_i dz}{z-a_i} \right) \left( \sum_{j=1}^N
\frac{\beta_j dz}{\ol{z}-\ol{b}_j} \right),
$$
and explicit calculations show that their correlation functions
indeed exhibit factorization into ``conformal blocks''.

Such holomorphic factorization certainly cannot be expected in the
theory at the finite values of the coupling constant $\la$, because
the action contains the term $\la^{-1} g^{a\ol{b}} p_a p_{\ol{b}}$
mixing holomorphic and anti-holomorphic fields (and a similar mixed
fermionic term). But for $\la=\infty$ this term disappears and we find
that both the Lagrangian and the Hamiltonian of our theory are equal
to sums of holomorphic and anti-holomorphic parts (see formula
\eqref{1D action}). Therefore naively one might expect that the space
of states of the theory is the tensor product of holomorphic and
anti-holomorphic sectors. However, what we find is a {\em direct
sum} of such tensor products. This is a precursor of the Quillen
anomaly familiar from two-dimensional conformal field theory.

Naively, the supercharge $Q$ is the de Rham differential $d$ naturally
acting on the spaces ${\mc H}^{\on{in}}_\al$, and the hamiltonian is
$H_{\on{naive}} = \{ Q,\imath_v \} = {\mc L}_v$. The cohomology of $Q$
on each ${\mc H}^{\on{in}}_\al$ is one-dimensional, occurring in degree
$2i$ and represented by the delta-form $\Delta_\al$. These are
therefore the BPS states of our theory, in agreement with the
expectation that the BPS states are identified with the cohomology of
$X$. Indeed, the ascending manifolds give us a decomposition of $X$
into even-dimensional cells, which therefore give a basis in the
homology of $X$. The forms $\Delta_\al$ give the dual basis in the
cohomology.

More precisely, we will see below that, just like in the case of
$X=\pone$, the spaces ${\mc H}^{\on{in}}$ are not canonically
isomorphic to the above direct sums of the spaces ${\mc
H}^{\on{in}}_\al$. Rather, there are canonical filtrations with the
consecutive subquotients isomorphic to ${\mc
H}^{\on{in}}_\al$. Because of that, the hamiltonian is not
diagonalizable; it is equal to $H_{\on{naive}} = {\mc L}_v$ plus
off-diagonal terms mixing the spaces ${\mc H}^{\on{in}}_\al$ with
${\mc H}^{\on{in}}_\beta$ corresponding to the strata $X_\beta$ of
lower dimension which are in the closure of $X_\al$. However, this
mixing occurs only within the subspaces of differential forms of a
fixed degree and fixed eigenvalue with respect to ${\mc L}_v$. Because
these subspaces are finite-dimensional, all Jordan blocks are finite
and their length is bounded by $\on{dim}_{\C} X + 1$.\footnote{we will
see in \secref{gen kahler} that the maximal length of the Jordan
blocks may well be less than $\on{dim}_{\C} X + 1$}

Likewise, we will see that the supercharge $Q$ is equal to $d$ plus
correction terms mixing ${\mc H}^{\on{in}}_\al$ with ${\mc
H}^{\on{in}}_\beta$ corresponding to the strata $X_\beta$ in the
closure of $X_\al$. However, we will show in \secref{coh of Q} that
these correction terms do not change the cohomology of $Q$.

The space of ``out'' states has a similar structure, but with respect
to the stratification of $X$ by the descending manifolds $X^\al$. For
each stratum $X^\al$ we have the space ${\CH}^{\on{out}}_\al$ of {\em
delta-forms supported on} $X_\al$. In particular, it contains the
ground state $\Delta^\al$ constructed in \secref{ground}. Moreover,
the space ${\CH}^{\on{out}}_\al$ is generated from $\Delta^\al$ under
the action of differential operators defined in the neighborhood of
$X_\al$. Similarly to the ``in'' spaces, it exhibits {\em holomorphic
factorization}
$$
{\CH}^{\on{out}}_\al = {\mc F}^{\on{out}}_\al \otimes
\ol{\mc F}^{\on{out}}_\al,
$$
where ${\mc F}^{\on{out}}_\al$ (resp. $\ol{\mc F}^{\on{out}}_\al$) is
the space of holomorphic (resp., anti-holomorphic) delta-forms
supported on $X^\al$. Finally, the space of ``out'' states is
isomorphic to
$$
{\mc H}^{\on{out}} \simeq \bigoplus_{\al \in A} {\mc
       H}^{\on{out}}_\al = \bigoplus_{\al \in A} {\mc
       F}^{\on{out}}_\al \otimes \ol{\mc F}^{\on{out}}_\al.
$$

In reality, this direct sum decomposition is not canonical. Instead,
there is a canonical filtration with the spaces ${\mc
H}^{\on{out}}_\al$ appearing as consecutive quotients, and the
hamiltonian is $-{\mc L}_v$ plus non-diagonal terms, as for ${\mc
H}^{\on{in}}$. Nevertheless, there is a canonical pairing between
${\mc H}^{\on{in}}$ and ${\mc H}^{\on{out}}_\al$, as expected on
general grounds. We will discuss all this in the next section.

\section{The structure of the space of states}    \label{structure}

In the previous section we have determined, in the first
approximation, the spaces ${\mc H}^{\on{in}}$ and ${\mc H}^{\on{out}}$
of ``in'' and ``out'' states of our quantum mechanical model in the
limit $\la = \infty$. In this section we will give a more precise
description of these spaces. We will show that states are naturally
interpreted as {\em distributions} (or {\em currents}) on our manifold
$X$. Because some of these distributions require regularization
(reminiscent of the Epstein-Glaser regularization \cite{EG} familiar
in quantum field theory), the action of the Hamiltonian on them
becomes non-diagonalizable. We compute this action, as well as the
action of the supercharges, in terms of the so-called {\em
Grothendieck-Cousin operators} associated to the stratification of our
manifold by the ascending and descending manifolds. We also compute
the cohomology of the supercharges using the Grothendieck-Cousin
complex \cite{Kempf}.

In \secref{action} we will realize the evaluation observables of our
model as linear operators acting on the spaces of states. We will then
be able to obtain the correlation functions as matrix elements of
these operators and to test our predictions by comparing these matrix
elements with the integrals over the moduli of the gradient
trajectories which were obtained in the path integral approach, as
explained in \secref{cor fns as int}.

\ssec{States as distributions}    \label{str in space}

The answer we have given for the space of states ${\mc H}^{\on{in}}$
in formula \eqref{space in} requires some explanations. Consider for
example the case of $X=\pone$. We have claimed that
\begin{equation}    \label{sp st}
{\mc H}^{\on{in}} \simeq {\mc H}_{\C_0} \oplus {\mc H}_\infty,
\end{equation}
where ${\mc H}_{\C_0}$ is the space of differential forms on the
one-dimensional cell $\C_0 = \pone \bs \infty$ and ${\mc H}_\infty$ is
the space of delta-forms supported at $\infty$. These delta-forms are
naturally functionals, i.e., distributions (more precisely, currents)
on the space of differential forms on $\pone$. For example, the ground
state $\delta^{(2)}(w,\ol{w}) dw d\ol{w}$ is the functional
whose value on a function $f$ on $\pone$ is equal to $f(\infty)$, and
it is equal to zero on all differential forms of positive degree.

Thus, the subspace ${\mc H}_{\C_0}$ appears to be realized in the
space of differential forms, while the subspace ${\mc H}_\infty$ is
realized in the space of distributions or currents, that is
functionals on the space of differential forms. This is puzzling
because the two spaces appear to be of different nature. The second
puzzle is that elements of ${\mc H}_{\C_0}$ are well-defined only on
the subset $\C_0 \subset \pone$ and, with the exception of the
constant functions, have poles at $\infty$. Therefore the integrals of
the products of such elements with an observable of our theory, which
is {\em a priori} an arbitrary smooth differential form on $X$, is not
well-defined. But integrals of this type naturally appear as the
one-point correlation functions of our theory at finite values of
$\la$.

For example, let $\wh{\omega}$ be the observable of our theory which
corresponds to a smooth two-form $\omega$ on $\pone$, and consider the
one-point function represented by the matrix element
$$
\langle _\infty\Psi_{\on{vac}}| e^{(t-t_f)H} \wh{\omega} e^{(t_i-t)H}|
_0\Psi_{n,\ol{n},0,0} \rangle = q^{E_{n,\ol{n}}} \int_{\pone} \;
{}_\infty\Psi^{(0)}_{0,0} \; \; \omega \; \;
_0\Psi^{(0)}_{n,\ol{n},0,0},
$$
where $q=e^{t_i-t}$ and $E_{n,\ol{n}}$ is the eigenvalue of $H$ on
$_0\Psi_{n,\ol{n},0,0}$. We have argued that in the limit $\la \to
\infty$ the ``out'' state corresponding to $_\infty\Psi_{\on{vac}}$
becomes equal to $1$, while the ``in'' state corresponding to
$_0\Psi_{n,\ol{n},0,0}$ becomes equal to $z^n
\ol{z}^{\ol{n}}$. Therefore to make sense of the theory at
$\la=\infty$ we should be able to compute integrals of the form
\begin{equation}    \label{int omega}
\int_{\pone} z^n \ol{z}^{\ol{n}} \; \omega,
\end{equation}
for a general smooth two-form $\omega$ on $\pone$.

Unfortunately, these integrals generally diverge for $n,\ol{n}>0$. But
this discussion leads us to an important idea: it suggests that a proper
definition of the state corresponding to $_0\Psi_{n,\ol{n},0,0}$ in
the limit $\la \to \infty$ assumes that we can evaluate the integrals
of the form \eqref{int omega}. Therefore it is only natural to view
these states not as functions on $\pone$, but as distributions! This
at least allows us to treat ${\mc H}_{\C_0}$ and ${\mc H}_\infty$ on
equal footing. It now becomes clear that our space of states ${\mc
H}^{\on{in}}$ with its decomposition \eqref{sp st} should be
considered as a subspace of the space of distributions (or currents)
on the space of smooth differential forms on $\pone$.

Viewing states as distributions is most natural from the point of
view of the path integral. Recall formula \eqref{in st} describing
states as path integrals. Now, given a differential form $\omega$ on
$X$, set
$$
\langle \omega|{\mc O}_n(t_n) \ldots {\mc O}_1(t_1) \vac =
\underset{X}\int \omega \wedge \; \underset{x(t): (-\infty,0] \to X;
\, x(0) = x} \int {\mc O}_1(t_1) \ldots {\mc O}_n(t_n) e^{-S}.
$$
Thus, the state ${\mc O}_n(t_n) \ldots {\mc O}_1(t_1) \vac$ is
naturally interpreted as a linear functional on differential forms,
i.e., a distribution.

The delta-forms supported at $\infty$ are legitimate
distributions. But what kind of distribution can we associate to the
function $z^n \ol{z}^{\ol{n}}$ which has a pole at $\infty \in \pone$?
This question has a well-known answer in the theory of generalized
functions, as we now explain following \cite{Hoermander}, Sect.~3.2
and \cite{GelfandShilov}, Sect. B1.

First of all, let us recall that by definition a {\em distribution} on
$\pone$ is a continuous linear functional on the space of smooth
functions on $\pone$, equipped with the topology induced by the norm
$\Vert f \Vert = \on{sup} |f(x)|$.\footnote{for a general manifold
$X$, distributions are continuous linear functionals on the space of
smooth functions on $X$ with compact support, but on a compact
manifold $X$ the ``compact support'' condition is vacuous} In what
follows we will use the term ``distribution'' in more general sense,
as a continuous linear functional on the space of differential forms
on $\pone$ (a more common term for such an object is ``current''). We
denote the space of such distributions by $D(\pone)$.

Next, we define the space ${\mc S}(\C_0)$ of {\em Schwartz functions}
on $\C_0 = \pone \bs \infty \subset \pone$: its elements are smooth
functions $f$ on $\C_0$ such that $z^n \ol{z}^{\ol{n}} \pa_z^m
\pa_{\ol{z}}^{\ol{m}} f$ is bounded on $C_0$ (and hence well-defined
at $\infty$) for all $n,\ol{n},m,\ol{m} \in \Z_{\geq 0}$. Such a
function therefore extends to a smooth function on $\pone$; moreover,
it necessarily decays as $z,\ol{z} \to \infty$. Define a topology on
the space ${\mc S}(\C_0)$ induced by the family of semi-norms
$$
f \mapsto |z^n \ol{z}^{\ol{n}} \pa_z^m \pa_{\ol{z}}^{\ol{m}} f|.
$$
A {\em tempered distribution} on $\C_0$ is by definition a continuous
linear functional on ${\mc S}(\C_0)$.

We define in the same way the space ${\mc S}\Omega(\C_0)$ of Schwartz
differential forms on $\C_0$. We will call continuous linear
functionals on ${\mc S}\Omega(\C_0)$ ``tempered distributions on
differential forms on $\C_0$''.

Now observe that for all $n,\ol{n} \in \Z_{\geq 0}$ the monomial $z^n
\ol{z}^{\ol{n}}$ defines a continuous linear functional
$\varphi_{n,\ol{n}}$, hence a tempered distribution on differential
forms on $\C_0$ by the formula
\begin{equation}    \label{zn as dist}
\varphi_{n,\ol{n}}(\omega) = \int_{\C_0} z^n \ol{z}^{\ol{n}} \; \omega,
\qquad \omega \in {\mc S}\Omega(\C_0).
\end{equation}
The integral converges because of the condition imposed on elements of
${\mc S}\Omega(\C_0)$ (note that it is non-zero only if $\omega$ is a
two-form).

Thus, we are now in the following situation: we have the subspace
${\mc S}\Omega(\C_0) \subset \Omega(\pone)$ and a continuous linear
functional $\varphi_{n,\ol{n}}$ on ${\mc S}\Omega(\C_0)$ defined by
formula \eqref{zn as dist}. Can we extend this functional to the
larger space $\Omega({\pone})$?

It turns out that we can, but there are many possible extensions and
there is no canonical choice among them, unless $n=0$ or $\ol{n}=0$.
The good news, however, is that any two possible extensions differ by
a distribution supported at $\infty$. Therefore, even though the span
of all functionals $\varphi_{n,\ol{n}}, n,\ol{n} \in \Z_{\geq 0}$, is
not canonically defined as a subspace of $D(\pone)$, the span of these
functionals together with the functionals $\pa_w^{m}
\pa_{\ol{w}}^{\ol{m}} \delta^{(2)}(w,\wb)$ is well-defined.

We have an analogous statement for the $i$-form versions of these
spaces, where $i=1,2$. Thus, we obtain that the sum ${\mc H}_{\C_0} +
{\mc H}_\infty$ is a well-defined subspace of the space of all
distributions on $\pone$.

We now {\em define} the space ${\mc H}^{\on{in}}$ of ''in'' states of
the $\pone$ model in the $\la=\infty$ limit as this subspace of
$D(\pone)$. This way we resolve the first puzzle pointed out at the
beginning of this section (the fact that ${\mc H}_{\C_0}$ and ${\mc
H}_\infty$ seem to be objects of different nature). But we have not
yet explained how to extend the linear functionals
$\varphi_{n,\ol{n}}$ to $D(\pone)$ and make sense of the integrals
\eqref{int omega} for arbitrary smooth differential forms $\omega \in
\Omega(\pone)$. We will explain that in the next section.

\ssec{Regularization of the integrals in the case of $\pone$}
\label{expl reg}

A particular extension of the tempered distribution
$\varphi_{n,\ol{n}}$ to a distribution on $\pone$ is constructed by
introducing a ``cutoff'': for $\omega \in \Omega(\pone)$, consider the
integral
\begin{equation}    \label{partie finie}
\underset{|z|<\ep^{-1}}\int z^n \ol{z}^{\ol{n}} \; \omega.
\end{equation}
which is well-defined for any positive real $\ep$. One can show that
as a function in $\ep$ it may be uniquely represented in the form
\begin{equation}    \label{C log}
C_0 + \sum_{i>0} C_i \ep^{-i} + C_{\log} \log \ep + o(1),
\end{equation}
where the $C_i$'s and $C_{\log}$ are some numbers (see
\cite{Hoermander}, pp. 70-71). Therefore one defines, following
Hadamard, the {\em partie finie} of the above integral as the constant
coefficient $C_0$ obtained after discarding the terms with negative
powers of $\ep$ and $\log \ep$ in the integral \eqref{partie finie}
and taking the limit $\ep \to 0$. We denote it by
$$
\underset{|z|<\ep^{-1}}\pf z^n \ol{z}^{\ol{n}} \; \omega = C_0.
$$
A similar regularization has also been used in quantum field theory,
in particular, in the works of Epstein and Glaser \cite{EG}.

It is clear that if $\omega \in {\mc S}\Omega(\C_0)$, then
$$
\underset{|z|<\ep^{-1}}\pf z^n \ol{z}^{\ol{n}} \; \omega =
\underset{\pone}\int z^n \ol{z}^{\ol{n}} \; \omega,
$$
so we indeed obtain an extension of $\varphi_{n,\ol{n}}$ to a
distribution on $\pone$. We will denote it by $\wt\varphi_{n,\ol{n}}$.
Note that the distribution we obtain takes non-zero values only on
two-forms on $\pone$.

The problem with this definition is that it is not canonical. Indeed,
we do not have a canonical coordinate on $\pone$, because we are only
given points $0$ and $\infty$, so our coordinate $z$ is only defined
up to multiplication by a non-zero scalar. If we rescale our
coordinate $z \mapsto az$, then the functional $\wt\varphi_{n,\ol{n}}$
will change as we will now integrate over the region $|z|<a\ep^{-1}$,
and the integral will be different due to the presence of the
logarithmic term in \eqref{C log}. However, the resulting change will
amount to a distribution supported at $\infty \in \pone$, so that the
span of these distributions and the distributions supported at
$\infty$ is a canonically defined subspace of $D(\pone)$.

Let us compute the values of the distributions
$\wt\varphi_{n,\ol{n}}$ in some examples. Let
$$
\omega = \sum_{{\al}} \frac{\omega_{\al}}{z{\zb} + R_{\al}} dz d{\zb},
\qquad R_{\al} \in \C \bs \R_{\leq 0},
$$
where the numbers $\omega_\al$ satisfy the condition
$$
\sum_\al \omega_\al = 0,
$$
which ensures that $\omega$ is well-defined at $\infty$. Writing
$z=\sqrt{x} e^{i\theta}$, we find that
\begin{align}    \notag
\wt\varphi_{n,\ol{n}}(\omega) &= \underset{|z|<\ep^{-1}}\pf z^n
\ol{z}^{\ol{n}} \; \omega = \sum_\al \int_0^{2\pi} d\theta
e^{i\theta(n-\ol{n})} \left[ \int_0^{\ep^{-2}} \frac{\omega_\al
x^{(n+\ol{n})/2}}{x+R_\al} dx \right]_{\ep^0} \\ \label{Ral} &= 2\pi
(-1)^{n+1} \delta_{n,\ol{n}} \sum_\al \omega_\al R_\al^n \log R_\al.
\end{align}

The distributions $\wt\varphi_{n,\nb,\ph,\phb}$ corresponding to the
basis elements $z^n \ol{z}^{\ol{n}} (dz)^\ph (d\ol{z})^{\ol{\ph}}$
of ${\mc H}_{\C_0}$ are defined in the same way. These distributions
take non-zero values on differential forms of degree
$(1-\ph,1-\ol\ph)$. Again, they are not canonically defined, but
their span together with the span of $(\ph,\ol\ph)$-forms in ${\mc
H}_\infty$ will be well-defined. This span is the
$(\ph,\ol\ph)$-form part of our space of ``in'' states ${\mc
H}^{\on{in}}$.

\ssec{Action of the hamiltonian}    \label{action ham}

Having defined the space ${\mc H}^{\on{in}}$ as a particular subspace
of the space of distributions on $\pone$, we can now find explicitly
the action of the Hamiltonian $H={\mc L}_v$, where $v = z\pa_z +
\ol{z}\pa_{\ol{z}}$. Actually, we will compute separately the action
of ${\mc L}_\xi$ and ${\mc L}_{\ol{\xi}}$, where $\xi=z\pa_z,
\ol\xi=\ol{z}\pa_{\ol{z}}$. We will see that due to the non-canonical
nature of the decomposition \eqref{sp st} these operators act
non-diagonally, with Jordan blocks.

Consider first the action of ${\mc L}_\xi$ and ${\mc L}_{\ol\xi}$ on
the subspace ${\mc H}_\infty \subset {\mc H}^{\on{in}}$ (which is a
canonical subspace of ${\mc H}^{\on{in}}$). This subspace has the
following basis:
$$
|n,\nb,\ph,\phb\rangle_\infty := \frac{(-1)^{n+\nb}}{n!\ol{n}!}
\pa_w^n \pa_{\ol{w}}^{\ol{n}} \delta^{(2)}(w,\ol{w}) (dw)^\ph
(d\wb)^{\phb}.
$$
By definition,
\begin{equation}    \label{states as dist}
|n,\nb,\ph,\phb\rangle_\infty(\omega_{w\wb} (dw)^r (d\wb)^{\ol{r}}) =
(-i)^{\ph} i^{\phb} \delta_{r,1-\ph} \delta_{\ol{r},1-\phb} \pa_w^n
\pa_{\wb}^{\nb} \omega_{w\wb}|_{w=0}.
\end{equation}
We find that
\begin{align*}
{\mc L}_\xi \cdot |n,\nb,\ph,\phb\rangle_\infty &= (n+1-\ph)
|n,\nb,\ph,\phb\rangle_\infty, \\
{\mc L}_\xi \cdot |n,\nb,\ph,\phb\rangle_\infty &= (\nb+1-\phb)
|n,\nb,\ph,\phb\rangle_\infty.
\end{align*}

Next, we consider the subspace ${\mc H}_{\C_0}$. It has the following
basis:
$$
|n,\nb,\ph,\phb\rangle_{\C_0} := \wt\varphi_{n,\nb,\ph,\phb},
$$
where $\wt\varphi_{n,\nb,\ph,\phb}$ is the distribution defined at the
end of the previous section:
$$
\wt\varphi_{n,\nb,\ph,\phb}(\omega) = \underset{|z|<\ep^{-1}}\pf z^n
\ol{z}^{\ol{n}} (dz)^{\ph} (d\ol{z})^{\phb} \; \wedge \omega.
$$
These elements, and their span, are not canonically defined, but
depend on a particular ``partie finie'' regularization of the above
integral defined above.

Let us compute ${\mc L}_\xi \cdot |n,\nb,0,0\rangle_{\C_0}$. By
definition, this is the distribution, whose value on a two-form
$\omega$ on $\pone$ is equal to
$$
\wt\varphi_{n,\ol{n}}(-{\mc L}_\xi \omega) = -
\underset{|z|<\ep^{-1}}\pf z^n \ol{z}^{\ol{n}} \; {\mc L}_\xi \omega.
$$
Writing ${\mc L}_\xi$ as $\{ d,\imath_\xi \}$ and using Stokes
formula, we find that in addition to the differentiation of $z^n$,
which results in multiplication by $n$, there is also a boundary term,
which is the $\ep^0$-coefficient in the expansion of the integral
$$
\underset{|w|=\ep}\int w^{-n} w^{-\nb} \; \imath_\xi \omega
$$
in power series in $\ep^{\pm 1}$ and $\log \ep$ (the change of sign
here is due to the change of orientation of the circle $|z|=\ep^{-1}$
under the change of variables $z \mapsto w=z^{-1}$).

Writing $\xi = - w \pa_w$, $\omega = \omega_{w\ol{w}} dw d\wb = i
\omega_{w\ol{w}} dw \wedge d\wb$ and
$w=\ep e^{i\theta}$, we find that this boundary term is equal to
$$
\left[ \int_0^{2\pi} w^{1-n} \ol{w}^{1-n} \omega_{w\wb} d\theta
    \right]_{\ep^0} =
    \begin{cases} \left. - \frac{2\pi}{(n-1)! (\ol{n}-1)!} \pa_w^{n-1}
    \pa_{\wb}^{\nb-1} \omega_{w\wb} \right|_{w=0}, & \on{if }
    n,\ol{n}>0, \\ 0, & \on{if } n=0 \on{ or } \nb = 0. \end{cases}
$$

Thus, we obtain the following formula
$$
{\mc L}_\xi \cdot |n,\nb,0,0\rangle_{\C_0} = n
|n,\nb,0,0\rangle_{\C_0} - 2\pi
|n-1,\nb-1,0,0\rangle_{\infty}.
$$
Likewise, we obtain
\begin{align}    \label{L xi}
{\mc L}_\xi \cdot |n,\nb,\ph,\phb\rangle_{\C_0} &= (n+\ph)
|n,\nb,\ph,\phb\rangle_{\C_0} - 2\pi
|n+2\ph-1,\nb+2\phb-1,\ph,\phb\rangle_{\infty}, \\ \label{L olxi}
{\mc L}_{\ol\xi} \cdot |n,\nb,\ph,\phb\rangle_{\C_0} &= (\nb+\phb)
|n,\nb,\ph,\phb\rangle_{\C_0} - 2\pi
|n+2\ph-1,\nb+2\phb-1,\ph,\phb\rangle_{\infty}.
\end{align}

Here we use the convention
$$
|n,\ol{n},\ph,\phb\rangle_\infty \equiv 0, \qquad \on{if} \quad n<0
\quad \on{or} \quad \nb<0.
$$

Thus, we find that the operators ${\mc L}_\xi$ and ${\mc
L}_{\ol{\xi}}$, and hence the Hamiltonian ${\mc L}_v$, have Jordan
blocks of length two. The generalized eigenspace of the operators
${\mc L}_\xi$ and ${\mc L}_{\ol\xi}$ corresponding to the eigenvalues
$n+\ph \geq 0$ and $\nb+\phb \geq 0$ on the space of
$(\ph,\phb)$-forms in ${\mc H}^{\on{in}}$ is two-dimensional, spanned
by the vectors $|n,\ol{n},\ph,\phb\rangle_{\C_0}$ and
$|n+2\ph-1,\ol{n}+2\ph-1,\ph,\phb\rangle_\infty$. The former is an
eigenvector, and the second is a generalized eigenvector which is
adjoint to it.

In particular, the indeterminacy of the vector
$|n,\ol{n},\ph,\phb\rangle_{\C_0}$ with $n+2p-1\geq 0$ and
$\nb+2\phb-1\geq 0$ is contained in the two-dimensional subspace of
${\mc H}^{\on{in}}$ spanned by it and
$|n+2\ph-1,\ol{n}+2\phb-1,\ph,\phb\rangle_\infty$.

This could actually be seen from the outset. As we have explained, the
indeterminacy comes from the fact that in the definition of the
distribution $\wt\phi_{n,\ol{n}}$ as the ``partie finie'' of the
integral \eqref{partie finie} we use the ``cutoff'' $|z|<\ep^{-1}$,
and so if we replace $\ep$ by $\wt\ep=a^{-1} \ep$, then the ``partie
finie'' of the integral will get shifted by the term $- C_{\log} \log
a$, where $C_{\log}$is defined in formula \eqref{C log}. But the term
$C_{\log}$ is easy to evaluate explicitly. Indeed, we may split the
integral \eqref{partie finie} into the sum of two: over $|z|<1$ and
$1<|z|<\ep^{-1}$. The former converges and hence cannot contribute to
$C_{\log}$. On the other hand, writing $\omega$ in the form
$\omega_{w\wb} dw d\wb$, as before, we rewrite the latter as
$$
\underset{\ep<|w|<1}\int w^{-n} \wb^{-\nb} \omega_{w\wb} dw
d\wb.
$$
The $\log \ep$ contribution to this integral is equal to
$$
\left. 4\pi \log \ep \frac{1}{(n-1)!(\nb-1)!} \pa_w^{n-1}
\pa_{\wb}^{\nb-1} \omega_{w\wb} \right|_{w=0}.
$$
Thus, by changing the ``cutoff'' in the definition of
$\wt\phi_{n,\ol{n}}$, we replace $\wt\phi_{n,\ol{n}}$ by its linear
combination with $\frac{1}{(n-1)!(\ol{n}-1)!} \pa_w^{n-1}
\pa_{\ol{w}}^{\ol{n}-1} \delta^{(2)}(w,\ol{w}) dw \wb$, for
$n,\nb>0$.

Note however that the extra term is equal to $0$ if $n=0$ or
$\ol{n}=0$, that is for purely holomorphic or anti-holomorphic
functions on $\C_0$. This shows that the distributions
$\wt\varphi_{n,0}$ and $\wt\varphi_{0,\nb}$ corresponding to the
states $|n,0,0,0\rangle_{\C_0}$ and $|0,\nb,0,0\rangle_{\C_0}$,
respectively, are actually well-defined. In fact, it is easy to see
that they have canonical regularizations which are eigenvectors of
${\mc L}_\xi$ and ${\mc L}_{\ol\xi}$. The vectors
$|n,0,0,1\rangle_{\C_0}$ and $|0,\nb,1,0\rangle_{\C_0}$ are also
eigenvectors.

One can analyze the indeterminacy of $(\ph,\phb)$-forms in a similar
fashion, reproducing the above result that it is confined to the
two-dimensional subspaces of ${\mc H}^{\on{in}}$ spanned by
$|n,\ol{n},\ph,\phb\rangle_{\C_0}$ and
$|n+2\ph-1,\ol{n}+2\ph-1,\ph,\phb\rangle_\infty$. We will denote this
subspace by ${\mc H}^{\on{in}}_{n,\ol{n},\ph,\phb}$.

\ssec{Action of the supercharges}    \label{supercharges}

Next, we analyze the action of the supercharges. Recall that we have
two of them: $Q = d$, the de Rham differential, and $Q^* = 2\imath_v$,
the contraction with the vector field $2v = 2(z\pa_z + \zb
\pa_{\zb})$. As in the case of the Hamiltonian, we split each of them
into the sum of holomorphic and anti-holomorphic terms: $Q = \pa +
\ol{\pa}$ and $\imath_v = \imath_\xi + \imath_{\ol\xi}$. These
operators naturally act on the space of distributions on $\pone$. For
instance, we have
$$
\langle \pa \varphi,\omega \rangle = - \langle \varphi, \pa \omega
\rangle,
$$
and so on. As we will see, these operators preserve the subspace
${\mc H}^{\on{in}}$. We wish to compute the corresponding action of
these operators on ${\mc H}^{\on{in}}$.

The operators $\imath_\xi$ and $\imath_{\ol\xi}$ are the easiest to
compute. Their action is given by the standard formulas on both
${\mc H}^{\on{in}}_\infty$ (considered as the space of delta-forms
supported at $\infty \in \pone$) and ${\mc H}^{\on{in}}_{\C_0}$
(considered as the space of polynomial differential forms on $\C_0$).

The action of $\pa$ and $\ol\pa$ on ${\mc H}^{\on{in}}_\infty$ is also
the obvious one. However, due to boundary terms similar to the ones
arising in the above calculation of the Hamiltonian, the action of
$\pa$ and $\ol\pa$ on the subspace ${\mc H}^{\on{in}}_{\C_0}$ has
correction terms which belong to ${\mc H}^{\on{in}}_\infty$.

To see how this works, let us compute explicitly $\pa \cdot |n,\nb,0,0
\rangle_{\C_0}$. It follows from the definition that this is the
distribution, whose value on a one-form $\omega = \omega_{\ol{w}}
d\ol{w}$ on $\pone$ is equal to
$$
\wt\varphi_{n,\ol{n}}(-\pa \omega) = - \underset{|z|<\ep^{-1}}\pf z^n
\ol{z}^{\ol{n}} \; \pa \omega
$$
Using the Stokes formula, we find that this integral has two terms:
one corresponds to the obvious action of $\pa$ on $z^n \zb^{\nb}$,
sending it to $n z^{n-1} \zb^{\nb} dz$, and the other is the boundary
term
$$
\left[ \int_{|w|=\ep} w^{-n} \ol{w}^{-\nb} \omega_{\wb} d\wb
\right]_{\ep^0} = \left. -\frac{2\pi i}{n! (\ol{n}-1)!} \pa_w^{n}
\pa_{\wb}^{\nb-1} \omega_{w\wb} \right|_{w=0}.
$$
Therefore we obtain that
$$
\pa |n,\nb,0,0 \rangle_{\C_0} = n|n-1,\nb,1,0 \rangle_{\C_0}
+ 2\pi |n,\nb-1,1,0 \rangle_\infty
$$
(see formula \eqref{states as dist}).

Similarly, we find that
$$
\ol\pa |n,\nb,0,0 \rangle_{\C_0} = n|n,\nb-1,0,1 \rangle_{\C_0}
+ 2\pi |n-1,\nb,0,1 \rangle_\infty,
$$
and obtain analogous formulas for differential forms of higher
degrees. These formulas may be summarized as follows.

Let us use holomorphic factorization and realize the spaces ${\mc
    H}^{\on{in}}_{\C_0}$ and ${\mc H}_\infty$ as the tensor products
\begin{align*}
{\mc H}^{\on{in}}_{\C_0} &= \C[z] \otimes
\Lambda[dz] \otimes \C[\zb] \otimes \Lambda[d\zb], \\
{\mc H}^{\on{in}}_\infty &= \left( \C[w,w^{-1}]/\C[w] \right) \otimes
\Lambda[dw] \otimes \left( \C[\wb,\wb^{-1}]/\C[\wb] \right) \otimes
\Lambda[d\wb].
\end{align*}
In this realization our basis elements of ${\mc H}^{\on{in}}_\infty$
correspond to
$$
|n,\nb,\ph,\phb \rangle_\infty = \frac{(-1)^{n+\nb}}{n! \nb!} \pa_w^n
\pa_{\wb}^{\nb} \delta^{(2)} (dw)^{\ph} (d\wb)^{\phb} = w^{-n-1}
(dw)^{\ph} \otimes \wb^{-\nb-1} (d\wb)^{\phb}.
$$

This realization is convenient, because we can use the natural linear
maps
\begin{align*}
\delta: \; & \C[z] \otimes \Lambda[dz] \to \left( \C[w,w^{-1}]/\C[w]
\right) \otimes \Lambda[dw], \\ \ol\delta: \; & \C[\zb] \otimes
\Lambda[d\zb] \to \left( \C[\wb,\wb^{-1}]/\C[\wb] \right) \otimes
\Lambda[d\wb],
\end{align*}
obtained using the composition
$$
\C[z] \to \C[z,z^{-1}] = \C[w,w^{-1}] \to \C[w,w^{-1}]/\C[w]
$$
(we recall that $w=z^{-1}$). As we will see in \secref{action of Q},
these are the simplest examples of the Grothendieck-Cousin operators.

Then the formulas for $\pa$ and $\ol\pa$ are expressed in terms of
these operators as follows:
\begin{align*}
\pa(\Psi \otimes \ol\Psi) &= \pa_{\on{naive}}(\Psi \otimes \ol\Psi) +
2\pi \; \delta \left(\frac{dw}{w} \wedge \Psi \right) \otimes
\ol\delta(\ol\Psi), \\ \ol\pa(\Psi \otimes \ol\Psi) &=
\ol\pa_{\on{naive}}(\Psi \otimes \ol\Psi) + 2\pi \; \delta(\Psi)
\otimes \ol\delta \left(\frac{d\wb}{\wb} \wedge \ol\Psi \right).
\end{align*}

Using this formula and the Cartan formula ${\mc L}_v = \{ d,\imath_v
\}$, we obtain the following formula for the chiral and anti-chiral
components of the Hamiltonian:
\begin{align*}
{\mc L}_\xi &= {\mc L}_{\xi,\on{naive}} - 2 \pi \; \delta \otimes
\ol\delta, \\ {\mc L}_{\ol\xi} &= {\mc L}_{\ol\xi,\on{naive}} - 2 \pi
\; \delta \otimes \ol\delta,
\end{align*}
so that the Hamiltonian is given by the formula
$$
H = \frac{1}{2} \{ Q,Q^* \} = H_{\on{naive}} - 4\pi \; \delta
\otimes \ol\delta.
$$
This agrees with formulas \eqref{L xi} and \eqref{L olxi}.

\ssec{The space of states as a $\la \to \infty$ limit}

The two-dimensional subspaces ${\mc H}^{\on{in}}_{n,\ol{n},\ph,\phb}$
are thus the building blocks of the space of ``in'' states of our
model at the point $\la=\infty$. Using these spaces, we can now
clarify the result of the semi-classical analysis of the low-lying
eigenvectors of the Hamiltonian $\wt{H}_\la = {\mc L}_v -
\frac{1}{2\la} \Delta$.  This Hamiltonian commutes with the operator $P
= {\mc L}_{z\pa_z-\zb\pa_{\zb}}$ and we will consider their joint
eigenstates. For simplicity, we will restrict ourselves to the
$0$-forms.

According to the semi-classical analysis, for each $n,\ol{n} \geq 0$,
there is a two-dimensional space of eigenstates of $\wt{H}_\la$ and
$P$ in the space of functions on $\pone$ with the eigenvalues
$n+\ol{n}+O(\la^{-1})$ and $n-\ol{n}$, respectively. These eigenstates
are obtained by multiplying by $e^{\la f}$ the eigenstates of the
conjugated hermitean operator $H_\la$, given by formula \eqref{Hla for
pone}, with the same eigenvalues.  Let ${\mc H}^{\la}_{n,\nb}$ be the
corresponding two dimensional space of functions on $\pone$. Now
observe that each smooth function $\Psi$ on $\pone$ defines a
distribution by the formula
$$
\Psi(\omega) = \int_{\pone} \Psi \omega, \qquad \omega \in
\Omega^2(\pone).
$$
Therefore ${\mc H}^{\la}_{n,\nb}$ defines a two-dimensional subspace
of the space $D(\pone)$ of distributions on differential forms on
$\pone$, depending on $\la$.

\medskip

{\em We conjecture that the limit of the subspace ${\mc
H}^{\la}_{n,\nb} \subset D(\pone)$ as $\la \to \infty$ is the subspace
${\mc H}^{\on{in}}_{n,\ol{n},0,0}$ introduced above.}

\medskip

We have a similar conjecture for the $(\ph,\phb)$-form subspaces ${\mc
H}^{\on{in}}_{n,\ol{n},\ph,\phb}$.

Thus, we conjecture that our subspace ${\mc H}^{\on{in}} \subset
D(\pone)$ naturally appears as the $\la \to \infty$ limit of the span
of the low-lying eigenstates of the hamiltonian $\wt{H}_\la$, viewed
as distributions. Note that the eigenstates of $\wt{H}_\la$ are
obtained by multiplying the eigenstates of the hermitean operator
$H_\la$ by the function $e^{\la f}$ (which are differential forms on
$\pone$ viewed as distributions).

It is natural to ask whether we can develop a perturbation theory
for the eigenstates at finite $\la$ around the eigenstates
$|n,\ol{n},\ph,\phb\rangle_{\C_0}$ and
$|n+2\ph-1,\ol{n}+2\ph-1,\ph,\phb\rangle_{\infty}$ of the theory at
$\la=\infty$. This will be discussed in \secref{pert}.

\ssec{Definition of the ``out'' space and the pairing}
\label{pairin}

We now briefly describe the space of ``out'' states in the same
way. We have previously said that it is isomorphic to the direct sum
$$
{\mc H}^{\on{out}} \simeq {\mc H}^{\on{out}}_{\C_\infty} \oplus {\mc
H}^{\on{out}}_0.
$$
Now we realize ${\mc H}^{\on{out}}$ as a subspace of the space
$D(\pone)$ of distributions on $\pone$, following Sects. \ref{str in
space} and \ref{expl reg}. The subspace ${\mc H}^{\on{out}}_0$ is by
definition the space of distributions on the differential forms on
$\pone$ supported at the point $0 \in \pone$. It is spanned by the
eigenstates
$$
_0\langle n,\nb,\ph,\phb| = \frac{(-1)^{n+\nb}}{n! \nb!} \pa_z^n
\pa_{\zb}^{\nb} \delta^{(2)}(z,\zb) (dz)^{\ph} (d\zb)^{\phb}.
$$
The subspace ${\mc H}^{\on{out}}_{\C_\infty}$ is not canonical, but we
choose as its basis vectors
$$
_{\C_\infty} \langle n,\nb,\ph,\phb|
$$
the distributions defined by the formula
$$
_{\C_\infty} \langle n,\nb,\ph,\phb|\omega\rangle =
\underset{|w|<\ep^{-1}}\pf w^n \wb^{\nb} (dw)^{\ph}
(d\wb)^{\phb} \wedge \omega.
$$
The above integral needs to be regularized because the differential
form has a pole at $w=\infty$ (or $z=0$). We use the ``partie finie''
regularization introduced in \secref{expl reg}. If we change the
regularization, then the corresponding distribution will get shifted
by a distribution supported at $0$.

Therefore the sum of the spaces ${\mc H}^{\on{out}}_{\C_\infty}$ and
${\mc H}^{\on{out}}_0$ is a well-defined subspace of the space
$D(\pone)$ of distributions on $\pone$, and this is by definition the
space ${\mc H}^{\on{out}}$ of ``out'' states of our model.

The Hamiltonian is $-{\mc L}_v = - {\mc L}_\xi - {\mc
L}_{\ol\xi}$. The states $_0\langle n,\nb,\ph,\phb|$ are eigenvectors:
\begin{align*}
_0\langle n,\nb,\ph,\phb| \cdot - {\mc L}_\xi &= (n+1-\ph) \; _0\langle
n,\nb,\ph,\phb|, \\ _0\langle n,\nb,\ph,\phb| \cdot - {\mc L}_{\ol\xi}
&= (\nb+1-\phb) \; _0\langle n,\nb,\ph,\phb|.
\end{align*}
The states $_{\C_\infty} \langle n,\nb,\ph,\phb|$ are generalized
eigenvectors (unless $n=0, \ph=0$ or $\nb=0, \phb=0$) which satisfy
\begin{align*}
_{\C_\infty}\langle n,\nb,\ph,\phb| \cdot - {\mc L}_\xi &= (n+\ph)
\; _{\C_\infty}\langle n,\nb,\ph,\phb| + 2\pi \;
_0\langle n+2\ph-1,\nb+2\phb-1,\ph,\phb|, \\
_{\C_\infty}\langle n,\nb,\ph,\phb| \cdot - {\mc L}_{\ol\xi} &=
(\nb+\phb) \; _{\C_\infty}\langle n,\nb,\ph,\phb| + 2\pi \;
_0\langle n+2\ph-1,\nb+2\phb-1,\ph,\phb|.
\end{align*}

In particular, we see that, just as in the case of ``in'' states, the
``mixing'' between ${\mc H}^{\on{out}}_{\C_\infty}$ and ${\mc
H}^{\on{out}}_0$ is confined to the two-dimensional generalized
eigenspaces of ${\mc L}_\xi$ and ${\mc L}_{\ol\xi}$.

As in the case of ``in'' states, we conjecture that the subspace ${\mc
H}^{\on{out}} \subset D(\pone)$ naturally appears as the $\la \to
\infty$ limit of the span of the low-lying eigenstates of the
hermitean operator $H_\la$ by the function $e^{-\la f}$ (which are
differential forms on $\pone$, viewed as distributions).

On general grounds (see \secref{way out}) we expect that there is a
canonical pairing
$$
{\mc H}^{\on{in}} \times {\mc H}^{\on{out}} \to \C.
$$
Explicitly, it is defined as follows: let
$\Psi \in {\mc H}^{\on{in}}$ and $\Phi \in {\mc H}^{\on{out}}$. Then
$$
\langle \Phi,\Psi \rangle = \int_{\pone} \Phi \wedge \Psi.
$$
Some comments are necessary here, because {\em a priori} $\Psi$ and
$\Phi$ are distributions and so it is not clear what the above
integral means. However, $\Psi$ and $\Phi$ are distributions of a very
special kind, and for those this integral is
well-defined. Intuitively, this is because $\Psi$ may be viewed a
distribution in a small neighborhood of $\infty$ and a function
elsewhere, while $\Phi$ is a distribution in a small neighborhood of
$0$ and a function elsewhere.

More precisely, we write $\Phi = \Phi_0 + \Phi_\infty, \Psi = \Psi_0 +
\Psi_\infty$, where $\Phi_0 = c_{|z|\leq 1} \Phi$, $\Phi_\infty =
c_{|z|\geq 1} \Phi$ and $c_{|z|\leq 1}, c_{|z|\geq 1}$ are the
characteristic functions of the discs $|z|\leq 1$ and $|z|\geq 1$. We
then split the above integral into the sum
$$
\int_{|z|\leq 1} \Phi_0 \wedge \Psi_0 + \int_{|z|\geq 1} \Phi_\infty
\wedge \Psi_\infty.
$$
In the first summand $\Phi_0$ is a smooth function, whereas $\Psi_0$
is a distribution. Therefore their pairing is well-defined. The second
summand is well-defined for the same reason, with the roles of $\Phi$
and $\Psi$ reversed.

This pairing is especially easy to describe when either $\Psi \in {\mc
H}^{\on{in}}_\infty$ or $\Phi \in {\mc H}^{\on{out}}_0$. First of all,
if both inclusions are satisfied, then they are distributions
supported at two different points on $\pone$: $\infty$ and $0$,
respectively. Therefore the pairing between them is equal to $0$.

If $\Phi \in {\mc H}^{\on{out}}_0$, then
$$
\langle \Phi,\Psi \rangle = \int \Phi \wedge \on{pr}(\Psi) =
\Phi(\on{pr}(\Psi)),
$$
the evaluation of the distribution $\Phi$ on the projection
$\on{pr}(\Psi)$ of $\Psi$ onto ${\mc H}^{\on{in}}_{\C_0} \simeq
\Omega(\C_0)$, which is a differential form well-defined in the
neighborhood of $0 \in \pone$.

Likewise, if $\Psi \in {\mc H}^{\on{out}}_\infty$, then
$$
\langle \Phi,\Psi \rangle = \int \on{pr}(\Phi) \wedge \Psi =
\Psi(\on{pr}(\Phi)),
$$
the evaluation of the distribution $\Psi$ on the projection
$\on{pr}(\Phi)$ of $\Phi$ onto ${\mc H}^{\on{in}}_{\C_\infty} \simeq
\Omega(\C_\infty)$, which is a differential form well-defined in the
neighborhood of $\infty \in \pone$.

We find from this description that
\begin{multline}    \label{dual bases}
_0\langle m,\ol{m},r,\ol{r}|n,\nb,\ph,\phb\rangle_{\C_0} = \;
_{\C_\infty}\langle m,\ol{m},r,\ol{r}|n,\nb,\ph,\phb\rangle_\infty \\
= (-i)^{\ph} i^{\phb} (-1)^{\ph\ol{r}} \delta_{n,m}
\delta_{\nb,\ol{m}} \delta_{\ph,1-r} \delta_{\phb,1-\ol{r}}.
\end{multline}

Finally, we compute the pairing between the states
$|n,\nb,\ph,\phb\rangle_{\C_0}$ and $_{\C_\infty}\langle
m,\ol{m},r,\ol{r}|$. According to the above definition, it is equal to
the ``partie finie'' of the integral
$$
(-1)^{\ph\ol{r}} \underset{\ep_1 < |z| < \ep_2^{-1}}\int z^{n-m-2}
\zb^{\nb-\ol{m}-2} (dz)^{\ph+r} (d\zb)^{\phb+\ol{r}},
$$
which is obtained by discarding all terms that contain $(\ep_1)^{-1},
\log \ep_1, (\ep_2)^{-1}, \log \ep_2$ and setting $\ep_1,\ep_2=0$ in
the remainder. It is easy to see that the result is always zero.

Note however that if we chose different regularization schemes for
defining these states (which would shift them by the ``delta-like''
states), then the pairing between them would change.

Thus, we find that the bases
$$
\{ |n,\nb,\ph,\phb\rangle_{\C_0},
|n,\nb,\ph,\phb\rangle_\infty \} \quad \on{and} \quad \{ \; _0\langle
n,\ol{n},\ph,\phb|, \; _{\C_\infty}\langle n,\ol{n},\ph,\phb| \}
$$
are dual to each other in ${\mc H}^{\on{in}}$ and ${\mc
H}^{\on{out}}$, up to a power of $i$ (see formula \eqref{dual
bases}).

\ssec{The general case}    \label{gen case}

We now briefly discuss how to generalize the above results to the case
of a general K\"ahler manifold $X$ and a holomorphic vector field
$\xi$ satisfying the conditions of \secref{ground}. Recall that we
have two stratifications of $X$, defined in formula
\eqref{stratification}. In the first approximation we defined the
space of ``in'' states as the direct sum given by formula \eqref{space
in}. Each summand ${\mc H}^{\on{in}}_\al$ consists of delta-forms
supported on the stratum $X_\al$. However, this decomposition is not
canonical. We now apply the same reasonings as in the case of
$X=\pone$ to define the space ${\mc H}^{\on{in}}$ of ``in'' states as
a canonical subspace of the space of distributions on differential
forms on $X$.

This space has a canonical filtration ${\mc H}^{\on{in}}_{\leq i},
i=0,\ldots,\on{dim}_\C X$, such that each consecutive quotient ${\mc
H}^{\on{in}}_{\leq i}/{\mc H}^{\on{in}}_{\leq (i-1)}$ is isomorphic to
the direct sum
\begin{equation}    \label{consecutive}
\bigoplus_{\on{dim}_{\C} X_\al = i} {\mc H}^{\on{in}}_\al.
\end{equation}
We construct ${\mc H}^{\on{in}}_{\leq i}$ by induction on $i$.  The
space ${\mc H}^{\on{in}}_{\leq 0}$ is by definition the space ${\mc
H}^{\on{in}}_{\al_{\on{max}}}$ for the critical point
$x_{\al_{\on{max}}}$ such that the corresponding stratum
$X_{\al_{\on{max}}}$ consists of a single point, $x_{\al_{\on{max}}}$,
which happens when it is the absolute maximum of the Morse
function. Thus, ${\mc H}^{\on{in}}_{\leq 0}$ is the space of
distributions on $X$ supported at this point.\footnote{here, as above,
we use the term ``distribution'' to mean functionals on the space of
all smooth differential forms, and not just smooth functions}

Now suppose that we have already constructed the subspace ${\mc
H}^{\on{in}}_{\leq (i-1)}$. Let $X_\al$ be a stratum of complex
dimension $i$. We construct a new space ${\mc H}^{\on{in}}_{\leq
(i-1),\al}$ which is an extension
\begin{equation}    \label{ext al}
0 \to {\mc H}^{\on{in}}_{\leq (i-1)} \to {\mc H}^{\on{in}}_{\leq
    (i-1),\al} \to {\mc H}^{\on{in}}_\al \to 0
\end{equation}
as follows.

Let $B_\al \subset \ol{X}_\al \bs X_\al$ be the union of the strata
$X_\beta$ of complex dimension $(i-1)$ that belong to the closure
$\ol{X}_\al$ of $X_\al$. We denote the set of $\beta$'s appearing in
its decomposition by $A_\al$. We define ${\mc S}_\al \Omega(X)$ as the
space of smooth differential forms on $X$ which decay very fast along
$B_\al$. More precisely, we define it as the intersection of the
spaces ${\mc S}_{\al\beta} \Omega(X), \beta \in A_\al$, which are
constructed as follows.

Consider the union $X_\al \bigsqcup X_\beta$, where $\beta \in
A_\al$. It is isomorphic to a fibration over $\pone$, with fibers
being vector spaces isomorphic to $X_\beta$. The stratum $X_\beta$ is
embedded as the fiber at $\infty \in \pone$ and the stratum $X_\al$ as
its complement, the preimage of $\C_0 = \pone \bs \infty$. There is a
section $\pone \to X_\al \bigsqcup X_\beta$ such that the image of
$\infty \in \pone$ is the critical point $x_\beta$, the image of $0
\in \pone$ is the critical point $x_\al$, and the image of $\C^\times
\subset \pone$ is the intersection of $X_\al$ with the descending
manifold $X^\beta$ corresponding to $x_\beta$. Moreover, the
$\C^\times$-action on $X$ restricted to $X_\al \bigsqcup X_\beta$
lifts the standard $\C^\times$-action on $\pone$.

The function $w$ on $\pone$ which has a zero of order one at $\infty$
and a pole of order one at $0$ pulls back to a function on $X_\al
\bigsqcup X_\beta$ which we denote by $w_{\al\beta}$. Note that
$X_\beta$ is the zero set of $w_{\al\beta}$ and that $\xi \cdot
w_{\al\beta} = a_{\al\beta} w_{\al\beta}$, where $a_{\al\beta}$ is a
negative real number.

Now the condition on $\omega \in \Omega(X)$ to belong to ${\mc
S}_{\al\beta} \Omega(X)$ is that after we apply to $\omega$ any
sequence of Lie derivatives with respect to vector fields defined in a
neighborhood of $X_\al \bigsqcup X_\beta$, the restriction of the
resulting form to $X_\al \bigsqcup X_\beta$ will tend to zero as
$w_{\al\beta} \to 0$ faster than any polynomial in $w_{\al\beta}$.

Now we define ${\mc S}_\al \Omega(X)$ as the intersection of the
spaces ${\mc S}_{\al\beta} \Omega(X), \beta \in A_\al$.

Any vector $\Psi$ in the space ${\mc H}^{\on{in}}_\al$ of delta-forms
on $X_\al$ gives rise to a linear functional on ${\mc S}_\al
\Omega(X)$ whose value on $\omega \in {\mc S}_\al \Omega(X)$ is given
by the integral
$$
\langle \Psi,\omega \rangle = \int_X \Psi \wedge \omega.
$$
The integral converges because of the conditions we imposed on
$\omega$.

Now we wish to extend this functional to a distribution on all smooth
differential forms on $X$. Such an extension is constructed by
introducing a ``cutoff'', following the example of $\pone$ discussed
in \secref{expl reg}. Namely, to define its value on $\omega \in
\Omega(X)$ we take the integral of $\Psi \wedge \omega$ over $X$ minus
the union of subsets in $X_\al$ defined by the inequalities
$|w_{\al\beta}|<\ep$, for all $\beta \in A_\al$. Then we take the
``partie finie'' of the corresponding integral, i.e., discard all
terms involving negative powers of $\ep$ and $\log \ep$ and set
$\ep=0$ in the remainder.

The resulting functional is regularization dependent because it
depends on the choice of the functions $w_{\al\beta}$. But the
difference of two possible regularizations is a distribution which
belongs to the previously constructed space ${\mc H}^{\on{in}}_{\leq
(i-1)}$. Therefore we obtain an extension ${\mc H}^{\on{in}}_{\leq
(i-1),\al}$ as in \eqref{ext al}.

Finally, we define ${\mc H}^{\on{in}}_{\leq i}$ as the sum of the
extensions \eqref{ext al} over all $\al \in A$ such that $\dim_{\C}
X_\al = i$. This completes the inductive construction of the space of
``in'' states as a well-defined canonical subspace of the space $D(X)$
of all distributions (on differential forms) on $X$.

Recall the hamiltonian $\wt{H}_\la = {\mc L}_v + \frac{1}{2\la} \Delta$
which is the regularized version of our Hamiltonian ${\mc L}_v$. Its
``low-lying'' eigenfunctions are by definition the eigenfunctions
whose eigenvalues are equal to $C + O(\la^{-1})$, where $C$ is a
constant. We recall that each of these eigenfunctions is equal to
$e^{\la f}$ times an eigenfunction of the hermitean Hamiltonian
$H_\la$ given by formula \eqref{ham}.

\medskip

{\em We conjecture that the space ${\mc H}^{\on{in}}$
appears as the limit of the span of the low-lying eigenfunctions of
the Hamiltonian $\wt{H}_\la$ (considered as distributions).}

\medskip

The space ${\mc H}^{\on{out}}$ of ``out'' states is defined in the
same way, by using the opposite stratification by the submanifolds
$X^\al, \al \in A$.

Finally, we define a pairing
$$
{\mc H}^{\on{in}} \times {\mc H}^{\on{out}} \to \C
$$
$$
\langle \Phi,\Psi \rangle \mapsto \int_X \Phi \wedge \Psi.
$$
To see that this integral makes sense, we argue in the same way as in
the case of $\pone$ (see \secref{pairin}).

\ssec{Action of the supercharges and the Hamiltonian}    \label{action
    of Q}

The supercharges of our theory: $Q=d$, the de Rham differential, and
$Q^* = 2\imath_v$, and the Hamiltonian, $H={\mc L}_v = {\mc L}_\xi +
{\mc L}_{\ol\xi}$, act naturally on the space ${\mc H}^{\on{in}}$, and
their adjoints act on ${\mc H}^{\on{out}}$. However, this action is
rather complicated because elements of the space ${\mc H}^{\on{in}}$
are constructed by a non-trivial regularization procedure. This
procedure distorts the action of these operators and as the result
they acquire correction terms. We have analyzed this in the case of
$X=\pone$ in \secref{action ham} and we have seen that as the result
of these correction terms the Hamiltonian is non-diagonalizable. The
same happens for general $X$ and $\xi$ satisfying the assumptions of
\secref{ground}. From now on we will assume in addition that $X$ is a
projective algebraic variety.

We will now describe a model for the space of ``in'' states in which
the action of the supersymmetry charges and the Hamiltonian are given
by very transparent formulas. This generalizes the formulas in the
case of $X=\pone$ presented in \secref{supercharges}. The key
ingredients are the {\em Grothendieck-Cousin} (GC) boundary operators
which act between the spaces of delta-forms supported on the strata
$X_\al$ and $X_\beta$ of our decomposition, with $X_\beta$ being a
codimension one stratum in the closure of $X_\al$. We will write
$X_\al \succ X_\beta$ if this is the case. The construction of of
these operators is explained in detail in \cite{Kempf}, Sect. 7.

The GC operators act between the spaces of {\em local cohomology} of a
sheaf ${\mc F}$ on $X$ with support on locally closed submanifolds of
$X$. Let us recall the definition of the functor of local
cohomology. Let $Z$ be a closed submanifold of $X$. We will say that a
section $s$ of a sheaf ${\mc F}$ on $X$ is supported on $Z$ if its
restriction to $X \bs Z$ is equal to $0$. Let $\Gamma_Z(X,{\mc F})$ be
the space of sections of ${\mc F}$ supported on $Z$. Thus, we obtain a
functor ${\mc F} \mapsto \Gamma_Z(X,{\mc F})$. It is left exact, but
not right exact. We will denote its higher derived functors by
$H^i_Z({\mc F})$. More generally, let $Y$ be a locally closed subset
of $X$ such that $Y = Z' \bs Z$, where $Z \subset Z'$ are two closed
subsets of $X$. Then we denote by $H^i_Y({\mc F})$ the higher derived
functors of $\Gamma_Z(X\bs Z,{\mc F}|_{X\bs Z})$.

Using standard technique of homological algebra, we then obtain
boundary maps
$$
H^i_Y({\mc F}) \to H^{i+1}_Z({\mc F})
$$
for any sheaf ${\mc F}$ on $X$ and a pair of closed subsets $Z
\subset Z'$ of $X$ and $Y = Z' \bs Z$ (see \cite{Kempf}, Sect. 7, for
the precise definition). These are the GS operators that we will need.

Consider the special case when $Y$ is an open subset of $X$ and $Z$ is
a divisor. In this case the GC operator may be described in very
concrete terms. For simplicity suppose that $Z$ is a smooth divisor in
a smooth (possibly non-compact) algebraic variety $X$ such that $Z$
and $Y = X \bs Z$ are affine algebraic varieties. Suppose that our
sheaf ${\mc F}$ is a holomorphic vector bundle ${\mc E}$ on $X$. Since
$Y$ is open and dense in $X$, we have $H^i_Y({\mc E}) = H^i(Y,{\mc
E}|_Y)$. From now on we will simply write ${\mc E}$ for ${\mc
E}|_Y$. Since $Y$ is assumed to be affine, $H^i(Y,{\mc E}) = 0$ for
$i>0$ and $H^0(Y,{\mc E})$ is the space of (regular) holomorphic
sections of ${\mc E}$ on $Y$. On the other hand, consider $H^1_Z({\mc
E})$, the first local cohomology of ${\mc E}$ with support on $Z$. To
define it, choose another smooth divisor $Z_1$ in $X$ such that $Z
\cap Z_1 = \emptyset$ and $X \bs Z_1$ is affine. Then we have the
following exact sequence
$$
0 \to H^0(X \bs Z_1,{\mc E}) \to H^0(X \bs (Z \sqcup Z_1),{\mc E}) \to
H^1_Z({\mc E}) \to 0,
$$
which allows us to define $H^1_Z({\mc E})$ as the quotient
$$
H^1_Z({\mc E}) \simeq H^0(X \bs (Z \sqcup Z_1),{\mc E})/H^0(X
\bs Z_1,{\mc E}).
$$
This definition is independent of the choice of $Z_1$ satisfying the
above conditions.

Now we define the GC operator corresponding to the pair $(Y,Z)$,
$$
H^0(Y,{\mc E}) \to H^1_Z({\mc E}),
$$
as the composition
\begin{equation}    \label{cg 1}
H^0(Y,{\mc E}) \to H^0(X \bs (Z \sqcup Z_1),{\mc E}) \to
H^1_Z({\mc E}).
\end{equation}
Informally, it corresponds to taking the polar part along $Z$ of
meromorphic sections of ${\mc E}$ defined on a small neighborhood of
$Z$, which are allowed to have poles only along $Z$.

For example, if $X=\pone,Z=\infty,Z_1=0$, so that $Y = \C_0 = \pone
\bs \infty$, and ${\mc E} = {\mc O}$ is the trivial line bundle, then
\begin{equation}    \label{ident for pone}
H^0(Y,{\mc E}) = \C[w^{-1}], \qquad H^1_Z({\mc O}) =
\C[w,w^{-1}]/\C[w],
\end{equation}
where $w$ is a function on $\pone$ which vanishes at $\infty$ to order
one and has a pole of order one at $0$.\footnote{note that we have
already encountered this space in our discussion of ``holomorphic
delta-functions'' in \secref{back to pone}} The corresponding GC
operator \eqref{cg 1} is just the natural map
\begin{equation}    \label{GC pone}
\C[w^{-1}] \to \C[w,w^{-1}]/\C[w],
\end{equation}
obtained by composing the maps
$$
\C[w^{-1}] \to \C[w,w^{-1}] \qquad \on{and} \qquad \C[w,w^{-1}] \to
\C[w,w^{-1}]/\C[w].
$$

Now let us return to our situation. So we have a projective algebraic
variety $X$ of complex dimension $n$ with a stratification by smooth
locally closed strata $X_\al, \al \in A$, isomorphic to
$\C^{n_\al}$. Let ${\mc E}$ be a holomorphic vector bundle. Then for
each stratum $X_\al$ of complex dimension $i$ one defines the local
cohomology groups $H^{n-i}_{X_\al}({\mc E})$ of ${\mc E}$ with support
on $X_\al$. One can show that
$$
H^{n-i}_{X_\al}({\mc E}) = 0, \qquad i \neq n_\al.
$$
so the local cohomology is non-trivial only in dimension $n-n_\al$,
which is the codimension of $X_\al$.

How to relate this discussion to our space of ``in'' states? Recall
from \secref{hol fact} that we have holomorphic factorization
$$
{\mc H}^{\inn}_\al = \CF^{\inn}_\al \otimes \ol\CF^{\inn}_\al,
$$
where $\CF^{\inn}_\al$ and $\ol\CF^{\inn}_\al$ are the spaces of
holomorphic and anti-holomorphic delta-forms supported on $X_\al$,
respectively. The point is that $\CF^{\inn}_\al$ is precisely the local
cohomology $H^{n-n_\al}_{X_\al}({\mc E})$, for ${\mc E} =
\Omega_{X,\on{hol}}$, the sheaf of holomorphic differential forms on
$X$ (and similarly for $\ol\CF^{\inn}_\al$):
\begin{equation}    \label{is local coh}
\CF^{\inn}_\al = H^{n-n_\al}_{X_\al}(\Omega_{X,\on{hol}}).
\end{equation}

Now for each pair of strata such that $X_\al \succ X_\beta$ (which
means that $X_\beta$ is a codimension one stratum in the closure of
$X_\al$) there is a canonical GC operator
$$
\delta_{\al\beta}: H^{n-n_\al}_{X_\al}({\mc E}) \to
H^{n-n_\beta}_{X_\beta}({\mc E}).
$$
Therefore we obtain canonical maps
$$
\delta_{\al\beta}: \CF^{\inn}_\al \to \CF^{\inn}_\beta
$$
for all $\al,\beta \in A$ such that $X_\al \succ X_\beta$.

Likewise, we have anti-holomorphic analogues of these maps:
$$
\ol\delta_{\al\beta}: \ol\CF^{\inn}_\al \to \ol\CF^{\inn}_\beta
$$
for $X_\al \succ X_\beta$.

Now we use these maps to write formulas for $Q,Q^*$ and $H$. Actually,
all of these operators decompose into sums of holomorphic and
anti-holomorphic parts:
$$
Q = d = \pa + \ol\pa, \qquad Q^* = 2\imath_v = 2(\imath_\xi +
\imath_{\ol\xi}),
$$
$$
H = {\mc L}_v = {\mc L}_\xi + {\mc L}_{\ol\xi},
$$
and we will write separate formulas for these parts.

Let us choose, as in \secref{gen case}, a function $w_{\al\beta}$ on
$X_\al \bigsqcup X_\beta$ such that $X_\beta$ is the divisor of zeros
of $w_{\al\beta}$. Once we choose these coordinates, we obtain
particular regularizations of our integrals, as explained in the
previous section, and hence we may identify ${\mc H}^{\inn}$ with the
direct sum $\bigoplus_{\al \in A} \CH^{\inn}_\al$. This gives us a
concrete realization of the space of ``in'' states, which is more
convenient for computations than its more abstract definition as a
subspace of the space of distributions on $X$. We now describe the
action of the supersymmetry charges and the Hamiltonian on the space
of states in this realization.

Let
$$
\Psi = (\Psi_\al \otimes \ol\Psi_\al) \in \bigoplus_{\al \in A}
\CF^{\inn}_\al \otimes \ol\CF^{\inn}_\al \simeq {\mc H}^{\inn}.
$$
In the same way as in the case of $\pone$ (see \secref{supercharges})
we obtain the following formulas for the action of these operators on
\begin{align}    \label{pa}
\pa \Psi &= \pa_{\on{naive}} \Psi + 2\pi \; \sum_{\beta; X_\al\succ
X_\beta} \delta_{\al\beta} \left(\frac{dw_{\al\beta}}{w_{\al\beta}}
\wedge \Psi_\al \right) \otimes
\ol\delta_{\al\beta}(\ol\Psi_\al), \\ \label{olpa} \ol\pa \Psi &=
\ol\pa_{\on{naive}} \Psi + 2\pi \; \sum_{\beta; X_\al\succ X_\beta}
\delta_{\al\beta} (\Psi_\al) \otimes \ol\delta_{\al\beta}
\left(\frac{d\wb_{\al\beta}}{\wb_{\al\beta}} \wedge \ol\Psi_\al
\right),
\end{align}
\begin{align}    \label{Lxi}
{\mc L}_\xi &= \{ {\mc L}_{\xi},\imath_{\xi} \} = {\mc
L}_{\xi,\on{naive}} + 2 \pi \; \sum_{\beta; 
X_\al\succ X_\beta} a_{\al\beta} \; \delta_{\al\beta} \otimes
\ol\delta_{\al\beta}, \\
\label{Lolxi} {\mc L}_{\ol\xi} &= \{ {\mc L}_{\ol\xi},\imath_{\ol\xi}
\} = {\mc L}_{\ol\xi,\on{naive}} + 2 \pi \; \sum_{\beta; X_\al\succ
X_\beta} a_{\al\beta} \; \delta_{\al\beta} \otimes
\ol\delta_{\al\beta},
\end{align}
where the numbers $a_{\al\beta}$ are defined by the formula $\xi \cdot
w_{\al\beta} = a_{\al\beta} w_{\al\beta}$, so that $\imath_\xi \cdot
\frac{dw_{\al\beta}}{w_{\al\beta}} = a_{\al\beta}$ (see
\secref{gen kahler} below where we discuss explicitly the example of
$X=\ptwo$). Therefore we find that
$$
H = \{ d,\imath_v \} = H_{\on{naive}} + 4\pi \; \sum_{X_\al\succ
X_\beta} a_{\al\beta} \; \delta_{\al\beta} \otimes
\ol\delta_{\al\beta}.
$$
The operators $\imath_\xi$ and $\imath_{\ol\xi}$ have no correction
terms.

The fact that $\pa^2 = 0$ is the consequence of a non-trivial property
of the GC operators: suppose we have four strata $X_\al, X_{\beta'},
X_{\beta''},X_\ga$, such that $X_\al \succ X_{\beta'} \succ X_\ga$ and
$X_\al \succ X_{\beta''} \succ X_\ga$. Then we have
\begin{equation}    \label{square zero}
\delta_{\beta'\gamma} \circ \delta_{\alpha\beta'} =
\delta_{\beta''\gamma} \circ \delta_{\alpha\beta''}
\end{equation}
(see \cite{Kempf}). The fact that $\ol\pa^2 = 0$ is proved in
the same way.

While the identification of ${\mc H}^{\inn}$ with the direct sum
$\bigoplus_{\al \in A} \CH^{\inn}_\al$ depends on the choice of the
coordinates $w_{\al\beta}$, we see that the formulas for the operators
$\pa$ and $\ol{\pa}$ depend on their logarithmic derivatives, and the
formulas for the Hamiltonian and its holomorphic and anti-holomorphic
parts only depend on the eigenvalues of $\xi$ on these coordinates.

Thus, we find that, as in the $\pone$ model, the hamiltonian is
non-diagonalizable. It has off-diagonal terms which are given by
the GC operators.

\medskip

\noindent {\em Remark.} We want to stress that the appearance of
Jordan blocks in the Hamiltonian is tied up with the assumptions we
have made in \secref{ground} about the manifold $X$ being K\"ahler and
the gradient vector field of the Morse function $f$ coming from a
$\C^\times$-action. If we allow more general Morse functions, then the
spectrum of the Hamiltonian in our model may (and generically will) be
non-degenerate, and hence the Hamiltonian will be diagonalizable. This
is related to the well-known property of the ``partie finie''
regularization: the functions $x^{-\alpha}$, where ${\alpha}$ is not a
positive integer, have canonical extensions to homogeneous
distributions on the line, unlike the functions with ${\alpha} \in
{\mathbb Z}_{>0}$ which we discussed above (see
\cite{Hoermander,GelfandShilov}).

\ssec{Cohomology of the supercharges}    \label{coh of Q}

A natural application of the above formula for the supercharge $Q$ is
to use it to compute its cohomology on the space ${\mc H}^{\inn}$ of
states of our theory and check that it coincides with
$H^\bullet(X,\C)$.

We compute this cohomology by using a spectral sequence. Consider the
filtration ${\mc H}^{\inn}_{\leq i}$ introduced in \secref{gen
case}. According to formulas \eqref{pa} and \eqref{olpa}, the
supercharge $Q$ preserves this filtration. Therefore we may compute
the cohomology of $Q$ by using the spectral sequence associated to
this filtration. The $0$th term of this spectral sequence is
\begin{equation}    \label{zeroth}
\bigoplus_{i\geq 0} {\mc H}^{\inn}_{\leq i}/\CH^{\inn}_{\leq (i-1)} =
\bigoplus_{\al \in A} \CH^{\inn}_\al
\end{equation}
Let us compute the $0$th differential. We find that the second terms
in formulas \eqref{pa} and \eqref{olpa} map ${\mc H}^{\inn}_{\leq
i}$ to $\CH^{\inn}_{\leq (i-1)}$. Therefore the corresponding
differential on $\bigoplus_{\al \in A} \CH^{\inn}_\al$
is just the de Rham differential.

It is easy to see from the description of $\CH^{\inn}_\al$ given in
\secref{hol fact} that the cohomology of the de Rham differential
acting on $\CH^{\inn}_\al$ is one-dimensional, occurring in
cohomological degree $(n-n_\al,n-n_\al)$ and spanned by the delta-form
$\Delta_\al$. Therefore the first term of our spectral sequence is
spanned by the delta-forms $\Delta_\al, \al \in A$. We have
$$
{\mc H}^{\inn}_\al \simeq {\mc F}^{\inn}_\al \otimes \ol{\mc
  F}^{\inn}_\al,
$$
where ${\mc F}^{\inn}_\al$ and $\ol{\mc F}^{\inn}_\al$ are the spaces
of holomorphic and anti-holomorphic delta-forms on $X_\al$,
respectively (see \secref{hol
fact}). Using local cohomology, we may express them as follows:
$$
\CF^{\inn}_\al = H^{n-n_\al}_{X_\al}(\Omega_{X,\on{hol}}), \qquad
\ol\CF^{\inn}_\al = H^{n-n_\al}_{X_\al}(\Omega_{X,\on{anti-hol}}).
$$
With respect to this tensor product decomposition, we have
$$
\Delta_\al = \Delta_\al^{\on{hol}} \otimes
\Delta_\al^{\on{anti-hol}},
$$
where $\Delta_\al^{\on{hol}}$ and $\Delta_\al^{\on{anti-hol}}$ are
generating vectors of ${\mc F}^{\inn}_\al$ and $\ol{\mc
F}^{\inn}_\al$, respectively, considered as ${\mc D}$-modules.

According to formulas \eqref{pa}--\eqref{olpa}, the differential
$d_1$ of the first term of our spectral sequence is given by a
linear combination of the CS operators:
\begin{align*}
d_1(\Delta_\al) = 2\pi \; \sum_{\beta; X_\al\succ X_\beta} &\left(
\delta_{\al\beta} \left(\frac{dw_{\al\beta}}{w_{\al\beta}} \wedge
\Delta^{\on{hol}}_\al \right) \otimes
\ol\delta_{\al\beta}\left(\Delta^{\on{anti-hol}} \right) \right. \\ &+
\left.  \delta_{\al\beta} \left(\Delta^{\on{hol}}_\al \right) \otimes
\ol\delta_{\al\beta}\left( \frac{d\wb_{\al\beta}}{\wb_{\al\beta}}
\wedge\Delta^{\on{anti-hol}} \right) \right).
\end{align*}
It follows from the definition that $\Delta^{\on{hol}}_\al$ and
$\Delta_\al^{\on{anti-hol}}$ extend to all strata $X_\beta$ of
codimension $1$ in the closure of $X_\al$. Therefore
$\delta_{\al\beta}(\Delta^{\on{hol}}_\al) = 0$ and
$\ol\delta_{\al\beta}(\Delta_\al^{\on{anti-hol}}) = 0$, and so
$d_1(\Delta_\al) = 0$. Hence we conclude that $d_1$, as well
as all higher differentials of our spectral sequence, are all equal to
zero. Therefore the cohomology is spanned by the delta-forms
$\Delta_\al$. These form the dual basis to the homology basis
represented by the even-dimensional cycles $X_\al$.

Therefore we conclude that the cohomology of $Q$ acting on
$\CH^{\inn}_\al$ coincides with the cohomology of the de Rham
differential and is isomorphic to the cohomology of $X$,
$H^\bullet(X,\C)$, as expected.

Let us recall that our supercharge $Q = d$ has a canonical
decomposition $Q = \pa + \ol\pa$. Therefore it is also interesting to
compute the cohomology of the differentials $\pa$ and $\ol\pa$
separately. The operators $\pa$ and $\ol\pa$ are quantum mechanical
analogues of the left and right moving supercharges in two-dimensional
sigma models. According to \cite{Kapustin,Witten:cdo} (see also
\cite{AiB,Tan}), the cohomology of the right moving supercharge is
a chiral algebra which is closely related to the chiral de Rham
complex \cite{MSV} of $X$. The cohomology of $\ol\pa$ may be thought
of as a ``baby version'' of this chiral algebra. The explicit formulas
of the previous section give us an effective tool for computing this
cohomology.

This tool is the {\em Grothendieck-Cousin resolution} (GC resolution
for short). This is a complex $C^\bullet({\mc E}) = \bigoplus_{i\geq
0} C^i({\mc E})$, defined for a holomorphic vector bundle $\CE$ on
$X$, whose cohomology coincides with the cohomology of ${\mc E}$,
considered as a coherent sheaf on $X$, $H^\bullet(X,{\mc E})$. The
$i$th term $C^i({\mc E})$ of the complex is equal to
$$
C^i({\mc E}) = \bigoplus_{\dim X_\al = i} H^{n-i}_{X_\al}({\mc E}),
$$
where $H^{n-i}_{X_\al}({\mc E})$ is the local cohomology of ${\mc E}$
with support on $X_\al$ that was introduced in \secref{action of Q}.

The differential $\delta^i: C^i({\mc E}) \to C^{i+1}({\mc E})$ is
given by the alternating sum of the GC operators $\delta_{\al\beta} =
\delta_{\al\beta}^{\CE}: H^{n-n_\al}_{X_\al}({\mc E}) \to
H^{n-n_\al+1}_{X_\beta}({\mc E})$ introduced in \secref{action of Q}:
\begin{equation}    \label{GC diff}
\delta^i = \sum_{\beta; X_\al \succ X_\beta} \ep_{\al\beta}
\delta_{\al\beta},
\end{equation}
where $\ep_{\al\beta} = \pm 1$ are signs chosen so as to ensure that
$\delta^{i+1} \circ \delta^i = 0$. The existence of such signs follows
from the fact that the GC operators satisfy the identity \eqref{square
zero} if we have $X_\al \succ X_{\beta'} \succ X_\ga$ and $X_\al \succ
X_{\beta''} \succ X_\ga$.

Under our assumptions on the stratification $X = \bigsqcup_{\al \in A}
X_\al$ the $j$th cohomology of the complex $C^\bullet({\mc E})$
coincides with $H^j(X,{\mc E})$, see \cite{Kempf}.

Let us see how this works in the simplest example of $X=\pone$ and the
stratification $\pone = \C_0 \bigsqcup \infty$.

Consider first the case when ${\mc E}$ is the trivial line bundle on
$\pone$. The corresponding complex $C^\bullet({\mc O})$ looks
as follows:
$$
H^0(\C_0,{\mc O}) \to H^1_\infty({\mc O}).
$$
We have already determined these spaces in formula \eqref{ident
for pone}. The resulting complex is \eqref{GC pone}, where the
differential is obtained by composing the maps
$$
\C[w^{-1}] \to \C[w,w^{-1}] \qquad \on{and} \qquad \C[w,w^{-1}] \to
\C[w,w^{-1}]/\C[w].
$$
It is easy to see that the differential is surjective, and its kernel
is one-dimensional spanned by the constants in $\C[w^{-1}]$. This
coincides with the cohomology of ${\mc O}$ on $\pone$.

More generally, suppose that ${\mc E}$ is the line bundle ${\mc O}(n),
n \in \Z$. Then we can trivialize this line bundle on $\C_0$ and
$\C_\infty$, and the transition function is $w^n$. We still have the
identifications \eqref{ident for pone}, but now the GC operator is the
map \eqref{GC pone} obtained by composing the embedding $\C[w^{-1}]
\to \C[w,w^{-1}]$, multiplication by $w^n$, and the projection
$\C[w,w^{-1}] \to \C[w,w^{-1}]/\C[w]$. As the result, the kernel and
the cokernel of the GC operator change: if $n\geq 0$, then the kernel
is $(n+1)$-dimensional, spanned by $1,w^{-1},\ldots,w^{-n}$, and the
cokernel is zero. If $n<-1$, then the kernel is zero, and the cokernel
is spanned by $w^{-1},\ldots,w^{n+1}$. If $n=-1$, then both kernel and
cokernel are zero. Again, we find the agreement with the cohomology
$H^i(\pone,{\mc O}(n))$.

Now we use the GC complex to compute the cohomology of the anti-chiral
supercharge $\ol\pa$. We compute this cohomology using the spectral
sequence associated to the same filtration that we used in the above
computation of the cohomology of $Q$. The $0$th term of the spectral
sequence is again \eqref{zeroth} and the $0$th differential is the
operator $\ol\pa$ acting on this space. Now recall that each of the
summands in the direct sum \eqref{zeroth} factorizes into the tensor
product
$$
\CH^{\inn}_\al = {\mc F}^{\inn}_\al \otimes \ol{\mc F}^{\inn}_\al.
$$

The operator $\ol\pa$ acts along the second factor $\ol{\mc
F}^{\inn}_\al$. It is easy to see that its cohomology is
one-dimensional, spanned by the generator $\De^{\on{anti-hol}}_\al$ of
$\ol{\mc F}^{\inn}_\al$, which is in cohomological dimension
$n-n_\al$. Therefore we obtain that the first term of the spectral
sequence is concentrated in the $0$th row, and the terms in this row
are isomorphic to
$$
E^{i,0}_1 \simeq \bigoplus_{\al \in A; i=n-n_\al} {\mc F}^{\inn}_\al
\otimes \Delta^{\on{anti-hol}}_\al.
$$
According to formula \eqref{olpa}, the action of the
differential $\ol\pa_1$ of the first term of the spectral sequence is
given by the formula
$$
\ol\pa_1(\Psi_\al \otimes
\Delta^{\on{anti-hol}}_\al) = 2\pi \; \sum_{\beta; X_\al\succ X_\beta}
\delta_{\al\beta} (\Psi_\al)
\otimes \ol\delta_{\al\beta} \left(
\frac{d\wb_{\al\beta}}{\wb_{\al\beta}} \wedge
\Delta^{\on{anti-hol}}_\al \right).
$$
However, it follows from the definitions that we can normalize the
states $\Delta^{\on{anti-hol}}_\al$ in such a way that
$$
\ol\delta_{\al\beta} \left( \frac{d\wb_{\al\beta}}{\wb_{\al\beta}}
\wedge \Delta^{\on{anti-hol}}_\al \right) =
\ep_{\al\beta} \Delta^{\on{anti-hol}}_\beta,
$$
where the sign $\ep_{\al\beta} = \pm 1$ is due to the fact that we
obtain $\Delta^{\on{anti-hol}}_\beta$ by multiplying
$\Delta^{\on{anti-hol}}_\al$ with $d\wb_{\al\beta}$, and so for fixed
$\beta$ and varying $\al$ we obtain different signs in general.

Let us now identify the $(i,0)$ group of the first term of the
spectral sequence with
$$
E_1^{i,0} = \bigoplus_{\al \in A; i=n-n_\al} {\mc F}^{\inn}_\al
$$
by $\Psi_\al \otimes \Delta^{\on{anti-hol}}_\al \mapsto
\Psi_\al$. Recall that according to formula \eqref{is local coh} we
have
$$
\CF^{\inn}_\al = H^{n-n_\al}_{X_\al}(\Omega_{X,\on{hol}}),
$$
where $\Omega_{X,\on{hol}} = \bigoplus_{j \geq 0}
\Omega^j_{X,\on{hol}}$ is the sheaf of holomorphic differential
forms. Thus,
$$
E_1^{i,0} = \bigoplus_{\al \in A; i=n-n_\al}
H^{n-n_\al}_{X_\al}(\Omega_{X,\on{hol}})
$$
is precisely the $i$th term of the GC complex associated with
$\Omega_{X,\on{hol}}$. Moreover, we find that the first differential
$\ol\pa^i_1: E^{i,0}_1 \to E^{i+1,0}_1$ of our spectral sequence is
given by the following formula: for $\Psi = (\Psi_\al) \in E^{i,0}_1$
we have (up to the inessential factor of $2\pi$)
$$
\ol\pa_1(\Psi) = \sum_{\beta; X_\al\succ X_\beta}
\ep_{\al\beta} \delta_{\al\beta} (\Psi_\al).
$$
This is precisely the differential \eqref{GC diff} of the GC complex
$C^\bullet(\Omega_{X,\on{hol}})$ associated with the sheaf
$\Omega_{X,\on{hol}}$. Since the GC complex computes the cohomology of
this sheaf, we find that the cohomology of $\ol\pa_1$ is equal to
$$
H^i(X,\Omega_{X,\on{hol}}).
$$
Recall that the first term of our spectral sequence has only one
row. Therefore the spectral sequence collapses in the first term, and
we find that this is in fact the answer for the cohomology of $\ol\pa$
on $\CH^{\inn}$.

Since $X$ is K\"ahler, we have
$$
H^k(X,\C) \simeq \bigoplus_{i,j \geq 0} H^{j,i}, \qquad H^{j,i} =
H^i(X,\Omega^j_{X,\on{hol}}).
$$
Thus, the cohomology of the space $\CH^{\inn}$ of ''in'' states with
respect to the anti-chiral supercharge $\ol\pa$ is isomorphic to the
cohomology of $X$, but with the cohomological grading coming from the
anti-holomorphic cohomological degrees of differential forms. However,
the holomorphic cohomological degree is preserved by $\ol\pa$, so we
could consider it as an extra grading on our complex. Thus, we find
that the cohomology of the anti-chiral supercharge $\ol\pa$ coincides
with the cohomology of the full supercharge $Q$. In \secref{nonSUSY}
we will generalize this result to a class of non-supersymmetric
models.

\section{Action of observables on the space of states}
\label{action}

In the previous section we described the space of states of our
quantum mechanical model in the limit $\la=\infty$. We have seen that
there are actually two spaces: ${\mc H}^{\inn}$ and ${\mc
H}^{\on{out}}$ and that their structure is dramatically different from
the usual structure in hermitean quantum mechanics. The reason is that
before we pass to the limit $\la \to \infty$ we make a violent
transformation of the states: the ``in'' states are multiplied by
$e^{\la f}$, and the ``out'' states are multiplied by $e^{-\la f}$. In
contrast, the {\em observables} of our theory are getting conjugated
by the function $e^{\la f}$. There is a large class of observables,
namely, all evaluation observables introduced in \secref{cor fns as
int} (corresponding to smooth differential forms on $X$), which
commute with $e^{\la f}$. Those remain intact in the limit $\la \to
\infty$. This immediately leads to the question: how do these
observables act on the spaces of states, ${\mc H}^{\inn}$ and ${\mc
H}^{\on{out}}$?

This is the question that we take up in this section,
first in the case of $\pone$ and then in the general case. We will see
that analytic properties of the observables play an important role in
the limit $\la \to \infty$. We will also see that factorization of the
correlation functions over intermediate states leads to some
non-trivial identities on analytic differential forms.

\ssec{The case of $\pone$}

The spaces of states in this case have been described in great detail
in the previous sections. The space ${\mc H}^{\inn}$ has a basis
consisting of the states
\begin{equation}    \label{in once more}
|n,\nb,\ph,\phb\rangle_{\C_0} \qquad \on{and} \qquad
|n,\nb,\ph,\phb\rangle_\infty,
\end{equation}
and the space ${\mc H}^{\on{out}}$ has a basis consisting of the
states
\begin{equation}    \label{out once more}
_0 \langle n,\nb,\ph,\phb| \qquad \on{and} \qquad {}_{\C_\infty}\langle
n,\nb,\ph,\phb|.
\end{equation}

Recall that we have defined the vectors in ${\mc H}^{\inn}_{\C_0}$ and
${\mc H}^{\on{out}}_{\C_\infty}$ by using particular regularizations
of the integrals of the differential forms $z^n \zb^{\nb} (dz)^{\ph}
(d\zb)^{\phb}$ and $w^n \wb^{\nb} (dw)^{\ph} (d\wb)^{\phb}$,
respectively, as explained in \secref{expl reg} (recall that $z$ is a
coordinate at $0 \in \pone$, and $w=z^{-1}$ is a coordinate at
$\infty$). We have seen in \secref{pairin} that if we choose the
``cutoffs'' appearing in these regularized integrals in a compatible
way ($|z|<\ep^{-1}$ in the first case, $|w|<\ep^{-1}$ in the second
case), then we have
$$
_{\C_\infty}\langle n,\nb,\ph,\phb|n,\nb,\ph,\phb\rangle_{\C_0} = 0,
$$
and so the bases \eqref{in once more} and \eqref{out once more} are
dual to each other (up to powers of $i$). We will use this property in
what follows. If one were to choose regularizations of ``in'' and
``out'' states independently, then some of our formulas below would
need to be modified to account for that.

The action of the evaluation observable $\wh{\omega}$ corresponding to
a smooth differential form on $\pone$ on ${\mc H}^{\inn}$ and ${\mc
H}^{\on{out}}$ may be found from the matrix coefficients
$$
\langle \Psi^{\out}|\wh\omega|\Psi^{\inn} \rangle.
$$
It follows from our construction that this matrix coefficient is equal
to the integral
$$
\int \Psi^{\out} \wedge \omega \wedge
\Psi^{\inn},
$$
understood in the same way as in \secref{pairin}.

This allows us to compute explicitly the action of the evaluation
observables. For example, consider the case when $\omega$ is a smooth
$(r,\ol{r})$-differential form on $\pone$, and write
$$
\omega = \omega_{z\zb}(z,\zb) (dz)^r (d\zb)^{\ol{r}} =
\omega_{w\wb}(w,\wb) (dw)^r (d\wb)^{\ol{r}},
$$
so that
$$
\omega_{w\wb}(w,\wb) = (-1)^{r+\ol{r}}
\omega_{z\zb}(w^{-1},\wb^{-1}) w^{-2r} \wb^{-2\ol{r}}
$$
(we recall our convention \eqref{d2z}).

Then we find from the above formula that
$$
\wh{\omega} \vert n,\nb,\ph,\phb \rangle_{{\C}_{0}} = (-1)^{p\ol{r}}
\sum_{m,\ol{m} = 0}^{\infty} \left. \frac{1}{m! {\ol{m}}!} {\p}_z^{m}
{\p}_{\zb}^{\ol{m}} (\omega_{z\zb} z^n \zb^{\nb}) \right|_{z=0} \;
\vert m,\ol{m},\ph+r,\phb+\ol{r} \rangle_{{\C}_{0}}
$$
$$
+ (-1)^{\ph+\phb+p\ol{r}} \sum_{m,\ol{m} = 0}^{\infty}
\underset{|w|<\ep^{-1}}\pf \omega_{w\wb} \ w^{m-n-2\ph}
{\wb}^{\ol{m}-\nb-2\phb} dw d\wb \; \; \vert
m,\ol{m},\ph+r,\phb+\ol{r} \rangle_{\infty}
$$
and
$$
\wh{\omega} \vert n,\nb,\ph,\phb \rangle_\infty = (-1)^{p\ol{r}}
\left. \frac{1}{n! {\ol{n}}!} \sum_{m=0}^{n} \sum_{\ol{m}=0}^{\ol{n}}
{\p}_w^{n} {\p}_{\wb}^{\ol{n}} (\omega_{w\wb} w^{m} \wb^{\ol{m}})
\right|_{w=0} \; \vert m,\ol{m},\ph+r,\phb+\ol{r} \rangle_\infty.
$$
Here we use the convention that a state is equal to zero if at least
one of its indices takes a value that is not allowed.

Likewise, we find that
$$
_{\C_\infty} \langle n,\nb,\ph,\phb| \wh{\omega} = (-1)^{p\ol{r}}
\sum_{m,\ol{m} = 0}^{\infty} \left. \frac{1}{m! {\ol{m}}!} {\p}_w^{m}
{\p}_{\wb}^{\ol{m}} (\omega_{w\wb} w^n \wb^{\nb}) \right|_{w=0} \;
{} _{\C_\infty} \langle m,\ol{m},\ph+r,\phb+\ol{r}|
$$
$$
+ (-1)^{\ph+\phb+p\ol{r}} \sum_{m,\ol{m} = 0}^{\infty}
\underset{|z|<\ep^{-1}}\pf \omega_{z\zb} \ z^{m-n-2\ph}
{\zb}^{\ol{m}-\nb-2\phb} dz d\zb \; \;
{}_0 \langle m,\ol{m},\ph+r,\phb+\ol{r}|
$$
and
$$
_0 \langle n,\nb,\ph,\phb| \wh{\omega}= (-1)^{p\ol{r}}
\left. \frac{1}{n! {\ol{n}}!} \sum_{m=0}^{n} \sum_{\ol{m}=0}^{\ol{n}}
{\p}_z^{n} {\p}_{\zb}^{\ol{n}} (\omega_{z\zb} z^{m} \zb^{\ol{m}})
\right|_{z=0} \; {}_0 \langle m,\ol{m},\ph+r,\phb+\ol{r}|.
$$

The right hand sides of these formulas, as well as the other formulas
that appear below, are in general infinite linear combinations of our
states. This means that they are really vectors in the completions of
the spaces ${\mc H}^{\inn}$ and ${\mc H}^{\out}$, which are the direct
products of the (finite-dimensional) generalized eigenspaces with
respect to ${\mc L}_\xi$ and ${\mc L}_{\ol\xi}$. However, these naive
completions are too big, and we would like to define some more
reasonable subspaces whose elements possess some analytic
properties. Here is a possible definition of such a completion
$\wt{\mc H}^{\inn}_{\C_0}$ of ${\mc H}^{\inn}_{\C_0}$: take the space
of all analytic differential forms on $C_0$ which grow not faster than
a polynomial at $\infty$. It contains the monomials $z^n \zb^{\nb}
dz^p d\zb^{\phb}$ that we have considered previously and the products
of $z^n \zb^{\nb}$ with analytic differential forms on $\pone$. The
expansion of such a form at $z=0$ gives rise to a (possibly infinite)
linear combination of our monomial states. The completion of ${\mc
H}^{\inn}$ is then defined as the sum of ${\mc H}^{\inn}_\infty$ and
the subspace of the space of distributions on $\pone$ spanned by all
possible regularizations of elements of $\wt{\mc
H}^{\inn}_{\C_0}$. This completion of ${\mc H}^{\inn}$ contains, in
particular, all finite linear combinations of the derivatives of the
$\delta$-forms supported at $\infty$ as well as some of their infinite
linear combinations.

The completion of ${\mc H}^{\out}$ is defined in a similar
way. However, we note that the pairing between ${\mc H}^{\inn}$ and
${\mc H}^{\out}$ does not extend to a pairing between their
completions. Instead, we have a weaker property. For example, for a
two-form $\omega$ in the completion of ${\mc H}^{\out}$ and a
zero-form $f$ in the completion of ${\mc H}^{\inn}$ the pairing
$\langle \omega,f \rangle$ will in general be a divergent infinite
sum. But the pairing $\langle \omega,\phi^*(q)(f) \rangle$, where
$\phi^*(q)(f)(z,\ol{z}) = f(qz,\ol{q}\ol{z})$ should converge if
$0<|q|<\delta$ for sufficiently small $\delta>0$ (which depends on
$\omega$ and $f$). We will observe a similar phenomenon in the next
section when we discuss factorization of the correlation
functions.

\ssec{Correlation functions and their factorization over intermediate
   states} \label{their factorization}

We now compare the above formulas for the matrix elements of
evaluation observables attached to differential forms on $\pone$ with
the exact expression \Ref{corrf} for the correlation functions of
these observables.

We recall that in general correlation functions are labeled by pairs
of critical points $x_-,x_+$ corresponding to the choice of the ``in''
and ``out'' vacua. The corresponding path integral is given by the
integral over the moduli space ${\mc M}_{x_-,x_+}$ of gradient
trajectories of the differential forms on $X$, pulled back to ${\mc
M}_{x_-,x_+}$ via the evaluation maps.

In our case $X=\pone$, there are two critical points: $0$ and
$\infty$, and there are three moduli spaces. Two of them, ${\mc
M}_{0,0}$ and ${\mc M}_{\infty,\infty}$, consist of a single
point. They correspond to constant maps (taking the value $0$ and
$\infty$ in $\pone$, respectively). The only non-trivial moduli space
is ${\mc M}_{0,\infty}$ which corresponds to the only possible
instanton transition: gradient trajectories going from $x_-=0$ to
$x_+=\infty$. This moduli space is isomorphic to $\C^\times$, which is
naturally isomorphic to the subset $\pone \bs \{ 0,\infty \}$ under
the evaluation at $t=0$ map $\on{ev}_0$. Thus, it has a natural
compactification isomorphic to $X=\pone$.

Consider the two-point function of the evaluation observables
$\wh\omega$ and $\wh{F}$, where $\omega$ is a smooth two-form on
$\pone$ and $F$ is a smooth function on $\pone$. Let us insert
$\wh\omega$ at the time $t_1 \in \R$ and $\wh\omega$ at the time
$t_2<t_1$. Then the corresponding correlation function is equal to
\begin{equation}    \label{cor 2}
_\infty \langle \wh\omega(t_1) \wh{F}(t_2) \rangle_0 =
\int_{\pone} \omega \; \phi(e^{-t})^*(F),
\end{equation}
where $(\phi(e^{-t})^*(F))(z,\ol{z}) = F(qz,q\ol{z})$ with
$q=e^{-t} \in \R^\times$ and $t=t_1-t_2$.

Actually, it would be convenient to break $H = {\mc L}_v$ into the sum
$H = {\mc L}_\xi + {\mc L}_{\ol\xi}$ where $\xi = z\pa_z$ and $\ol\xi
= \zb \pa_{\zb}$ and to allow $t_1,t_2$ and $t=t_1-t_2$ to be complex:
\begin{equation}    \label{cor 2 comp}
_\infty \langle \wh\omega(t_1) \wh{F}(t_2) \rangle_0 =
\int_{\pone} \omega \phi(e^{-t})^*(F),
\end{equation}
where $(\phi(e^{-t})^*(F))(z,\ol{z}) = F(qz,\ol{q}\ol{z})$ with
$q=e^{-t}, \ol{q} = e^{-\ol{t}}$. If $t$ is real, then we recover
formula \eqref{cor 2}.

The right hand side of formula \eqref{cor 2 comp} is the answer that
we obtain from the Lagrangian, or path integral, formulation of
the model. On the other hand, we may compute the same correlation
function from our Hamiltonian formulation. In the hamiltonian
realization the vacuum ``in'' state corresponding to the critical
point $0$ is
$$
\vac_{\C_0} = |0,0,0,0 \rangle_{\C_0},
$$
and the covacuum ``out'' state corresponding to the critical point
$\infty$ is
$$
_{\C_\infty} \covac = {}_{\C_\infty} \langle 0,0,0,0|.
$$
Therefore the same correlation function should be equal to the matrix
element
\begin{equation}    \label{ham 2}
_{\infty} \langle \wh\omega(t) \wh{F}(0) \rangle_0 = {}
_{\C_\infty} \covac \wh\omega e^{-t{\mc L}_\xi - \ol{t} {\mc
L}_{\ol\xi}} \wh{F} \vac_{\C_0}.
\end{equation}
We evaluate this matrix element using the formulas for the action of
the observables on the states obtained in the previous section for the
action of ${\mc L}_\xi$ and ${\mc L}_{\ol\xi}$ on the states obtained
in \secref{action ham}.

We have
\begin{align*}
\wh{F} \vac_{\C_0} &= \sum_{m,\ol{m} = 0}^{\infty} \left. \frac{1}{m!
{\ol{m}}!} {\p}_z^{m} {\p}_{\zb}^{\ol{m}} F \right|_{z=0} \; \vert
m,\ol{m},0,0 \rangle_{{\C}_{0}} \\ &+ \sum_{m,\ol{m} = 0}^{\infty}
\underset{|w|<\ep^{-1}}\pf F \ w^{m} {\wb}^{\ol{m}} dw \wedge d\wb \;
\; \vert m,\ol{m},0,0 \rangle_{\infty}.
\end{align*}
Next, we find using formulas \eqref{L xi} and \eqref{L olxi} that
\begin{align*}
e^{-t{\mc L}_\xi - \ol{t} {\mc L}_{\ol\xi}} \wh{F} \vac_{\C_0} &=
\sum_{m,\ol{m} = 0}^{\infty} \left. \frac{q^m \ol{q}^{\ol{m}}}{m!
{\ol{m}}!} {\p}_z^{m} {\p}_{\zb}^{\ol{m}} F \right|_{z=0} \; \vert
m,\ol{m},0,0 \rangle_{{\C}_{0}} \\ &- 2\pi \log(q\ol{q})
\sum_{m,\ol{m} = 1}^{\infty} \left. \frac{q^m \ol{q}^{\ol{m}}}{m!
{\ol{m}}!} {\p}_z^{m} {\p}_{\zb}^{\ol{m}} F \right|_{z=0} \; \vert
m-1,\ol{m}-1,0,0 \rangle_\infty \\ &+ \sum_{m,\ol{m} = 0}^{\infty}
q^{m+1} \ol{q}^{\ol{m}+1} \underset{|w|<\ep^{-1}}\pf F \ w^{m}
{\wb}^{\ol{m}} dw \wedge d\wb \; \; \vert m,\ol{m},0,0
\rangle_{\infty}.
\end{align*}

On the other hand, writing $\omega = \omega_{w\wb} dw d\wb =
\omega_{z\zb} dz d\zb$, we obtain that
$$
_{\C_\infty} \covac \wh\omega = \sum_{m,\ol{m} = 0}^{\infty}
\left. \frac{1}{m! {\ol{m}}!} {\p}_w^{m} {\p}_{\wb}^{\ol{m}}
\omega_{w\wb} \right|_{w=0} \; {} _{\C_\infty} \langle
m,\ol{m},1,1|
$$
$$
+ \sum_{m,\ol{m} = 0}^{\infty}
\underset{|z|<\ep^{-1}}\pf \omega_{z\zb} \ z^{m}
{\zb}^{\ol{m}} dz d\zb \; \;
{}_0 \langle m,\ol{m},1,1|.
$$

Therefore the right hand side of \eqref{ham 2} is equal to
\begin{align*}
{} _{\C_\infty} \covac \wh\omega e^{-t{\mc L}_\xi - \ol{t} {\mc
L}_{\ol\xi}} \wh{F} \vac_{\C_0} &= \sum_{m,\ol{m} = 0}^\infty
\underset{|z|<\ep^{-1}}\pf \omega_{z\zb} \ z^{m} {\zb}^{\ol{m}} dz
d\zb \; \cdot \left. \frac{q^m \ol{q}^{\ol{m}}}{m! {\ol{m}}!}
{\p}_z^{m} {\p}_{\zb}^{\ol{m}} F \right|_{z=0} \\ &+ q \ol{q}
\sum_{m,\ol{m} = 0}^{\infty} \left. \frac{q^m \ol{q}^{\ol{m}}}{m!
{\ol{m}}!} {\p}_w^{m} {\p}_{\wb}^{\ol{m}} \omega_{w\wb} \right|_{w=0}
\cdot \underset{|w|<\ep^{-1}}\pf F \ w^{m} {\wb}^{\ol{m}} dw
d\wb \\ &- 2\pi \log(q\ol{q}) \sum_{m,\ol{m} = 1}^{\infty}
\left. \frac{1}{(m-1)! (\ol{m}-1)!} {\p}_w^{m-1} {\p}_{\wb}^{\ol{m}-1}
\omega_{w\wb} \right|_{w=0} \cdot \left. \frac{q^m \ol{q}^{\ol{m}}}{m!
{\ol{m}}!} {\p}_z^{m} {\p}_{\zb}^{\ol{m}} F \right|_{z=0}.
\end{align*}

Combining this formulas with formula \eqref{cor 2 comp} obtained using
the path integral,
$$
{} _{\C_\infty} \covac \wh\omega e^{-t{\mc L}_\xi - \ol{t} {\mc
L}_{\ol\xi}} \wh{F} \vac_{\C_0} = \int_{\pone} \omega F(qz,\ol{q}\zb),
$$
we arrive at the following identity:
\begin{align}    \notag
\int_{\pone} \omega F(qz,\ol{q}\zb) &= \sum_{m,\ol{m} = 0}^\infty
\underset{|z|<\ep^{-1}}\pf \omega_{z\zb} \ z^{m} {\zb}^{\ol{m}} dz
  d\zb \; \cdot \left. \frac{q^m \ol{q}^{\ol{m}}}{m! {\ol{m}}!}
{\p}_z^{m} {\p}_{\zb}^{\ol{m}} F \right|_{z=0} \\ \label{identity1} &+
q\ol{q} \sum_{m,\ol{m} = 0}^{\infty} \left. \frac{q^m
\ol{q}^{\ol{m}}}{m!  {\ol{m}}!} {\p}_w^{m} {\p}_{\wb}^{\ol{m}}
\omega_{w\wb} \right|_{w=0} \cdot \underset{|w|<\ep^{-1}}\pf F \ w^{m}
{\wb}^{\ol{m}} dw  d\wb \\ \notag &- 2\pi \log(q\ol{q})
\sum_{m,\ol{m} = 1}^{\infty} \left. \frac{1}{(m-1)! (\ol{m}-1)!}
{\p}_w^{m-1} {\p}_{\wb}^{\ol{m}-1} \omega_{w\wb} \right|_{w=0} \cdot
\left. \frac{q^m \ol{q}^{\ol{m}}}{m!  {\ol{m}}!} {\p}_z^{m}
{\p}_{\zb}^{\ol{m}} F \right|_{z=0}.
\end{align}

Note that the identity \eqref{identity1} may also be rewritten in the
following way:
\begin{align}    \notag
\int_{\pone} \omega F(qz,\ol{q}\zb) &= \sum_{m,\ol{m} = 0}^\infty
\underset{|z|<\ep^{-1}}\pf \omega_{z\zb} \ z^{m} {\zb}^{\ol{m}} dz
  d\zb \; \cdot \left. \frac{q^m \ol{q}^{\ol{m}}}{m! {\ol{m}}!}
{\p}_z^{m} {\p}_{\zb}^{\ol{m}} F \right|_{z=0} \\ \label{identity2} &+
q \ol{q} \sum_{m,\ol{m} = 0}^{\infty} \left. \frac{q^m
\ol{q}^{\ol{m}}}{m!  {\ol{m}}!} {\p}_w^{m} {\p}_{\wb}^{\ol{m}}
\omega_{w\wb} \right|_{w=0} \cdot \underset{|w|<q^{-1}\ep^{-1}}\pf F \
w^{m} {\wb}^{\ol{m}} dw  d\wb.
\end{align}
The terms with $\log|q|^2$, which appeared in the third term of
\eqref{identity1} are now hidden in the definition of the ``partie
finie'' regularization of the integral in the second term: instead of
the ``cutoff'' $|w|<\ep^{-1}$ as in \eqref{identity1} we are now using
the ``cutoff'' $|w|<q^{-1}\ep^{-1}$.

These identities express the factorization of the two point
correlation functions over intermediate states. Indeed, assuming for
simplicity that $t$ is real, we rewrite it as follows:
\begin{multline}    \label{factoriz}
\covac \wh\omega e^{-tH} \wh{F} \vac = \sum_\nu \langle
0|\wh\omega|\Psi_\nu \rangle \langle \Psi_\nu^*| e^{-tH} \wh{F} |0
\rangle = \\ = \sum_{\mu,\nu} \langle
0|\wh\omega|\Psi_\nu \rangle \langle \Psi_\nu^*| e^{-tH} | \Psi_\mu
\rangle \langle \Psi_\mu^* \wh{F} |0
\rangle,
\end{multline}
where $\{ \Psi_\mu \}$ and $\{ \Psi^*_\mu \}$ are dual bases of the
spaces of ``in'' and ``out'' states, respectively. Identities of this
type are taken for granted in conventional (i.e., CPT-invariant)
quantum mechanics, where the space of states is a Hilbert space, so
that we may take as $\{ \Psi_\mu \}$ and $\{ \Psi^*_\mu \}$ a complete
orthonormal basis. However, our model is not CPT-invariant, and so we
do not have the structure of Hilbert space on the space of states.
Instead, we have two distinct spaces of ``in'' and ``out'' states and
a pairing between them. Because of that, identity \eqref{identity1} is
more subtle, as we will see in the next section. It requires that $F$
and $\omega$ be analytic, and also the equality should be understood
in the sense that the right hand side is the $q,\ol{q}$-power series
expansion of the left hand side in the domain $0<|q|<\delta$ for some
$\delta$ which depends on $F$ and $\omega$. Before discussing these
subtleties, we point out some salient features of this identity.

The most important feature of the identity \eqref{identity1} is the
appearance of the logarithms of $q$ and $\ol{q}$ in the right hand
side. If the operators ${\mc L}_\xi$ and ${\mc L}_{\ol\xi}$ and the
Hamiltonian were diagonalizable, then the right hand side would be a
power series in $q = e^{-t{\mc L}_\xi}$ and $\ol{q} = e^{-t{\mc
L}_{\ol{\xi}}}$. But identity \eqref{identity1} shows that there are
also the terms involving $\log q$ and $\log \ol{q}$. This means that
the operators ${\mc L}_\xi$ and ${\mc L}_{\ol\xi}$, and the
Hamiltonian $H = {\mc L}_\xi + {\mc L}_{\ol\xi}$, are not
diagonalizable, but have Jordan blocks.

This leads us to the following conclusion: the logarithmic nature of
our model, which manifests itself in the fact that the Hamiltonian has
Jordan block structure, can be detected from, and is in fact {\em
dictated} by the correlation functions of evaluation
observables. These correlation functions are given by a simple and
explicit formula \eqref{corrf}. Applying this formula to various
observables of our model, we find logarithmic terms in $q$ and
$\ol{q}$, which implies that the Hamiltonian is not diagonalizable.

However, it is important to stress that in order to observe these
``logarithmic effects'' at least one of our observables should be a
{\em non-BPS} observable, i.e., not be annihilated by the supercharge
$Q$. If both $\wh{F}$ and $\wh\omega$ were BPS observables, i.e.,
$Q$-closed, then the one-point functions appearing in the right hand
side of formula \eqref{factoriz} would be non-zero only when the
intermediate states $\Psi_\mu, \Psi_\nu$ are the BPS states (i.e., the
vacuum states). On such states the Hamiltonian is diagonalizable, so
we would not be able to observe the logarithmic terms. Since $Q$ acts
as the de Rham differential and $\omega$ is a two-form, we find that
$\wh\omega$ is $Q$-closed. However, $\wh{F}$ is not $Q$-closed if $F$
is non-constant (if it were constant, then the logarithmic terms in
\eqref{identity1} would indeed disappear).

Thus, we can discover the structure of the space of states of the
theory, and in particular, the existence of the Jordan blocks of the
Hamiltonian, {\em only} if we consider correlation functions of
non-BPS observables. This is one more reason for considering
correlation functions beyond the topological sector of the theory: one
simply cannot observe these important features of the model within the
topological sector.

One can write similar identities for $n$-point correlation functions
of evaluation observables with $n>2$.

\ssec{Analytic aspects of the identity}

We now analyze identity \eqref{identity1} in more detail. There are
two important aspects that we notice right away. The first one is that
it does not hold for differential forms/functions that are not
analytic at the points $0$ and $\infty$. Indeed, suppose that all
derivatives of $F$ vanish at the point $0$ (i.e., at $z=0$), and all
derivatives of $\omega_{w\wb}$ vanish at the point $\infty$ (i.e.,
when $w=0$). Then the right hand side of \eqref{identity1} is equal to
$0$, but the left hand side may well be non-zero.

Thus, from now on we will assume that $F$ and $\omega$ are {\em
analytic}. For the function $F$ this means that for each point $x \in
\pone$ there is a small neighborhood of $x$ in which $F$ is equal to
its Taylor series expansion in $z_x,\ol{z}_x$ (where $z_x$ is a
holomorphic coordinate at $x$). In other words, this means that the
real and imaginary components of the function $F$ are real-analytic
(this does not mean that $F$ is holomorphic!). For the two-form
$\omega$ this means that the functions $\omega_{w\wb}$ and
$\omega_{z\ol{z}}$ are analytic on $\C_\infty$ and $\C_0$,
respectively.

\medskip

{\em We conjecture that the identity \eqref{identity1} holds whenever
$F$ and $\omega$ are analytic on $\pone$, in the sense that the right
hand side is a $q,\ol{q}$-series expansion of the left hand side,
which converges in the punctured disc $0<|q|<\delta$ for some
$\delta>0$ depending on $F$ and $\omega$.}

\medskip

We also expect that the left hand side is an analytic function in
$q,\ol{q}$ on $\C^\times$.

In what follows we present some evidence for this conjecture. We start
by proving it for a large class of analytic functions and differential
forms which may be represented as follows:
$$
\omega = \sum_{{\al}}
\frac{{\varpi}_{\al}}{z\zb+R_{\al}} dz  d{\zb}, \qquad F =
\sum_{{\beta}} \frac{f_{\beta}}{z\zb+Q_{\beta}},
$$
where $R_\al, Q_\beta \in \C \bs \R_{\leq 0}$, and the condition
\begin{equation}    \label{zero sum}
\sum_{\al} {\varpi}_{\al} = 0
\end{equation}
is satisfied to ensure that $\omega$ is well-defined at $z=\infty$.

Let us compute the left hand side of the identity
\eqref{identity1}. To this end we need to compute the integral of the
form
$$
\int_{\pone} \frac{1}{z\zb+R_{\al}}
\frac{1}{|q|^2 z\zb+Q_{\beta}} dz  d\zb.
$$
By making a charge of variables $z=\sqrt{x} e^{i \theta}$, we rewrite
this as
$$
2 \pi \int_0^\infty \frac{1}{(x+R_\al)(|q|^2 x+Q_\beta)} dx.
$$
Next, we represent it as a contour integral of
$$
- i \log(-x) \frac{1}{(x+R_\al)(|q|^2 x+Q_\beta)} dx,
$$
over the contour that goes along the real axis from $+\infty$ to $\ep$,
then goes counter-clockwise around $0$ and then along the real axis
from $\ep$ to $\infty$, in the limit when $\ep \to 0$. The latter is
evaluated as the sum of residues of the integrand in the complex
plane, and we obtain the following answer:
\begin{equation}    \label{answer}
\int_{\pone} \omega F(qz,\ol{q}\zb) = - 2\pi \sum_{\al,\beta}
\varpi_\al f_\beta \frac{\log(|q|^2 R_\al/Q_\beta)}{Q_\beta - |q|^2
R_\al}.
\end{equation}

Now we compute the right hand side. Using formula \eqref{Ral}, we find
that
$$
\underset{|z|<\ep^{-1}}\pf \omega_{z\zb} \ z^{m} {\zb}^{\ol{m}} dz
  d\zb = 2\pi (-1)^{m+1} \delta_{m,\ol{m}} \sum_\al \varpi_\al
R_\al^m \log R_\al.
$$
Similarly, we obtain that
$$
\underset{|w|<\ep^{-1}}\pf F \ w^{m} {\wb}^{\ol{m}} dw
d\wb = 2\pi (-1)^{m+1} \delta_{m,\ol{m}} \sum_\beta f_\beta
Q_\beta^{-m-2} \log Q_\beta.
$$
On the other hand, we have
\begin{align*}
\left. \frac{1}{m!{\ol{m}}!} {\p}_w^{m} {\p}_{\wb}^{\ol{m}}
\omega_{w\wb} \right|_{w=0} &= (-1)^{m+1} \delta_{m,\ol{m}} \sum_\al
\varpi_\al R_\al^{m+1}, \\ \left. \frac{1}{m! {\ol{m}}!} {\p}_z^{m}
{\p}_{\zb}^{\ol{m}} F \right|_{z=0} &= (-1)^m \delta_{m,\ol{m}}
\sum_\beta f_\beta Q_\beta^{-m-1}.
\end{align*}
Substituting these expressions into the right hand side of formula
\eqref{identity1} and using the condition \eqref{zero sum}, we obtain
the following series
$$
- 2\pi \sum_{\al,\beta} \varpi_\al f_\beta \sum_{m=0}^\infty
Q_\beta^{-m-1} R_\al^m |q|^{2m} (\log R_\al - \log Q_\beta + \log
|q|^2).
$$
But this is precisely the $|q|$-series expansion of \eqref{answer},
which converges to \eqref{answer} for all non-zero $q$ inside the disc
of radius $\underset{\al,\beta}{\on{min}} \{ |Q_\beta|/|R_\al|
\}$. Note that this is true if and only if the condition \eqref{zero
sum} holds which is needed to make $\omega$ well-defined at
$z=\infty$. So it appears that the identity \eqref{identity1} somehow
``knows'' about this condition.

Thus, we discover an interesting phenomenon: the factorization
identity \eqref{identity1} should be understood in the analytic
continuation sense. Namely, the right hand side is the expansion of
the left hand side in powers of $q,\ol{q}$ (note, however, that it
also includes terms with $\log|q|^2$), which converges in the domain
$0 < |q| < \delta$ for some $\delta$ which depends on the choice of
$F$ and $\omega$. Moreover, we expect that the left hand side of
\eqref{identity1} is an analytic function in $q$ for all $q \in
\C^\times$.

In the same way as above we prove the identity \eqref{identity1} in
the more general case of $F$ and $\omega$ of the form
\begin{align}
\omega_{n,\ol{n}} &= \sum_{{\al}} {\varpi}_{\al}\frac{z^{n}
{\zb}^{\nb}}{z{\zb} + R_{\al}} dz  d{\zb}, \\
\label{basis} f_{m, {\ol{m}}} &= \sum_{{\beta}} f_{\beta} \frac{z^{m}
{\zb}^{\ol{m}}}{z{\zb} + Q_{\beta}},
\end{align}
where the $R_\al$'s and $Q_\beta$'s are in $\C \bs \R_{\leq 0}$. The
numbers $\varpi_\al$ and $R_\al$ (resp., $f_\beta$ and $Q_\beta$)
should also satisfy some conditions which ensure that $\omega$ (resp.,
$F$) is well-defined at $z=\infty$. In the same way as above we obtain
that the left hand side of \eqref{identity1} is given by the formula
\begin{equation*}
2\pi {\delta}_{n + m, {\nb} + {\ol{m}}} q^{m} \ol{q}^{\ol{m}}
(-1)^{n+m+1}\sum_{{\al}, {\beta}} {\varpi}_{\al} f_{\beta}
\frac{R_{\al}^{n+m} {\rm log} \left( R_{\al} \right) - \left(
Q_{\beta} |q|^{-2} \right)^{n+m} {\rm log} \left( Q_{\beta} |q|^{-2}
\right) }{Q_{\beta}- |q|^{2} R_{\al}},
\end{equation*}
and the right hand side of \eqref{identity1} is equal to the
$q,\ol{q}$-series expansion of the left hand side.

Thus, we have now checked the validity of the identity
\eqref{identity1} for a large class of functions and two-forms.

It is important to realize that we could reverse our
calculation. Namely, after computing the integral $\int_{\pone} \omega
F(qz,\ol{q}\zb)$ we could expand it in a power series in $q,\ol{q}$
and interpret the result as the formula for the factorization of the
two-point correlation function of $\omega$ and $F$ over intermediate
states, as in equation \eqref{factoriz}. This determines completely
the matrix elements of $F$ between the vacuum $\vac_{\C_0}$ and
generic ``out'' states, and matrix elements of $\omega$ between the
covacuum $_{\C_\infty}\covac$ and generic ``in'' states. Thus, we
could start with the two-point, and more generally, $n$-point
correlation functions of evaluation observables, which are given by
explicit integrals over moduli spaces of instantons, and use them to
{\em derive} the matrix elements of these observables acting on the
space of states. In particular, this way we find that the Hamiltonian
of our model is non-diagonalizable. Moreover, we can estimate the
maximal size of the Jordan blocks (the maximal power of the logarithm
of $q$ plus $1$). This points to an effective strategy for determining
matrix elements from the correlation functions, which can be applied
to more general models, such as two-dimensional sigma models and
four-dimensional Yang-Mills theory that we will discuss in Part
II. Again, we stress that in order to obtain non-trivial results we
must consider {\em non-BPS observables}.

To conclude this section, we point out another case when the identity
\eqref{identity1} obviously holds. Namely, suppose that $F$ is
analytic, but $\omega_{z,\zb}$ has compact support on the complex
plane $\C_0 = \pone \bs \infty$, i.e., away from the point $z=\infty$
(thus, $\omega$ is not analytic, and so this is not a special case of
our conjecture). Then there exist positive numbers $R$ and $r$ such
that $\omega_{z\zb} \equiv 0$ for all $z$ such that $|z|>R$, while $F$
is equal to its Taylor series expansion in the disc $|z| < r$. In this
case for all $0 < |q| < r/R$ we have
\begin{multline*}
\int_{\pone} \omega F(qz,\ol{q}\zb) = \int_{|z|<R} F(qz,\ol{q}\zb)
\omega_{z\zb} dz  d\zb \\ = \sum_{m,\ol{m}=0}^\infty
\left. \frac{q^m \ol{q}^{\ol{m}}}{m! {\ol{m}}!} {\p}_z^{m}
{\p}_{\zb}^{\ol{m}} F \right|_{z=0} \cdot \int_{\pone}
\omega_{z\ol{z}} z^m \zb^{\ol{m}} dz  d\zb.
\end{multline*}
Thus, the left hand side of \eqref{identity1} is equal to the
$q,\ol{q}$-series expansion of the first term in the right hand side,
while the second and the third terms vanish. Therefore the identity
\eqref{identity1} holds in this case.

In the same way we prove the identity with the roles of $F$ and
$\omega$ reversed, i.e., assuming that $F$ has compact support away
from $z=0$ while $\omega$ is analytic at $z=\infty$.

\medskip

\noindent{\em Remark.}  It is interesting to investigate what will
happen if we allow smooth but non-analytic observables in the spectral
decomposition of the correlation functions. For example, consider the
correlation functions of the evaluation observables whose support does
not contain the critical points of the Morse function.

Let us pass to the coordinates in which the gradient vector field
looks like a translation in one of them, say,
$$
v = {\partial}_{t}.
$$
For simplicity, suppose that the observables are independent of the
remaining coordinates. Then the correlation function reduces to a
one-dimensional integral over the $t$-line. For example, if we have
two observables, giving rise to a one-form $\omega = {\omega}(t) dt$
and a function $f(t)$ on the $t$-line, then the correlation function
will look like this:
$$
{\mathcal C}(q)=
\int_{-\infty}^{+\infty} {\omega} (t) f (t + {\rm log}( q{\ol{q}}))
dt.
$$
This integral converges because we have assumed that $f$ and $\omega$
have compact support on the complement to the set of critical points
(which in the model example consists of the points $t=\pm\infty$).
Clearly, in this case the decomposition of ${\mathcal C}(q)$ as a sum
of the contributions of the eigenstates of the Hamiltonian looks as
the integral
$$
{\mathcal C}(q) = \int \frac{dk}{2\pi} e^{ i k {\rm log}(q{\ol{q}})}
{\widehat{\mathcal C}}_{f, \omega}(k),
\label{spectral}
$$
with
$$
{\widehat{\mathcal C}}_{f, \omega}(k) = {\widehat f}(-k)
{\widehat\omega}(k)
$$
being the product of the Fourier transforms of $f$ and $\omega$.
Formula \Ref{spectral} implies that the spectrum of the Hamiltonian
contains a continuous part, with the eigenvalues given by
$$
E_{k} = i k,
$$
i.e., purely imaginary!

Thus, we are facing a dilemma: either these compactly supported
functions and forms require a new, infinite-dimensional, sector in the
space of states, or, by some sort of resummation, they are already
included in the space of states that we have constructed.

It is instructive to reconsider from this point of view the example of
the harmonic oscillator, i.e., the quantum mechanical model on
$\mathbb C$, with the quadratic Morse function. We have analyzed in
detail the $\la \to \infty$ limit of the full set of the eigenstates
of the Hamiltonian in \secref{flat space}, and we did not see any need
for the continuous imaginary spectrum. How could it be that the
functions with compact support not containing zero are included in the
space of states built from polynomials? Physically, the explanation is
the following. For $\la = \infty$ the evolution looks like a constant
velocity motion in the logarithmic coordinate $t$. However, once $\la$
becomes finite, there is an admixture of diffusion, caused by the term
$\frac{1}{2\la} {\Delta}$ in the Hamiltonian $\wt{H}_\la$. What is more
important, this diffusion takes place in the linear, as opposed to the
logarithmic, coordinates. Roughly speaking, the evolution during some
time $T$ spreads the initially localized object as $\sim
\sqrt{\frac{T}{\la}} \gg e^{-T}$, for large $T$. Thus, even if we
start with a distribution with compact support not containing zero,
the critical point, the diffusion will spread it so it will contain
zero.  Once this has happened, the resulting distribution can be
well-approximated by the Taylor series at zero, i.e., by the wave
functions from our space of states.

Another point worth mentioning is that the wave functions of the
Hamiltonian ${\mc L}_v$ corresponding to the imaginary eigenvalues are
not smooth at the critical points. For instance, in the case of the
$\pone$ model, where $v$ is the Euler vector field, these
eigenfunctions have the form $|z|^{ik}$. When we deform away from the
point $\la=\infty$, we add the term $\la^{-1} \Delta$ to the
Hamiltonian. If these eigenfunctions were present in the spectrum,
then we would be able to deform them to eigenfunctions of the deformed
Hamiltonian. But applying $\Delta$ to $|z|^{ik}$, we obtain a
function which has poles at the critical points, and this shows that
it cannot be deformed to a smooth eigenfunction of the Hamiltonian in
perturbation theory in $\la^{-1}$.

Let us mention, however, that in Part II of this article, when we
discuss the quantum mechanical models on non-simply connected
manifolds, we shall see some version of the "imaginary" space of
states. Its appearance (in a much more tame form, with discrete
spectrum) will be related to the existence of gradient trajectories
which go from "nowhere to nowhere", i.e., never terminate. But this is
only possible for Morse-Novikov, i.e. multivalued, functions.

\ssec{Interpretation as an expansion of the delta-form on the
   $q$-shifted diagonal}    \label{q-shifted}

The identity \eqref{identity1} expresses the factorization over
intermediate states of the two-point correlation function of
evaluation observables, one of which is a function and the other is a
two-form. But we could consider instead the correlation functions of
two one-forms, or to switch $F$ and $\omega$ (so that $\C^\times$ acts
on the two-form rather than the function). In each case we obtain a
similar identity.

It is instructive to think of all of these identities as expressing
the delta-form supported on the ``$q$-shifted diagonal'' in $\pone
\times \pone$ in terms of distributions along the first and the second
factors. More precisely, consider the submanifold
$$
\on{Diag}_q = \{ (x,y) \in \pone \times \pone \, | \, y = qx \}
\subset \pone \times \pone
$$
Note that $qx$ simply means the point obtained by acting on $x \in
\pone$ with $q \in \C^\times$. This is the $q$-shifted diagonal. Now
let $\Delta_q$ be the delta-form (of degree two) supported on
$\on{Diag}_q$. Note that $\Delta_q$ is precisely the kernel of the
evolution operator in our quantum mechanical model on $\pone$.

Observe that
\begin{equation}    \label{0,2}
\underset{\pone \times \pone}\int \Delta_q \wedge (\omega \otimes F) =
\underset{\pone}\int \omega \phi(q)^*(F),
\end{equation}
where $(\phi(q)^*(F))(z,\zb) = F(qz,\ol{q}\zb)$, is precisely the
two-point function appearing in the left hand side of the identity
\eqref{identity1}. Likewise, if we take two one-forms $\eta_1$ and
$\eta_2$ on $\pone$, then we have
\begin{equation}    \label{1,1}
\underset{\pone \times \pone}\int \Delta_q \wedge (\eta_1 \otimes
\eta_2) = \underset{\pone}\int \eta_1 \wedge \phi(q)^*(\eta_2),
\end{equation}
and similarly for a function and a two-form switched:
\begin{equation}    \label{2,0}
\underset{\pone \times \pone}\int \Delta_q \wedge (F \otimes \omega) =
\underset{\pone}\int F \phi(q)^*(\omega).
\end{equation}
In these formulas, given differential forms $\omega_1$ and $\omega_2$
on $\pone$, we denote by $\omega_1 \otimes \omega_2$ the corresponding
differential form on $\pone \times \pone$.

The identities discussed above correspond to an expansion of different
components of the distribution $\Delta_q$. More precisely, $\Delta_q$
is the sum of three components which are differential forms of degrees
$(0,2)$, $(1,1)$, and $(2,0)$ on $\pone \times \pone$. The expansion
of each of them gives rise to the three identities considered
above.

For example, consider the $(0,2)$ part of $\Delta_q$ which we will
denote by $\Delta_q^{(0,2)}$. This is the part which contributes to
the integral \eqref{0,2}. Let $_z\wt\varphi_{n,\ol{n}}$ and
$_w\wt\varphi_{n,\ol{n}}$ be the distributions on $\pone$ defined by
the formulas
\begin{align*}
_z\wt\varphi^\ep_{m,\ol{m}}(\omega) &= \underset{|z|<\ep^{-1}}\pf
\omega_{z\zb} \ z^{m} {\zb}^{\ol{m}} dz  d\zb, \qquad \omega \in
\Omega^2(\pone), \\ _w\wt\varphi^\ep_{m,\ol{m}}(F) &=
\underset{|w|<\ep^{-1}}\pf F \ w^{m} {\wb}^{\ol{m}} dw  d\wb,
\qquad F \in \Omega^0(\pone).
\end{align*}
Given two distributions $\varphi,\phi$ on $\pone$, we will denote by
$\varphi \otimes \phi$ the corresponding distribution on $\pone \times
\pone$.

Then the identity \eqref{identity1} may be rewritten as follows:
\begin{align}    \notag
\Delta_q^{(0,2)} &= \sum_{m,\ol{m}=0}^\infty q^{m} \ol{q}^{\ol{m}} \;
_z\wt\varphi^\ep_{m,\ol{m}} \otimes \frac{1}{m!\ol{m}!} (-\pa_z)^m
(-\pa_{\ol{z}})^{\ol{m}} \delta^{(2)}(z,\ol{z}) dz  d\zb \\
\label{Deltaq} &+ q \ol{q} \sum_{m,\ol{m}=0}^\infty q^{m}
\ol{q}^{\ol{m}} \; \frac{1}{m!\ol{m}!} (-\pa_w)^m
(-\pa_{\ol{w}})^{\ol{m}} \delta^{(2)}(w,\ol{w}) \otimes {}
_w\wt\varphi^\ep_{m,\ol{m}} \\ \notag &- 2 \pi \log(q\ol{q})
\sum_{m,\ol{m}=1}^\infty q^{m} \ol{q}^{\ol{m}}
\frac{\pa_w^{m-1} \pa_{\wb}^{\ol{m}-1}}{(m-1)!(\ol{m}-1)!}
\delta^{(2)}(w,\ol{w}) \otimes \frac{\pa_z^m
\pa_{\ol{z}}^{\ol{m}}}{m!\ol{m}!} \delta^{(2)}(z,\ol{z}) dz  d\zb.
\end{align}
One needs to be careful in interpreting this identity. Since we expect
the identity \eqref{identity1} to be true only for analytic functions,
it is natural to consider \eqref{Deltaq} as an identity in the space
of {\em hyperfunctions} on $\pone \times \pone$.\footnote{we thank
P. Schapira and K. Vilonen for discussions of our identity in the
context of hyperfunctions} We recall that the space of hyperfunctions
is the dual space to the space of analytic functions, equipped with an
appropriate topology. We will use the term ``hyperfunction'' in the
broader sense as an element of the dual space to the space of analytic
differential forms. The right hand side of \eqref{Deltaq} should be
understood as a $q,\ol{q}$-expansion of the left hand side, as in the
case of the identity \eqref{identity1}, which converges for
$0<|q|<\delta$ for some real $\delta>0$. It is clear that applying
\eqref{identity1} to the differential form $\omega \otimes F$ on
$\pone \times \pone$ we obtain \eqref{identity1}.

Likewise, there are similar identities on the components of degrees
$(1,1)$ and $(2,0)$ in $\Delta_q$. Those contribute to the integrals
\eqref{1,1} and \eqref{2,0}, respectively, and give rise to the
corresponding identities.

One may wonder whether there is a simplified analogue of the identity
\eqref{Deltaq} in the case of the target manifold $X=\C$. There is
indeed such an analogue, which is however more restrictive due to the
non-compactness of $\C$. Nevertheless, it is instructive to look at
this identity.

The analogue of the $q$-shifted diagonal in this case is just the line
$y=xq$ in $\C \times \C$, and the corresponding delta-form is just
$$
\Delta_q = \delta^{(2)}(y-xq,\ol{y}-\ol{q}\ol{x}) d(y-xq) \wedge
d(\ol{y}-\ol{q}\ol{x}),
$$
which is the kernel of the evolution operator of our model that we
discussed in \secref{evolution}. It appears as the $\la \to \infty$
limit of the kernel $K_{t,\ol{t}}$ of the model at finite $\la$ (see
formula \eqref{tends to}). Let us look at its $(0,2)$ component, which
reads
$$
\delta^{(2)}(y-xq,\ol{y}-\ol{q}\ol{x}) dy \wedge d\ol{y}.
$$
The naive Taylor series expansion of this distribution looks as
follows:
\begin{equation}    \label{case of C}
\sum_{m,\ol{m}=0}^\infty q^{m} \ol{q}^{\ol{m}} \; x^m \ol{x}^{\ol{m}}
\otimes  \frac{1}{m!\ol{m}!} (-\pa_y)^m
(-\pa_{\ol{y}})^{\ol{m}} \delta^{(2)}(y,\ol{y}) dy \wedge d\ol{y}.
\end{equation}

This formula may be interpreted in the following way. Let $\omega$
be a two-form on $\C$ with compact support and $F$ a function that is
analytic at $0 \in \C$. Then for sufficiently small $q$ the series
$$
\sum_{m,\ol{m}=0}^\infty q^{m} \ol{q}^{\ol{m}} \int x^m \ol{x}^{\ol{m}}
\omega \cdot \left. \frac{1}{m! \ol{m}!} \pa_y^m \pa_{\ol{y}}^{\ol{m}}
F \right|_{y=0},
$$
obtained by applying \eqref{case of C} to $\omega \otimes F$,
converges to the integral
$$
\underset{\C \times \C}\int \Delta_q \wedge (\omega \otimes F) =
\underset{\C}\int \omega \phi(q)^*(F).
$$

The $(1,1)$ and $(2,0)$ parts of the decomposition of $\Delta_q$ have
a similar structure and interpretation. The decomposition of $\Delta_q$
obtained this way may be viewed as the $\la \to \infty$ limit of the
decomposition of the kernel of the evolution operator $K_{t,\ol{t}}$
in terms of the orthonormal basis of eigenstates of the Hamiltonian,
which is given in formula \eqref{KTT}. However, because we do not have
the structure of Hilbert space on our space of states at $\la=\infty$,
this decomposition becomes more subtle: it has to be understood in the
sense of analytic continuation and the observables are required to
have some analytic properties.

Thus, we see that an analogue of the identity \eqref{identity1} in the
case of $X=\C$ may be obtained simply by applying a Taylor series
expansion to the delta-form supported on the $q$-shifted
diagonal. While it seems very easy to derive it, its validity is very
limited because $\C$ is not compact. Indeed, for the integral to
converge we need to make an additional assumption of compactness of
support of at least one of the objects involved, $F$ and $\omega$. On
the other hand, in order to make sense of this identity we need them
to be analytic. Thus, we have a clash between two seemingly
irreconcilable properties of differential forms on $\C$: analyticity
and compactness of support. The best we can do is to assume that $F$
is analytic and $\omega$ has compact support. The resulting identity
is not very useful, but it is still instructive to consider it as a
toy model for the corresponding identity in the case of $\pone$.

In the case of $X=\pone$ the structure of the identity is more
complicated, but it is applicable to a larger class of differential
forms. Now instead of one infinite sum we have three infinite
sums. The first two have the structure similar to that of the sum
appearing in the identity for $\C$. They correspond to the two
critical points of the Morse function on $\pone$ (or equivalently, the
fixed points of the $\C^\times$-action): $0$ and $\infty$. The third
term has to do with the non-diagonalizable nature of the
Hamiltonian. It would be interesting to understand its meaning from
the analytic point of view. On the positive side, all differential
forms on $\pone$ have compact support, so the convergence of the left
hand side of \eqref{identity1} is not an issue. Hence it makes sense
to impose the condition of analyticity on both $F$ and $\omega$.

In order to understand the generalization of these identities to other
K\"ahler manifolds, it is more convenient to work with the other
version \eqref{identity2} of our identity. This version also has an
interpretation in terms of the decomposition of the delta-form on the
$q$-shifted diagonal:
\begin{align}    \notag
\Delta_q^{(0,2)} &= \sum_{m,\ol{m}=0}^\infty q^{m} \ol{q}^{\ol{m}} \;
_z\wt\varphi^\ep_{m,\ol{m}} \otimes \frac{1}{m!\ol{m}!} (-\pa_z)^m
(-\pa_{\ol{z}})^{\ol{m}} \delta^{(2)}(z,\ol{z}) dz  d\zb \\
\label{Deltaq2} &+ q \ol{q} \sum_{m,\ol{m}=0}^\infty q^{m}
\ol{q}^{\ol{m}} \; \frac{1}{m!\ol{m}!} (-\pa_w)^m
(-\pa_{\ol{w}})^{\ol{m}} \delta^{(2)}(w,\ol{w}) \otimes {}
_w\wt\varphi^{q\ep}_{m,\ol{m}}.
\end{align}
The terms with $\log|q|^2$ have now disappeared at the cost of
changing the regularization in the second term: we now use
$_w\wt\varphi^{q\ep}_{m,\ol{m}}$ instead of
$_w\wt\varphi^{\ep}_{m,\ol{m}}$. The resulting identity has two terms,
corresponding to the fixed points $0$ and $\infty$ of the
$\C^\times$-action on $\pone$. There are also similar identities for
the $(1,1)$ and $(2,0)$ components of $\Delta_q$.

\ssec{Generalization to other K\"ahler manifolds}    \label{gen kahler}

We now briefly discuss how to generalize the results of the previous
section to more general K\"ahler manifolds, using ${\mathbb C}{\mathbb
P}^2$ as the main example. Points of $\ptwo$ will be represented by
triples $(z_1:z_2:z_3)$ of non-zero complex numbers, up to an overall
scalar multiple. Consider the $\C^\times$-action on $\ptwo$ generated
by the vector field $v=\xi+\ol\xi$, where
$$
\xi = z_1 \pa_{z_1} + \ga z_2 \pa_{z_2},
$$
where $\ga$ is a rational number such that $0 < \ga < 1$. If $\ga$
does not satisfy these inequalities, then the structure of the
ascending manifolds described below will be different. Rationality of
$\ga$ is needed to ensure that $\xi$ comes from a $\C^\times$-action.

The vector field $v$ is the gradient of the Morse function
$$
f = \frac{1}{2} \frac{2|z_1|^2 + (1+\ga) |z_2|^2}{|z_1|^2 + |z_2|^2 +
|z_3|^2}.
$$
Its critical points are $(0:0:1)$ of index $0$, $(0:1:0)$ of index $2$
and $(1:0:0)$ of index $4$. The corresponding ascending manifolds
$X_\al$ are $X_{(0:0:1)} = \{ (z_1:z_2:1) \} \simeq \C^2$,
$X_{(0:1:0)} = \{ (w_1:1:0) \} \simeq \C$, and $X_{(1:0:0)}$,
which is a point.

The coordinates $z_1,z_2$ are local coordinates at $(0:0:1)$
representing points $(z_1:z_2:1)$. Let us introduce local coordinates
$w_1,w_2$ at $(0:1:0)$ by representing nearby points as
$(w_1:1:w_2)$. Then we have
\begin{equation}    \label{from z to w}
w_1 = \frac{z_1}{z_2}, \qquad w_2 = \frac{1}{z_2}.
\end{equation}
We also choose local coordinates $u_1,u_2$ at $(1:0:0)$ by
representing nearby points as $(1:u_1:u_2)$, so that
$$
u_1 = \frac{z_2}{z_1}, \qquad u_2 = \frac{1}{z_1}.
$$

The space $\CH^{\inn}$ of ``in'' states is isomorphic to
$$
\CH^{\inn} \simeq \CH^{\inn}_{(0:0:1)} \oplus \CH^{\inn}_{(0:1:0)}
\oplus \CH^{\inn}_{(1:0:0)},
$$
where
\begin{align*}
\CH^{\inn}_{(0:0:1)} &= \C[z_1,z_2,\zb_1,\zb_2] \otimes
\Lambda[dz_1,dz_2,d\zb_1,d\zb_2], \\ \CH^{\inn}_{(0:1:0)} &=
\C[w_1,\pa_{w_2},\ol{w}_1,\pa_{\ol{w}_2}] \otimes
\Lambda[dw_1,\imath_{\pa_{w_2}},d\ol{w}_1,\imath_{\pa_{\ol{w}_2}}]
\cdot \delta^{(2)}(w_2,\ol{w}_2) d^2w_2, \\ \CH^{\inn}_{(1:0:0)} &=
\C[\pa_{u_1},\pa_{u_2},\pa_{\ol{u}_1},\pa_{\ol{u}_2}] \otimes
\Lambda[\imath_{\pa_{u_1}},\imath_{\pa_{u_2}},\imath_{\pa_{\ol{u}_1}},
\imath_{\pa_{\ol{u}_2}}] \cdot \delta^{(4)}(u_1,u_2,\ol{u}_1,\ol{u}_2)
d^2 u_1 \wedge d^2 u_2.
\end{align*}
The ground states are
$$
\Delta_{(0:0:1)} = 1, \qquad \De_{(0:1:0)} =
\delta^{(2)}(w_2,\ol{w}_2) d^2 w_2,
$$
$$
\De_{(1:0:0)} = \delta^{(4)}(u_1,u_2,\ol{u}_1,\ol{u}_2)
d^2 u_1 \wedge d^2 u_2.
$$
Note that each space exhibits holomorphic factorization
$\CH^{\inn}_\al = \CF^{\inn}_\al \otimes \ol\CF_\al$.

We realize $\CH^{\inn}$ as a subspace in the space of distributions on
differential forms on $\ptwo$. The subspace $\CH^{\inn}_{(1:0:0)}$ is
a canonical subspace, which consists of distributions supported at the
point $(1:0:0)$. The subspaces $\CH^{\inn}_{(0:1:0)}$ and
$\CH^{\inn}_{(0:0:1)}$ are not canonical. The definition of elements
of these subspaces depends on a particular choice of regularization of
divergent integrals, as in the case of $\pone$ which we have studied
in great detail earlier. Changing regularization would add to elements
of $\CH^{\inn}_{(0:1:0)}$ correction terms that lie in
$\CH^{\inn}_{(1:0:0)}$, and to elements of $\CH^{\inn}_{(0:0:1)}$
correction terms in $\CH^{\inn}_{(0:1:0)}$ and
$\CH^{\inn}_{(1:0:0)}$. Because of that, we only have a canonical
filtration with associated graded spaces being $\CH^{\inn}_{(1:0:0)}$,
$\CH^{\inn}_{(0:1:0)}$ and $\CH^{\inn}_{(0:0:1)}$.

This is reflected in the non-diagonalizability of the operators ${\mc
L}_\xi$ and ${\mc L}_{\ol{\xi}}$. According to our general formulas
\eqref{Lxi} and \eqref{Lolxi}, we have
\begin{align*}
{\mc L}_\xi &= {\mc L}_{\xi,\on{naive}} + 2 \pi ( - \ga \delta_{12}
\otimes \ol\delta_{12} + (\ga-1) \delta_{23} \otimes \ol\delta_{23}),
\\ {\mc L}_{\ol\xi} &= {\mc L}_{\ol\xi,\on{naive}} + 2 \pi ( - \ga
\delta_{12} \otimes \ol\delta_{12} + (\ga-1) \delta_{23} \otimes
\ol\delta_{23}).
\end{align*}
where $\delta_{12}: \CH^{\inn}_{(0:0:1)} \to \CH^{\inn}_{(0:1:0)}$ and
$\delta_{23}: \CH^{\inn}_{(0:1:0)} \to \CH^{\inn}_{(1:0:0)}$ are the
GC operators, and $\ol{\delta}_{12}$ and $\ol\delta_{23}$ are their
complex conjugates. Note that (in the notation of \secref{action of
Q}) we have $w_{12} = w_2$, $w_{23} = u_1$ and so $a_{12} = - \ga$,
$a_{23} = \ga - 1$.

As an example, we describe the action of $\delta_{12}$ on the subspace
$\C[z_1,z_2]$ of zero-forms in $\CF^{\inn}_{(0:0:1)}$. Given an
element of $\C[z_1,z_2]$, we rewrite it as a polynomial in
$w_1,w_2^{\pm 1}$, using the substitution \eqref{from z to w}. Then we
project it onto the quotient $\C[w_1,w_2^{\pm 1}]/\C[w_1,w_2^{-1}]$,
which we identify with the space $\C[w_1,\pa_{w_2}]$
by the formula
$$
w_1^n w_2^{-m} \mapsto w_1^n \frac{1}{(m-1)!}(-\pa_{w_2})^{m-1}.
$$
One defines the action of $\delta_{12}$ on differential forms of
degree greater than $0$ in the same way. The operator $\delta_{23}$ is
defined similarly.

What is the maximal length of the Jordan blocks of the operators ${\mc
L}_\xi$, ${\mc L}_{\ol\xi}$ and the Hamiltonian? In the case of
$\pone$ the space $\CH^{\inn}$ was an extension of two subspaces,
$\CH^{\inn}_{\C_0}$ and $\CH^{\inn}_\infty$, and the off-diagonal
parts of these operators were acting from one of them to the
other. Therefore the Jordan blocks had maximal length $2$. Now the
space $\CH^{\inn}$ is an extension of three spaces and the
off-diagonal parts of these operators act from the first to the second
and from the second to the third. Therefore {\em a priori} one could
expect Jordan blocks of length $3$. It comes as a bit of a surprise
when we learn that in fact the maximal Jordan blocks have length $2$.

There are two ways to see that. The first is to find the spectra of
the diagonal parts of ${\mc L}_\xi$ and ${\mc L}_{\ol\xi}$ on each of
the three subspaces of $\CH^{\inn}$. According to the above
description of $\CH^{\inn}_{(0:0:1)}$, the eigenvalues of ${\mc
L}_\xi$ on it (which are the same as the eigenvalues of ${\mc
L}_{\xi,\on{naive}}$) have the form $n_1 + \ga n_2$, where $n_1,n_2
\in \Z_{\geq 0}$. We also find that the eigenvalues on
$\CH^{\inn}_{(0:1:0)}$ have the form $m_1(1-\ga) + (m_2+1) \ga$, where
$m_1,\m_2 \in \Z_{\geq 0}$, and the eigenvalues on
$\CH^{\inn}_{(1:0:0)}$ have the form $(l_1+1)(1-\ga) + (l_2+1)$, where
$l_1,l_2 \in \Z_{\geq 0}$. We have similar formulas for the
eigenvalues of ${\mc L}_{\ol\xi}$. By inspection of these formulas we
find that for irrational values of $\ga$ there is an overlap of the
spectra between $\CH^{\inn}_{(0:0:1)}$ and $\CH^{\inn}_{(0:1:0)}$, and
between $\CH^{\inn}_{(0:1:0)}$ and $\CH^{\inn}_{(1:0:0)}$, but none
between $\CH^{\inn}_{(0:0:1)}$ and $\CH^{\inn}_{(1:0:0)}$. Even though
we are only allowed to take rational values of $\ga$, we expect the
operators ${\mc L}_\xi$ and ${\mc L}_{\ol\xi}$ to depend continuously
on $\ga$. Hence we find that they cannot have off-diagonal terms
acting from $\CH^{\inn}_{(0:0:1)}$ to $\CH^{\inn}_{(1:0:0)}$, and so
the maximal size of their Jordan blocks cannot be greater than $2$.

Another way to see that is to use the identity \eqref{square zero},
which in our case reads
\begin{equation}    \label{comp zero}
\delta_{23} \circ \delta_{12} = 0, \qquad \ol\delta_{23} \circ
\ol\delta_{12} = 0.
\end{equation}
Therefore
$$
(\delta_{12} \otimes \ol\delta_{12}) \circ (\delta_{23} \otimes
\ol\delta_{23}) = 0.
$$
It implies that the square of the off-diagonal parts of ${\mc
L}_\xi$ and ${\mc L}_{\ol\xi}$ (which are equal to each other) is
zero. This means that the Jordan blocks have length at most $2$.

This should be contrasted to the case of the model defined on the
manifold $\pone \times \pone$, with the vector field $\xi = z_1
\pa_{z_1} + z_2 \pa_{z_2}$. In that case we have four ascending
manifolds: one two-dimensional $X_1$, two one-dimensional, $X_2$ and
$X_3$, and one zero-dimensional, $X_4$. The off-diagonal part of ${\mc
L}_\xi$ now has four terms:
\begin{equation}    \label{sum of deltas}
\delta_{12} \otimes \ol{\delta}_{12} + \delta_{13} \otimes
\ol{\delta}_{13} + \delta_{24} \otimes \ol{\delta}_{24} + \delta_{34}
\otimes \ol{\delta}_{34}.
\end{equation}
The analogue of the identity \eqref{comp zero} reads as follows:
$$
\delta_{24} \circ \delta_{12} = \delta_{34} \circ \delta_{13}, \qquad
\ol\delta_{24} \circ \ol\delta_{12} = \ol\delta_{34} \circ
\ol\delta_{13}.
$$
However, it does not mean that the square of the sum \eqref{sum of
deltas} is equal to zero. It would be zero only if we would change one
of the four signs to minus (this is a good example of how the signs
$\ep_{\al\beta}$ discussed in \secref{action of Q} are
chosen).\footnote{Note also that the squares of the supersymmetry
charges $\pa$ and $\ol\pa$ are equal to zero, but this is because of
the presence of the differentials $dw_{\al\beta}$ and
$d\wb_{\al\beta}$ in formulas \eqref{pa} and \eqref{olpa}.} Thus,
because we now have two ``channels'' from the two-dimensional stratum
to the zero-dimensional stratum (via two intermediate one-dimensional
strata) we find that the square of the off-diagonal term of ${\mc
L}_{\xi}$ (and, likewise, ${\mc L}_{\ol\xi}$) is non-zero. Therefore
there are Jordan blocks of length three in the case of $\pone \times
\pone$.

For $\ptwo$ we can go from the two-dimensional stratum to the
zero-dimensional stratum in only one way, because there is only one
one-dimensional stratum. Therefore the identity \eqref{comp zero} on
the GC operators now implies that the squares of the off-diagonal
terms of the chiral and anti-chiral components of the Hamiltonian are
equal to zero. This analysis leads us to a non-trivial prediction: the
two-point correlation functions of evaluation observables of the
$\ptwo$ model can only contain $\log q$ and $\log \ol{q}$, but not
$(\log q)^2$ or $(\log \ol{q})^2$. (Those can appear if and only if the
Hamiltonian has Jordan blocks of length three, as formula
\eqref{factoriz} makes clear.) This prediction turns out to be in
perfect agreement with experiment.

Indeed, consider the largest moduli space of gradient trajectories,
corresponding to the trajectories going from $(0:0:1)$ to
$(1:0:0)$. This moduli space is naturally identified with an open
dense subset of $\ptwo$ and it is naturally compactified by
$\ptwo$. The correlation function of evaluation observables
corresponding to an analytic function $F$ on $\ptwo$ (inserted at the
time $0$) and an analytic four-form $\omega$ on $\ptwo$ (inserted at
the time $t$, which we allow to be complex, as before) is then given
by the integral
\begin{equation}    \label{int for ptwo}
\int_{\ptwo} \omega \; \phi(q)^*(F),
\end{equation}
where $q=e^{-t}$. At first glance, it is easy to produce examples of
$\omega$ and $F$ for which this integral contains $(\log q)^2$, for
example,
\begin{equation}    \label{omega and F}
\omega = \frac{d^2 z_1 \wedge d^2 z_2}{(z_1 \zb_1 + z_2 \zb_2 + R)^3},
\qquad F = \frac{1}{(z_1 \ol{z}_1 + Q)(z_2 \zb_2 + P)},
\end{equation}
where $P,Q,R$ are positive real numbers. However, the problem is that
the function $F$ is not smooth at the point $(0:1:0)$, because in
terms of the local coordinates $w_1,w_2$ around this point it reads
$$
F = \frac{w_2 \wb_2}{(\frac{w_1 \wb_1}{w_2 \wb_2} + Q)(1 + P w_2
   \wb_2)},
$$
and so it is clear that its second derivative with respect to $w_1$
and $\wb_1$ is not continuous. Likewise, $F$ is not smooth at the
point $(1:0:0)$. To make it smooth, we would need to pull it back to
the blow-up of $\ptwo$ at the points $(0:1:0)$ and $(1:0:0)$. The
resulting manifold is a Del Pezzo surface, and the pull-back of $F$ is
a legitimate observable on it. But the instanton picture is different
on this Del Pezzo surface than on $\ptwo$, and so it is not surprising
that $(\log q)^2$ appears in the correlation functions on this Del
Pezzo surface, even though it does not appear on $\ptwo$.

The forms \eqref{omega and F} are also legitimate observables in the
$\pone \times \pone$ model, where the appearance of $(\log q)^2$ in
the correlation functions of this model is to be expected due to the
existence of Jordan blocks of length $3$ in the action of the
Hamiltonian.

But the forms \eqref{omega and F} are {\em not} legitimate observables
in the $\ptwo$ model, because $F$ is not smooth, let alone analytic.
Therefore this calculation is not a valid counterexample to our claim
(based on the analysis of the off-diagonal terms in the Hamiltonian
action) that there are no terms with $(\log q)^2$. In fact, in all
examples of correlation functions given by the integrals \eqref{int
for ptwo}, where $\omega$ and $F$ are truly analytic on $\ptwo$, that
we have computed, we have observed the appearance of $\log q$, but not
of $(\log q)^2$. Thus, the correlation functions really distinguish
between $\ptwo$ and $\pone \times \pone$ (or Del Pezzo surface)
instantons, in agreement with our predictions.

Note that we expect the same phenomenon for the ${\mathbb C}{\mathbb
P}^n$ model (with a generic $\C^\times$ action) as well.  Here we
again have a single stratum in each dimension, so the same argument as
above again applies to show that the square of the off-diagonal part
of the Hamiltonian is equal to zero. This means that the maximal size
of the Jordan blocks of the Hamiltonian (and its chiral and
anti-chiral components) is $2$ (and not $n$, as one might have
thought). Therefore we expect the appearance of $\log q$ and $\log
\ol{q}$ in the correlation functions, but not their higher powers.

The action of evaluation observables on the spaces of ``in'' and
``out'' states of the $\ptwo$ model is obtained in the same way as in
the $\pone$ model. It is not difficult to write analogues of the
identity \eqref{identity2}, corresponding to factorization over
intermediate states in the $\ptwo$ model. In the most interesting case
of instantons propagating from $(0:0:1)$ to $(1:0:0)$, the left hand
side of the identity is given by the integral \eqref{int for
ptwo}. The right hand side is the sum of three terms, each
corresponding to one of the fixed points of the $\C^\times$-action. It
is a $q,\ol{q}$-series that should converge to the left hand side of
the identity inside a sufficiently small disc on the $q$ plane.

For a general K\"ahler manifold $X$, the compactification of the
largest instanton moduli space of gradient trajectories is $X$ itself,
and we have a similar identity in which the right hand side has terms
corresponding to the fixed points of the $\C^\times$. This identity
may be interpreted as a decomposition of the delta-form on $X \times
X$ supported on the $q$-shifted diagonal, as we explained in the case
of $\pone$ in \secref{q-shifted}. There are also similar identities
for other instanton moduli spaces. It would be very interesting to
find a general proof of these identities for all analytic evaluation
observables.

We conclude our discussion of the observables with the following
remark. So far we have considered the evaluation observables of our
theory which correspond to differential forms on $X$. As we pointed
out in \secref{other observables}, there are other observables in the
theory which correspond to differential operators on $X$. In
particular, those include global holomorphic and anti-holomorphic
differential operators. The algebras of such operators should be
viewed as the precursors of the chiral (and anti-chiral) de Rham
complex of the two-dimensional sigma model that we will discuss in
Part II. Differential operators on $X$ naturally act on the spaces of
states ${\mc H}^{\inn}$ and ${\mc H}^{\out}$ of our theory (it follows
from the definition that these spaces are ${\mc D}$-modules).

These operators are particularly important for computing the
perturbation theory expansion of the correlation functions of our
theory away from the point $\la=\infty$. This is because perturbation
to finite values of $\la$ is achieved by adding to the action the term
$\la^{-1} g^{a\ol{b}} p_a p_{\ol{b}}$, which corresponds, in the
Hamiltonian formalism, to a differential operator on $X$. Therefore in
order to compute the correlation functions in the theory defined for
finite values of $\la$ by perturbation theory in $\la^{-1}$ we need to
insert these operators in the correlation functions of the theory at
$\la=\infty$. We will discuss this in \secref{pert}.

\section{Various generalizations}    \label{generalizations}

In this section we comment on possible generalizations of the results
obtained above. We begin by discussing the perturbation theory around
the point $\la=\infty$. In the previous chapters we have exhibited the
structure of the space of states of the theory in the limit
$\la=\infty$ and discussed various methods for computing the
correlation functions. It would be highly desirable to use these
results to obtain information about the theories defined at finite
values of $\la$. In particular, we consider the question of how the
space of states of the model changes when we move away from the point
$\la=\infty$, first in the case when $X=\C$ and then for
$X=\pone$. Because the Hamiltonian is non-diagonalizable we cannot
apply the standard tools of quantum mechanics and the perturbation
theory turns out to be a more challenging task. We then discuss the
computation of the correlation functions in $\la^{-1}$ perturbation
theory. We present some evidence that the quantum mechanical models
for finite values of $\la$ may be studied by using this perturbation
theory.

Next, we consider some non-supersymmetric analogues of our models. We
discuss in particular the computation of the cohomology of the
anti-chiral supercharge $\ol\pa$ (in those models in which it exists;
they are one-dimensional analogues of the $(0,2)$ supersymmetric
two-dimensional sigma models). We make contact to the GC complexes of
arbitrary vector bundles on K\"ahler manifolds and results of Witten
\cite{Witten:hol} and Wu \cite{Wu} on holomorphic Morse theory.

We also discuss briefly the generalization in which a Morse functions
is replaced by a Morse-Bott function having non-isolated critical
points. More precisely, we will consider the situation where the Morse
function comes from a $\C^\times$-action on our K\"ahler manifold $X$
with non-isolated fixed points. We show how some of the features of
the models with Morse functions change in this more general situation.

Finally, we comment on the Morse-Novikov functions, which are
multivalued analogues of the Morse functions. They are particularly
important for applications to two-dimensional and four-dimensional
models.

\ssec{Perturbation theory around the point $\la = \infty$}
\label{pert}

We have described above the structure of the spaces of ``in'' and
``out'' states in the limit $\la \to \infty$ of our quantum mechanical
model. This structure is very different from the structure observed
at finite values of $\la$. It is natural to ask whether one can
relate the two pictures by some kind of perturbation theory. This
question is important because we would like to understand our models
at finite values of $\la$ using the results obtained at $\la=\infty$,
where the theory simplifies dramatically.

Note that here we have to deal with a somewhat unfamiliar situation,
where the problem does not have a Hilbert space formulation, so we
cannot use the hermitian inner product, as is customary in the quantum
mechanics, at least in all of its textbook examples.

We start with the case of the flat space $\C$. We recall from
\secref{flat space} that in this case the space of ``in'' states is
the space of differential forms on $\C$ if $\omega>0$, and the space
of distributions supported at $0 \in \C$ if $\omega<0$. In what
follows we will restrict ourselves to the subspace of $0$-forms.

In the former case the perturbation theory is very simple and
finite. Indeed, the Hamiltonian $\wt{H}_\la$ at finite $\la$ is
obtained from the Hamiltonian ${\mc L}_v, v=z\pa_z + \zb\pa_{\zb}$ at
$\la = \infty$ by adding the term $-\frac{2}{\la} \pa_z
\pa_{\zb}$. This extra term lowers the degree of a polynomial by $1$
in $z$ and by $1$ in $\zb$. Therefore, starting with a monomial $z^n
\zb^{\nb}$, which is an eigenvector of ${\mc L}_v$, we can obtain an
eigenvector of $\wt{H}_\la$ by adding monomials of lower degrees. The
resulting polynomial is closely related to the Hermite polynomials.

The perturbation theory in the second case is more subtle. In this
case we are trying to reproduce the eigenfunctions of $\wt{H}_\la$,
which look like $e^{-{\la}|\omega|z\zb}$ times a polynomial in $z,\zb$, as
linear combinations of the derivatives of the delta-function
$\delta^{(2)}(z,\zb)$. These linear combinations are sums of
infinitely many terms, which may be thought of as asymptotic
expansions of the eigenfunctions at $\la^{-1}=0$. That there are
infinitely many terms in the expansion is easy to see from the fact
that now the additional term $-\frac{2}{\la} \pa_z \pa_{\zb}$
appearing in the Hamiltonian increases the number of derivatives of
the delta-function.

The exact formulas for the eigenstates look as follows (in the
notation of \secref{flat space}):
\begin{align*}
\wt\Psi^{\on{in}}_{n,\nb} = \frac{1}{n!\ol{n}!} \pa_z^n
\pa_{\zb}^{\nb} (2\la e^{-\la z\zb}) &\sim
\sum_{k=0}^\infty \frac{1}{n! \nb! k!} \la^{-k} \pa_z^{n+k}
\pa_{\zb}^{\nb+k} \delta^{(2)}(z,\zb), \\ & \sim
\sum_{k=0}^\infty \frac{(n+k)!  (\nb+k)!}{n! \nb! k!} \la^{-k}
|\nb+k,n+k \rangle,
\end{align*}
where
$$
|m,\ol{m} \rangle = \frac{1}{m! \ol{m}!} \pa_z^{m}
\pa_{\zb}^{\ol{m}} \delta^{(2)}(z,\zb)
$$
and the right hand side is understood as the asymptotic expansion of
the left hand side (viewed as a distribution on $\C$) at
$\la=\infty$. The Borel summation of this series gives the left hand
side.

We shall now sketch some aspects of the perturbation theory in
${\la}^{-1}$ in the $\pone$ model. We start with the following simple
remark. Suppose we want to solve the following eigenvalue problem:
$$
H \ {\Psi} = E\ {\Psi}, \qquad H = H_{0} + \frac{1}{\la} H_{1},
$$
$$
{\Psi} = {\Psi}^{[0]} + \sum_{k=1}^{\infty} \frac{1}{{\la}^{k}}
{\Psi}^{[k]}, \qquad
E = E^{[0]} + \sum_{k=1}^{\infty} \frac{1}{\la^{k}} E^{[k]}.
$$
Then we have the following simple relations:
\begin{align} \notag
H_{0} \ {\Psi}^{[0]} &= E^{[0]} \ {\Psi}^{[0]} \\ \label{eq:perthth} (
H_{1} - E^{[1]} ) \ {\Psi}^{[0]} &= - ( H_{0} - E^{[0]} )\
{\Psi}^{[1]} \\ \notag ( - E^{[2]} ) \ {\Psi}^{[0]} + ( H_{1} -
E^{[1]} ) {\Psi}^{[1]} &= - ( H_{0} - E^{[0]} )\ {\Psi}^{[2]}, \\
\notag & . \; . \; .
\end{align}
In our case we have the additional subtlety of "almost" degenerate
perturbation theory.

Consider the Hamiltonian acting on functions, i.e. on zero-forms.  To
simplify our notation, we will write $|n,\nb\rangle_\infty$ and
$|n,\nb\rangle_{\C_0}$ for $|n,\nb,0,0\rangle_\infty$ and
$|n,\nb,0,0\rangle_{\C_0}$. We have
\begin{align} \label{H0}
H_{0} &= z {\p}_{z} + {\zb} {\p}_{\zb} = - ( w {\p}_{w} + {\wb}
{\p}_{\wb} ), \\ \label{H1}
H_{1} &= - 2 ( 1 + z{\zb})^{2} {\p}_{z} {\p}_{\zb} =
- 2 ( 1 + w{\wb})^{2} {\p}_{w} {\p}_{\wb}.
\end{align}
The subspace ${\CH}_{\infty}^{\on{in}}$ of the space of states is
preserved by $H_0$ and $H_{1}$:
\begin{align*}
H_0 |n,\nb\rangle_\infty &= (n+\nb+2) |n,\nb\rangle_\infty, \\ H_1
|n,\nb\rangle_\infty &= - 2\left( n+1\right)\left(\nb +1 \right)
\left( |n-1,\nb-1 \rangle_\infty + 2 |n,\nb,\rangle_\infty +
|n+1,\nb+1\rangle_\infty \right).
\end{align*}
The equations \Ref{eq:perthth} can be solved explicitly:
\begin{align*}
{\Psi}^{[0]} & = |n, \ol{n}\rangle_\infty =
\frac{1}{n!\ol{n}!}{\p}_{w}^{n} {\p}^{\ol{n}}_{\wb} {\delta}^{(2)}
\left( w, {\bar w} \right) \ , \\ {\Psi}^{[1]} & = \left(
n+1\right)\left( \nb +1 \right) \left( |n+1, \ol{n}+1\rangle_\infty -
|n-1, \ol{n}-1\rangle_\infty \right) \ , \\ {\Psi}^{[2]} & =
\frac{1}{2}\left(n+1\right)\left(\nb +1 \right) \left[ ( n +2) (
\ol{n}+2) \left( |n+2, \ol{n}+2 \rangle_\infty + 4 |n+1,
\ol{n}+1\rangle_\infty \right) \right. \\ & \qquad\qquad \qquad
\left. + n \ol{n} \left( |n-2, \ol{n}-2\rangle_\infty + 4 |n-1,
\ol{n}-1\rangle_\infty \right) \right] \\
\end{align*}
and
\begin{align*}
& \ E^{[0]}_{\infty} = n + \ol{n} +2 \\ 
& \ E^{[1]}_{\infty} = - 4 \left(n+1\right)\left(\nb +1 \right) \\
& \ E^{[2]}_{\infty} = 4 \left(n+1\right)\left(\nb +1 \right)  ( n +
\ol{n}+2 ) \\
\end{align*}
We conjecture that
$$
E^{[k]}_{\infty} = ( E^{[0]} )^{k+1} O\left( 1 \right).
$$
This would imply that the radius of convergence of the corresponding
perturbative expansion is approximately $\la^{-1} < (E^{[0]})^{-1}$,
which means that we understand well the eigenfunctions whose
eigenvalues are less than $\la$. This agrees with the qualitative
picture suggested by the semi-classical analysis.

The determination of the perturbation series for the ``in'' states
that belong to the subspace ${\CH}_{{\C}_{0}}^{\on{in}}$ is more
complicated because of the Jordan block structure of the
Hamiltonian. Indeed, according to the calculations of \secref{action
ham}, almost all of the states $|n, \ol{n}\rangle_{\C_0}$ in
${\CH}_{{\C}_{0}}^{\on{in}}$ are generalized eigenvectors of the
Hamiltonian:
\begin{equation}
H_{0} |n, \ol{n}\rangle_{\C_0} = ( n + \ol{n} ) |n,
\ol{n}\rangle_{\C_0} - 2\pi |n-1, \ol{n}-1\rangle_{\infty},
\label{eq:hzerozero}
\end{equation}
as we have seen in the previous section (the exception is the states
$|n,0\rangle_{\C_0}$ and $|0,n\rangle_{\C_0}$). We also find the
following formula for the action of $H_1$:
\begin{align*}
H_1 |n, \ol{n}\rangle_{\C_0} &= - 2n\nb \left( |n-1,\nb-1
\rangle_{\C_0} + 2 |n,\nb\rangle_{\C_0} + |n+1,\nb+1\rangle_{\C_0}
\right) \\ &+ 4\pi (n+\nb) \left( |n-1,\nb-1 \rangle_\infty +
2 |n,\nb\rangle_\infty + |n+1,\nb+1\rangle_\infty \right).
\end{align*}

The corresponding perturbation theory is unusual, because normally one
considers hermitean Hamiltonians which cannot have Jordan blocks. The
first question is whether the degeneracy of the eigenvalues is removed
and the Jordan block structure is broken in the $\la^{-1}$
perturbation theory. At first glance, it appears that the degeneracy
and the Jordan block structure should remain, because we know that the
difference between the two eigenvalues for finite $\la$ is of the
order $e^{-\la}$, which appears to be out of reach of the perturbation
theory. However, here we could in principle obtain the asymptotic
expansion of this difference. It would be very interesting to analyze
this perturbative expansion explicitly.

\ssec{Perturbative expansion for correlation functions}

It is important to understand to what extent the correlation functions
of the quantum mechanical models at finite values of $\la$ may be
reconstructed from the correlation functions at $\la=\infty$ by
perturbation theory. Here we consider the simplest non-trivial
example, which suggests that this may indeed be done successfully.

Let $X=\pone$ and consider the two-point functions of the form
$_\infty\langle \wh\omega(t_1) \wh{F}(t_2) \rangle_0$ as in
\secref{their factorization}. According to formula \eqref{cor 2}, the
value of this correlation function at $\la=\infty$ is equal to the
integral $\int_{\pone} \omega \; \phi(e^{-t})^*(F)$ and to the matrix
element
$$
_{\C_\infty} \covac \wh\omega e^{-tH_0} \wh{F} \vac_{\C_0}.
$$
For finite values of $\la$ this correlation function is given by the
formula
$$
_{\C_\infty} \covac \wh\omega e^{-t(H_0+\la^{-1}H_1)} \wh{F}
\vac_{\C_0},
$$
where $H_0$ and $H_1$ (acting on functions) are given by formulas
\eqref{H0} and \eqref{H1}. Now we use the expansion formula
\begin{align*}
_{\C_\infty} \covac \wh\omega e^{-t(H_0+\la^{-1}H_1)} \wh{F}
\vac_{\C_0} &= {}_{\C_\infty} \covac \wh\omega e^{-tH_0} \wh{F}
\vac_{\C_0} \\ &- \la^{-1} \int_0^t ds \; {}_{\C_\infty} \covac
\wh\omega e^{-sH_0} H_1 e^{-(t-s)H_0} \wh{F} \vac_{\C_0} + \ldots
\end{align*}
Together with formulas for $\wh{F} \vac_{\C_0}$ and $_{\C_\infty}
\covac \wh\omega$ found in \secref{their factorization} and the
formulas for the action of $H_0$ and $H_1$ on $\CH^{\inn}_{\C_0}$
found in the previous section, this gives us an explicit perturbative
$\la^{-1}$-expansion for the two-point correlation function
$_\infty\langle \wh\omega(t_1) \wh{F}(t_2) \rangle_0$.

The same analysis may be applied to $n$-point correlation functions of
evaluation observables. We find that each term of the corresponding
$\la^{-1}$-expansion is given by a finite integral of a matrix element
of these operators acting on the space of states at
$\la=\infty$. Thus, in principle all of the corresponding $\la^{-1}$
perturbation series are computable. It would be interesting to relate
these perturbation series to the actual correlation functions in the
theories at finite values of $\la$ computed by other methods.

\ssec{Comments on the non-supersymmetric case}    \label{nonSUSY}

Up to now we have considered the limit $\la = \infty$ of the
supersymmetric model of quantum mechanics defined by the action
\eqref{fourth action}. There are also analogous non-supersymmetric
models, and many of our results may be applied to those models as
well.

The simplest way to break supersymmetry is to consider fermions taking
values in vector bundles on $X$ that are different from the tangent
and cotangent bundles. Since we have assumed $X$ to be K\"ahler, we
have chiral fermions: $\psi^a, \pi_a$, taking values in the
holomorphic cotangent and tangent bundles on $X$, respectively, and
anti-chiral fermions, $\psi^{\ol{a}}, \pi_{\ol{a}}$, taking values in
the holomorphic cotangent and tangent bundles on $X$, respectively. We
may now stipulate that $\psi^a, \pi_a$ take values in vector bundles
$\CE$ and $\CE^*$, respectively, whereas $\psi^{\ol{a}}, \pi_{\ol{a}}$
take values in vector bundles $\ol\CE$ and $\ol\CE^*$,
respectively. The vector bundle $\ol{\CE}$ need not be complex
conjugate to $\CE$, thus allowing the possibility of ``heterotic''
quantum mechanical models, which are the precursors of the sigma
models appearing in heterotic string.

The bosonic part of the action remains the same as before:
$$
-i \int_I \left( p_a \left( \frac{dX^a}{dt} - v^a \right) +
p_{\ol{a}} \left(\frac{dX^{\ol{a}}}{dt} - \ol{v^a} \right) + \right)
dt,
$$
where $\xi = v^a \pa_{X^a}$ is a holomorphic vector field. We will
again assume that $\xi+\ol\xi$ is the gradient vector field of a Morse
function $f$ on $X$ and that $\xi$ comes from a holomorphic
$\C^\times$-action on $X$.

To write down the fermionic part of the action, we need to assume in
addition that the $\C^\times$-action on $X$ may be lifted to $\CE$ and
$\ol\CE$, i.e., that $\CE$ and $\ol\CE$ are $\C^\times$-equivariant
vector bundles on $X$. Then if we choose local trivialization of $\CE$
by local sections $\phi^i$ and a local trivialization of $\ol\CE$ by
local sections $\phi^{\ol{i}}$, the generator of the
one-dimensional Lie algebra of $\C^\times$ will act by the formula
$$
\phi^i \mapsto M_j^i \phi^j, \qquad \phi^{\ol{i}} \mapsto
\ol{M}_{\ol{j}}^{\ol{i}} \phi^{\ol{j}}.
$$
For example, in the case when $\CE$ and $\ol\CE$ are the holomorphic
and anti-holomorphic cotangent bundles of $X$, we have bases of sections
$dx^a$ and $dx^{\ol{a}}$, and so $M_b^a = \frac{\pa v^a}{\pa X^b}$,
$\ol{M}_{\ol{b}}^{\ol{a}} = \frac{\pa \ol{v^a}}{\pa X^{\ol{b}}}$. Now
the fermionic part of the action is
$$
i \int_I \left( \pi_i \left( \frac{D \psi^i}{Dt} - M_j^i \psi^j
\right) - \pi_{\ol{i}} \left( \frac{\ol{D}\psi^{\ol{i}}}{\ol{D}t} -
   \ol{M}_{\ol{j}}^{\ol{i}} \psi^{\ol{j}} \right) \right) dt.
$$
Here $D/Dt$ and $\ol{D}/\ol{D}t$ are the covariant derivatives
corresponding to chosen connections on $\CE$ and $\ol\CE$, which may
however be absorbed into the momenta $p_a$ and $p_{\ol{a}}$ in the
same way as in the supersymmetric model (see \secref{pi and grad}).

The definition of the corresponding path integral for general vector
bundles $\CE$ and $\ol\CE$ requires special care because the space of
fields does not carry a canonical integration measure as in the
supersymmetric case. We will discuss this question in Part III of this
article. Here we will only point out that our results on the
Hamiltonian description of the supersymmetric model have obvious
generalizations to the non-supersymmetric case.

We again have spaces of ``in'' and ``out'' states, $\CH^{\inn}$ and
$\CH^{\out}$. The space $\CH^{\inn}$ is isomorphic to the direct sum
of spaces $\CH^{\inn}_\al$ labeled by the fixed points $x_\al$ of the
$\C^\times$-action, as in the supersymmetric case. Each
$\CH^{\inn}_\al$ exhibits holomorphic factorization: $\CH^{\inn}_\al =
\CF^{\inn}_\al \otimes \ol\CF^{\inn}_\al$, where
$$
\CF^{\inn}_\al = H^{n-n_\al}_{X_\al}(\wedge^\bullet\CE), \qquad
\ol\CF^{\inn}_\al = H^{n-n_\al}_{X_\al}(\wedge^\bullet\ol\CE)
$$
(compare with formula \eqref{is local coh}). The Hamiltonian is the
vector field $v = \xi + \ol\xi$. The action of $\xi$ and $\ol\xi$ is
given by formulas similar to \eqref{Lxi} and \eqref{Lolxi}, in which
the GC operators are $\delta^{\mc E}_{\al\beta}: \CF^{\inn}_\al \to
\CF^{\inn}_\beta$ and $\ol\delta^{\ol{\mc E}}_{\al\beta}:
\ol\CF^{\inn}_\al \to \ol\CF^{\inn}_\beta$ for $X_\al \succ X_\beta$.

The space of ``out'' states has a similar structure, with the
ascending manifolds $X_\al$ replaced by the descending manifolds
$X^\al$. In addition, we need to replace $\CE$ by $\Omega^{\on{top}}
\otimes \CE^*$ and $\ol\CE$ by $\ol\Omega^{\on{top}} \otimes
\ol\CE^*$, where $\Omega^{\on{top}}$ and $\ol\Omega^{\on{top}}$ are
the line bundles of holomorphic and anti-holomorphic top forms,
respectively.

\subsection{Cohomology of the supercharge in ``half-supersymmetric''
  models}

An interesting special class of models arises if we let $\ol\CE$ be
the anti-holomorphic cotangent bundle on $X$, as in the supersymmetric
model. Then we retain the anti-chiral supercharge $\ol\pa$. The
corresponding ``half-supersymmetric'' models may therefore be viewed
as quantum mechanical analogues of the $(0,2)$ supersymmetric sigma
models (just like the fully supersymmetric model may be viewed as an
analogue of the $(2,2)$ supersymmetric sigma model). In such models it
is interesting to compute the cohomology of the supercharge $\ol\pa$,
which may be viewed as a ``baby version'' of the cohomology of the
right moving supercharge in the $(0,2)$ sigma models. This cohomology
has been studied in \cite{Witten:cdo}, where it was shown that it is
closely related to the chiral differential operators.

In the ``half-supersymmetric'' quantum mechanical model the
supercharge $\ol\pa$ is given by the same formula \eqref{olpa} as in
the supersymmetric case, except that we need to replace the GC
operators $\delta_{\al\beta}$ and $\ol\delta_{\al\beta}$ by
$\delta_{\al\beta}^{\CE}$ and $\ol\delta_{\al\beta}^{\ol\CE}$,
respectively.

The argument of \secref{coh of Q} for the computation of the
$\ol\pa$-cohomology in the supersymmetric case carries over verbatim
to this case, and we find that the first term of the corresponding
spectral sequence is just the GC complex
$C^\bullet(\wedge^\bullet\CE)$ of the vector bundle $\wedge^\bullet
\CE$ (see \secref{coh of Q} for the definition of this
complex). Actually, the grading on the exterior algebra
$\wedge^\bullet\CE$ is preserved by the differential, so the GC
complex decomposes into a direct sum of its subcomplexes
$C^\bullet(\wedge^p\CE), p=0,\ldots,\dim \CE$. This gives a second
grading on the cohomology of $\ol\pa$. This cohomology is therefore
equal to the direct sum of the (Dolbeault) cohomologies of the sheaves
of holomorphic sections of the vector bundles $\wedge^p\CE, p \geq 0$:
$$
H^i_{\ol\pa} = \bigoplus_{p\geq 0} H^i(X,\wedge^p\CE).
$$
Thus, we find an effective way for computing the cohomology of the
anti-chiral supercharge in the ``half-supersymmetric'' quantum
mechanical models. The result is that this cohomology is nothing but
the Dolbeault cohomology of the bundle $\wedge^\bullet\CE$, where the
chiral fermions take values.

This result is closely related to the work of E. Witten on the
holomorphic Morse theory. In \cite{Witten:hol} Witten has shown how to
adopt his approach to Morse theory from \cite{W:morse} (discussed in
\secref{recollections morse}), which allows one to compute the de Rham
cohomology of $X$ in terms of an instanton complex associated to a
Morse function, to the computation of the cohomology of the sheaf of
holomorphic sections of a $\C^\times$-equivariant vector bundle $\CE$
on $X$ (in other words, computing Dolbeault cohomology instead of de
Rham cohomology). It is assumed that $X$ is a K\"ahler manifold
equipped with a Morse function $f$ whose gradient satisfies the
conditions listed above. Witten has shown that this cohomology may be
obtained as the cohomology of a certain complex. The groups of this
complex are infinite-dimensional, but they are graded by the action of
$\C^\times$ and the corresponding graded components are
finite-dimensional. Witten has computed in \cite{Witten:hol} the
characters of the groups of this complex. This enabled him to write
down ``holomorphic Morse inequalities'' giving estimates on the
Dolbeault cohomology of $\CE$ in the same way as the usual Morse
inequalities give us estimates on the de Rham cohomology of $X$.

It was subsequently shown by S. Wu in \cite{Wu} that the characters
that Witten had computed were precisely the characters of the groups
of the GC complex $C^\bullet(\CE)$ of $\CE$. Therefore it was
suggested in \cite{Wu} that Witten's holomorphic instanton complex
should be interpreted as the GC complex of $\CE$. This is in agreement
with Witten's result, because we know that the cohomology of the GC
complex is equal to the Dolbeault cohomology of $\CE$. But the
connection of this complex to quantum mechanics still remained a
mystery.

But now we have found a natural explanation of the connection between
the GC complex and quantum mechanics. Namely, we have shown that the
GC complexes naturally appear in the framework of the
``half-supersymmetric'' quantum mechanical models in the limit
$\la=\infty$. We have found that the Dolbeault cohomology of a
holomorphic $\C^\times$-equivariant line bundle on a compact K\"ahler
manifold coincides with the cohomology of the supercharge $\ol\pa$ on
the space of ``in'' states of a particular model of this type: with
left fermions living in $\CE$ and right fermions living in the
anti-holomorphic cotangent bundle. Moreover, we have shown that the
computation of the cohomology of $\ol\pa$ naturally gives us the GC
complex of $\CE$ (and even of $\wedge^\bullet \CE$). This explains the
connection between ``holomorphic Morse theory'' and quantum mechanics.

We want to point out a particularly interesting class of
``half-supersymmetric'' models of this type. They are associated to a
flag variety $G/B$, where $G$ is a complex simple Lie group and $B$ is
its Borel subgroup. This flag variety has a natural Morse function:
the hamiltonian of the vector field corresponding to a generic element
of a maximal compact torus contained in $B$. Its critical points are
parameterized by the Weyl group $W$ of $G$. The corresponding
ascending manifolds are called the Schubert cells, which we denote by
$X_w, w \in W$. Its complex dimension is equal to the length of $w$,
denoted by $\ell(w)$. We can choose the Morse function in such a way
that they are the $B$-orbits on $G/B$. Note that if $G=SL_2$, then
$G/B \simeq \pone$, and the corresponding Morse function is the one we
have studied extensively in the earlier sections.

Suppose that $\CE$ is a line bundle on $G/B$. Then it corresponds to
an integral weight $\mu$ of the group $G$. Let us suppose that $\mu$
is dominant, and so can be realized as the highest weight of an
irreducible finite-dimensional representation $V_\mu$ of $G$. We
denote the corresponding line bundle by ${\mc E}_\mu$. The GC complex
of this line bundle is studied in detail in \cite{Kempf} (see also
\cite{FF:si,Wu}), where it is shown that this complex coincides with
the dual of the so-called Bernstein-Gelfand-Gelfand (BGG) resolution
of $V_\mu$.  The $i$th group of the
GC complex $C^\bullet(\CE_\mu)$ is equal to
$$
C^i(\CE_\mu) = \bigoplus_{\ell(w)=i} H^{\dim
   G/B-\ell(w)}(G/B,\CE_\mu).
$$

The group $G$ does not act on the GC complex $C^\bullet(\CE_\mu)$, but
the Lie algebra $\g$ does. Under this action
$$
H^{\dim G/B-\ell(w)}(G/B,\CE_\mu) \simeq M^*_{w(\mu+\rho)-\rho},
$$
the contragredient Verma module over $\g$ of highest weight
$w(\mu+\rho)-\rho$, where $\rho$ is the half-sum of positive roots of
$\g$. Thus, the $i$th group of the GC complex is
$$
C^i(\CE_\mu) = \bigoplus_{\ell(w)=i} M^*_{w(\mu+\rho)-\rho},
$$
and its cohomology is equal to $H^\bullet(G/B,\CE_\mu)$. According to
the Borel-Weil-Bott theorem, $H^0(G/B,\CE_\mu) \simeq V_\mu$ and
$H^i(G/B,\CE_\mu) = 0$, for $i>0$. Therefore the GC complex
$C^\bullet(\CE_\mu)$ is a {\em resolution} of the irreducible
representation $V_\mu$. It is dual to the BGG resolution, which is
well-known in representation theory \cite{Kempf}.\footnote{If $\mu$ is
not dominant, then the cohomology is either zero or it occurs in a
positive cohomological dimension; in this case the $\g$-modules
appearing in the complex are the twisted Verma modules, see
\cite{FF:si}}

According to the above discussion, this BGG resolution is naturally
realized in the context of a ``half-supersymmetric'' model on $G/B$ in
which left fermions take values in the line bundle $\CE_\mu$ and its
dual. The cohomology of the supercharge $\ol\pa$ in this model is
therefore equal to
$$
H^\bullet(G/B,\wedge^\bullet \CE_\mu) = H^0(G/B,{\mc O}) \oplus
H^0(G/B,\CE_\mu) \simeq  \C \oplus V_\mu.
$$
Thus, we realize irreducible representations of simple Lie groups as
$\ol\pa$-cohomologies of ``half-supersymmetric'' models on the flag
variety.

These results have interesting analogues in $(0,2)$ supersymmetric
two-dimensional sigma models, as we will see in Parts II and III of
this article.

\ssec{Comments on non-isolated critical points}

Up to now we have considered a Morse function $f$ on a K\"ahler
manifold $X$ of dimension $n$ and the corresponding gradient vector
field $v = \nabla f$ which, as we have assumed, decomposes into the
sum $\xi + \ol\xi$ of a holomorphic and anti-holomorphic vector fields
generating a $\C^\times$-action on $X$. The critical points of $f$ are
the fixed points of this $\C^\times$-action. The assumption that $f$
is a Morse function means that these points are isolated and
non-degenerate. In this section we discuss briefly what happens if we
are in a situation when the fixed points are not isolated and $f$ is a
Morse-Bott function.

Let $C_\al, \al \in A$, be the components of the fixed point set of
the $\C^\times$-action on $X$ (under our old assumptions, each $C_\al$
consisted of a single point). According to the results of
\cite{BB,CS}, in this case $X$ still has decompositions
\eqref{stratification}, with $X_\al$ and $X^\al$ defined in the same
way as before, by formulas \eqref{Xal1} and \eqref{Xal2}. However, in
this case each $X_\al$ is a $\C^\times$-equivariant holomorphic
fibration over $C_\al$, whose fibers are isomorphic to $\C^{n_\al}$,
where $n_\al$ is the number of positive eigenvalues of the Hessian of
$f$ at the points of $C_\al$. Moreover, locally over $C_\al$ the
bundle $X_\al$ is isomorphic to the subbundle of the normal bundle to
$C_\al \subset X$ spanned by the eigenspaces of the Hessian of the
function $f$ with positive eigenvalues. Likewise, $X^\al$ is also a
$\C^\times$-equivariant holomorphic bundle over $C_\al$ with fibers
isomorphic to $\C^{n-n_\al-\dim C_\al}$. Locally over $C_\al$ the
bundle $X^\al$ is isomorphic to the subbundle of the normal bundle to
$C_\al \subset X$ spanned by the eigenspaces of the Hessian of the
function $f$ with negative eigenvalues.

Consider, for example, the case of $X=\ptwo$ with the $\C^\times$
action $(z_1:z_2:z_3) \mapsto (qz_1:z_2:z_3)$, corresponding to the
vector field $v = z_1 \pa_{z_1} + \zb_1 \pa_{\zb_1}$. Then the fixed
point set has two components: the point $C_1=(1:0:0)$ and the
one-dimensional component $C_2 = \{ (0:z_2:z_3) \}$ isomorphic to
$\pone$. The corresponding strata $X_1$ and $X_2$ are the point
$(1:0:0)$ and its complement, respectively. Note that $X_2$ is a line
bundle over $\pone$ isomorphic to ${\mc O}(1)$, which is also
isomorphic to the normal bundle of $C_2 \subset \ptwo$. The strata
$X^1$ and $X^2$ are the plane $\{ (1:u_1:u_2) \}$ and $C_2 = \pone$,
respectively.

The description of the spaces of ``in'' and ``out'' states of this
model is similar to the one obtained previously in the Morse function
case. Namely, $\CH^{\inn}$ is isomorphic to the direct sum of the
spaces $\CH^{\inn}_\al, \al \in A$. Roughly speaking, each space
$\CH^{\inn}_\al$ is the space of $L_2$ differential forms on $C_\al$
extended in two ways: by polynomial differential forms in the bundle
directions of $X_\al$, and then by polynomials in the derivatives in
the transversal directions to $X_\al$ in $X$.

The ground states, on which the Hamiltonian ${\mc L}_v$ acts by zero,
correspond to differential forms on $C_\al$. Given such a form
$\omega_\al$, let $\wt\omega_\al$ be its pull-back to $X_\al$ under
the projection $X_\al \to C_\al$. Then $\wt\omega_\al$ defines a
``delta-like'' distribution supported on $X_\al$, whose value on
$\eta \in \Omega^\bullet(X)$ is equal to
$$
\int_{X_\al} \wt\omega_\al \wedge \eta|_{X_\al}
$$
(compare with formula \eqref{Deltaal}). We will use the same notation
$\wt\omega_\al$ for these distributions. While these are the ground
states of the model at $\la=\infty$, only those of them which
correspond to harmonic differential forms $\omega_\al \in
\Omega^\bullet(C_\al), \al \in A$, may be deformed to ground states
for finite values of $\la$.

Other elements of $\CH_\al$ are distributions obtained by applying to
the distributions $\wt\omega_\al$ Lie derivatives in the transversal
directions to $X_\al$ as well as multiplying them by differential
forms on $X_\al$ that are polynomial along the fibers of the
projection $X_\al \to C_\al$ (compare with formula \eqref{states in
Hal}). The definition of these distributions requires a regularization
similar to the one we used in the case of isolated critical
points. Because of this regularization, we obtain non-trivial
extensions between different spaces $\CH^{\inn}_\al$, and the action
of the Hamiltonian is not diagonalizable. However, the formula for the
Hamiltonian is more complicated than in the case of isolated fixed
points. Another difference with the case of isolated fixed points is
that we observe holomorphic factorization only in the fiber directions
of the maps $X_\al \to C_\al$, but not along the manifolds $C_\al$
themselves.

\ssec{Morse-Novikov functions}

In the analysis we have performed so far we worked with the
single-valued Morse functions $f$. Morse theory has a generalization
for non-simply connected manifolds, namely, the Morse-Novikov theory,
in which $f$ is multivalued and only its differential is
well-defined. However, according to the results of \cite{Frankel},
under the assumptions that we have made: that $X$ is a compact
K\"ahler manifold $X$ with a holomorphic vector field $\xi$ such that
its zeros are isolated and the set of zeros is non-empty, we have
$H_1(X,\Z)=0$. Therefore all closed one-forms on $X$ are exact. Let
$\beta$ be the one-form obtained by contraction of the vector field $v
= \xi + \ol\xi$ and the metric on $X$. Then $\beta = df$, where the
function $f$ is a Morse function on $X$ whose gradient vector field is
equal to $v$, and whose critical points are the zeros of $v$ and of
$\xi$. Therefore there is no need to consider the case of multivalued,
or Morse-Novikov, functions.\footnote{such functions exist if we allow
$X$ to be a real manifold, for example, a circle} However, for {\em
infinite-dimensional} K\"ahler manifolds, such as the loop space $LX$,
such functions do arise, and in fact it is necessary to study them in
order to understand two-dimensional sigma models. We will study this
in detail in Part II of this article.

\end{document}